\documentclass[a4paper,12pt]{report}

\usepackage[T1]{fontenc}
\usepackage{amsmath}      
\usepackage{amsfonts}     
\usepackage{amssymb}      

\usepackage{palatino}
\usepackage{fancyhdr}
\usepackage{epigraph}     

\usepackage{graphicx}     
\usepackage{float}        
\usepackage{caption}

\usepackage{breakcites}
\usepackage{booktabs}     
\usepackage{adjustbox}    
\usepackage{enumitem}

\usepackage{geometry}

\usepackage{xcolor}
\definecolor{cmyk_orange}{cmyk}{0, 0.5, 1, 0}
\definecolor{cmyk_blue}{cmyk}{1, 0.5, 0, 0} 

\graphicspath{figures/}

\setlist[itemize]{nosep}
\setlist[enumerate]{nosep}

\begin{document}
\pagenumbering{roman}

\thispagestyle{empty} 
\vspace*{0.3cm}
\begin{center} 
	{\huge \bf Occupational Tasks, Automation, \\and Economic Growth}\\
    \vspace{3mm}
    {\LARGE A modeling and simulation approach}
	\vspace{15mm}

    {\LARGE Georgios A. Tritsaris} \\ 
     \large 15 Jul 2025 

\end{center}
\clearpage

\thispagestyle{empty} 
\textbf{Abstract} \\
The Fourth Industrial Revolution commonly refers to the accelerating technological transformation that has been taking place in the 21st century, merging the physical with the virtual. Economic growth theories such as Romer-style models, which treat the accumulation of knowledge and its effect on production endogenously, remain relevant, yet they have been evolving to explain how the current wave of advancements in automation and artificial intelligence (AI) technology will affect productivity and different occupations. 

\qquad The work contributes to current economic discourse by developing a task-based framework that endogenously integrates knowledge accumulation with the structure of production, and  further incorporates frictions to describe how technological lock-in and the burden of knowledge generation and validation may bias economic trajectories. The interaction between production (or automation) and growth (or knowledge accumulation) is also described explicitly.

\qquad To uncover overarching patterns in how automation and AI shape economic outcomes, I rely on computer simulation of the developed analytical model and machine learning techniques for data analysis. The effect of the model's structural parameters on key variables such as the production output, wages, and labor shares of output is quantified, and based on the findings I briefly discuss suitable intervention strategies. A major, intuitive result is that wages and labor shares are not directly linked, instead they can, to a significant extent, be influenced independently through distinct policy levers. Generally, labor share depends sensitively on capital-labor ratio, while wages respond positively to larger knowledge stocks. 

\vspace{80mm}

\clearpage

\tableofcontents
\cleardoublepage

\pagenumbering{arabic}    
\setcounter{page}{1}  
\chapter{Introduction}\label{intro}
The Fourth Industrial Revolution commonly refers to the technological transformation that has been taking place in the 21st century, largely driven presently by advances in such diverse domains as automation, artificial intelligence (AI), biotechnology, and ubiquitous computing \cite{schwab_fourth_2024}. Unlike previous waves of accelerated industrialization, the current transformation appears capable of exacerbating divergences between capital and labor, both in their allocation to production tasks and their relative claims on economic output. 

\qquad These developments fuel debates at international, regional, and national levels, as well as within industry-specific forums. For example, in recent publications, the IMF has pointed to possible adverse effects of AI on employment \cite{cazzaniga_gen-ai_2024}, the OECD has emphasized an anticipatory approach to the governance of emerging technology \cite{oecd_framework_2024}, and the ILO highlighted a combination of profit sharing, capital taxation and a reduction in working time to fairly distribute the benefits of increased productivity \cite{ilo_economics_2018}. Moreover, a collaboration of leading universities, firms, and intergovernmental bodies proposed a systematic framework for mapping AI specializations in goods and services \cite{mishra_ai_2023}.

\qquad Economic growth theory that treats the accumulation of knowledge and its effect on production and labor endogenously remains relevant, yet a more recent strand has pivoted to conceptualize occupations as bundles of tasks in an attempt to illuminate how technological progress might affect different occupations in distinct ways \cite{acemoglu_simple_2025}. The complex ways new technologies interact with workers and with each other (e.g., manufacturing robotics increasingly controlled by AI) have made it difficult to identify key economic drivers of current developments with confidence: technologies, labor practices, and institutional norms co-evolve at the capital-labor interface within a broader socio-technical system in a way that defies simple causal explanations. 

\qquad Thus, important research questions remain, for instance: \textit{``under which conditions capital substitutes for labor instead of complementing it, what are the implications for wages and labor shares, and what are possible policy interventions?''} In this work, I attempt to tackle these questions by developing a task-based framework that endogenously integrates knowledge accumulation with the structure of production. The model further incorporates frictions to describe how technological lock-in and the burden of knowledge generation and validation may bias economic decisions. This contribution is intended to be novel as well as pedagogical.

\qquad While empirical studies typically aim at accurate insights into specific economies, the ability to generalize findings is limited by the peculiarities of each economic system. Computer simulation by contrast allows for a systematic exploration of wide parameter spaces and policy regimes, and offers a reliable baseline for empirical discussions. To identify overarching patterns in how automation and AI shape economic outcomes, I rely on forward numerical simulation of the developed analytical model and machine learning techniques for data analysis. The effect of the model's structural parameters on key variables such as production output, wages, and labor shares of output is quantified, and based on the findings I briefly discuss suitable intervention strategies. A key, intuitive result is that wages and labor shares are not directly linked, instead they can, to a significant extent, be influenced independently through distinct policy levers. Generally, labor share depends sensitively on capital-labor ratio, while wages respond positively to larger knowledge stocks. 

\qquad The work is structured as follows: recent developments are presented in light of the historical evolution of technology and related institutions in Chapter~\ref{ch:literature}. It is followed by a more in-depth discussion of canonical growth models in Chapter~\ref{ch:theory}, where key economic ideas are formalized and their implications briefly discussed. The two chapters together provide the necessary theoretical and conceptual foundation for the analytical models developed in Chapter~\ref{ch:modeling}. This chapter presents the various structural components that make the full model. Of particular importance are frictions (which constrain unbounded growth), and the coupling between production and growth. The model is numerically simulated in Chapter~\ref{ch:simulation}, wherein machine learning techniques are used to identify trends in the model's behavior with regards to input structural parameters. Based on the findings, simple policy interventions for influencing the capital-labor interface are briefly presented. Chapter~\ref{ch:development} takes a wider, development-oriented view to contextualize the preceding analysis. The work concludes with a brief summary of the main findings and implications in Chapter~\ref{ch:conclusions}.

\chapter{Background}\label{ch:literature}

\section{Historical Perspective}
\begin{figure}
    \centering
    \includegraphics[width=0.5\linewidth]{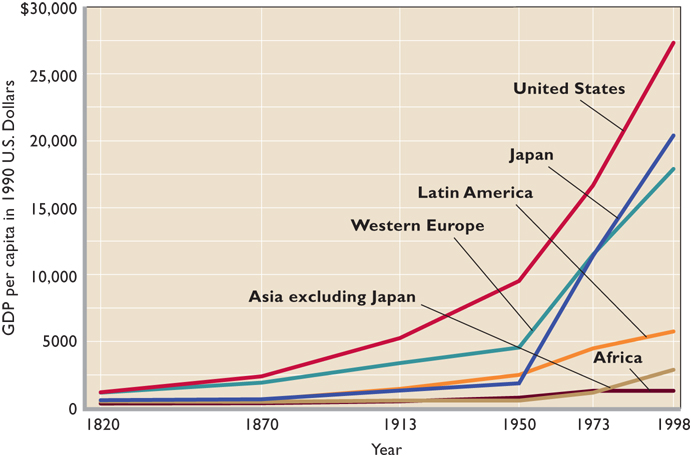}
    \caption{The ``great divergence'' in income levels around the world.}
    \label{fig:divergence}
\end{figure}

Economic development is a prominent theme in economic literature, concerned with the determinants of nations' productivity and economic growth (Figure \ref{fig:divergence}). Often through an evolutionary lens, it examines the enabling or constraining role of institutional structures, cultural influences, and geographic conditions that fundamentally shape economies. This chapter offers a historical account of technological and economic progress that will frame the rest of the discussion, based primarily on selected work in the political economy of innovation and development \cite{acemoglu_power_2024, mokyr_gifts_2005, warsh_knowledge_2007, clark_farewell_2010, north_institutions_1990}.

\paragraph{Roots of economic development.} Scholarly work on the roots of economic development attempts to provide an answer to the following question: 
\begin{center}
\textit{What are the fundamental causes of long-run economic growth?}
\end{center}
For example, in 2000, GDP per capita in the United States was more than \$34000. Compare this amount with the income per capita in a country like Mexico, \$8000, or China, \$4000.

\qquad Within this line of research, factor accumulation and innovation are viewed not as ultimate drivers but only as proximate causes of growth shaped by deeper forces. An integrative approach to understanding the fundamental determinants of economic development seeks to reconcile three major frameworks: 
\begin{enumerate}
    \item \textit{Geography.} It provides foundational conditions and constraints within which both institutions and cultures evolve. Stretching back to the Neolithic period, the path dependencies created or enabled by geographic and environmental features (e.g., disease environment) influence economic trajectories in significant and often enduring ways.
    \item \textit{Institutions.} They transform historical legacies and political power relationships into laws and regulations that directly impact economic activity. For example, disease environments not conducive to the settlement of European colonizers, favored extractive institutions in these regions. 
    \item \textit{Culture.} It permeates historical legacies and institutional logic. Values and norms influence the effectiveness of institutions, and therefore economic outcomes over time. 
\end{enumerate}

In more detail, in accordance with Olsson and Hibbs, agricultural productivity, most critical in early economic development, was heavily influenced by geographical conditions \cite{olsson_biogeography_2005}. Biogeographic endowments profoundly affected the social and economic structures that followed, since, roughly, the Neolithic revolution. 

\qquad However, this idea has been challenged by scholars such as Acemoglu in favor of the institutions hypothesis \cite{acemoglu_institutions_2004,acemoglu_institutions_2014}. As a case in point, in areas where European settlement during the colonial era was impeded by endemic diseases, extractive institutions were setup to funnel wealth back to colonizers' home countries. Settler institutions instead aimed to foster local economic development that facilitated the migration of Europeans \cite{acemoglu_colonial_2001, acemoglu_reversal_2002}. The advent of industrialization subsequently magnified preexisting deficiencies with long-term consequences for economic development.
 
\qquad The work of Alesina, Glaeser and co-authors extends the conversation beyond the role of economic and political institutions to include cultural underpinnings by examining the influence of cultural traits (e.g., individualism) on prosperity \cite{alesina_culture_2015, glaeser_institutions_2004}. 

\begin{figure}
    \centering
    \includegraphics[width=0.9\linewidth]{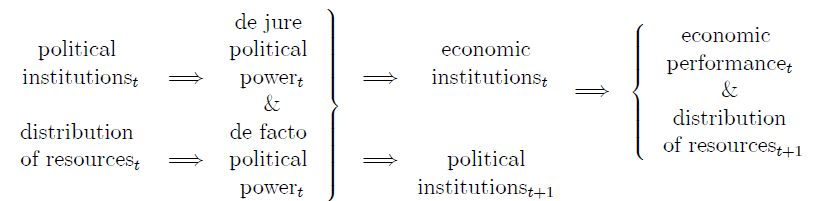}
    \caption{Endogeneity of economic outcomes and institutions \cite{acemoglu_institutions_2004}.}
    \label{fig:feedback}
\end{figure}

\qquad The discussion remains complicated as the three frameworks interact with each other. In this work, we align primarily with the institutions hypothesis: the development of strong institutions supports economic growth, while prosperous economic conditions, in turn, provide the resources and stability necessary for the development of robust institutions (Figure \ref{fig:feedback}). Consider for example a scenario in which a political elite offers incentives for entrepreneurial activity and technology development as a means for promoting economic growth. Lacking credible commitments to future policy (e.g., regarding taxation), those with political power have an incentive to ``hold up'' the entrepreneurs by raising taxes once investments are sunk, as a means for propagating their power.

\paragraph{Industrial revolution.}
Prior to the industrial revolution, most societies were primarily organized around subsistence farming and manorial economic systems. The household was the main unit of production alongside small guilds of limited specialization. These characteristics constrained significantly the scale of the economy within a Malthusian growth regime.

\qquad Knowledge was transmitted through experiential learning and apprenticeships. Literacy was generally low. Science was advancing, but out of the reach of common folk and therefore remained largely disconnected from production. The lack of closed feedback loop between knowledge generation and its application in production consequently impeded innovation. Moreover, weak property rights and state capacity to enforce them against the interests of incumbents (e.g., guilds), further obstructed the generation and diffusion of new ideas.

\qquad The advent of the \textbf{First} Industrial Revolution fueled technology's transformative impact on the organization of production and labor by mechanizing factories on an unprecedented scale, with commensurate gains in output. For instance, the introduction of \textit{Spinning Jenny} (1794) by James Hargreaves and the \textit{Power Loom }(1785) by Edmund Cartwright replaced manual weaving with mechanized textile production. James Watt's \textit{steam engine} (1769) and Stephenson’s \textit{locomotive} (1814) offered power and logistics infrastructure that further incentivized not only industrial manufacturing at large scale but also the development of thermodynamics as a scientific field.

\qquad Urbanization was accelerated, and long work hours and child labor became common. This rapid transformation of economic life in light of early \textit{laissez-faire} ideology and absence of protective measures for labor resulted in some cases to general discontent among the working class. The most characteristic case is the \textit{Luddite movement} (1811-1816), where active (and in cases destructive) resistance to real or perceived substitution of labor for capital took place.

\qquad The foundations of the \textit{digital age} were also laid during this era. Charles Babbage's \textit{Difference Engines}; and \textit{Analytical Engine} (with significant contributions by Ada Lovelace), conceptualized how machinery (or thinking machines) could be applied to ``the computation of astronomical and mathematical tables''. The formulae and data input were to be fed to the machine by punched cards, a technique that was already employed to control mechanical looms like the \textit{Jacquard loom} (1804).

\qquad The \textbf{Second} Industrial Revolution was defined by a wave of general-purpose technologies (GPT) such as electricity and the internal combustion engine, and new institutional arrangements \cite{bresnahan_general_1995}. For example, the \textit{Bessemer process} (1856) reduced the cost of steel production, which in turn enabled the construction of large-scale structures, from bridges to skyscrapers. One of the most representative innovations of this era however is Ford's \textit{moving assembly line} (1913), consolidating a trend in mass production based on the complete and consistent interchangeability of parts, which reduced the time to produce a vehicle $5-10 \times$.

\qquad As factory jobs became more specialized, routinization spurred the growth of labor unions and fostered collective bargaining movements. Labor reforms included social protection systems such as Germany's Bismarckian social insurance (1880s), minimum wage standards and limited working hours. Education reforms expanded access to schooling to prepare a workforce capable of operating within the contemporary industrial society, and firms such as Siemens and Bell Labs institutionalized R\&D as a strategic, organized function. Concurrently, a theory of knowledge spillovers was developed by the English economist Alfred Marshall who brought attention to ``\textit{thickly peopled industrial districts}''.

\qquad The implications of industrial automation continued to fuel workers' anxieties and imagination: Karel \v{C}apek's play \textit{R.U.R.} (Rossum's Universal Robots, 1920), which coined the term ``robot'', and Fritz Lang's film Metropolis (1927) remain enduring cultural representations of mechanized labor.

\qquad The \textbf{Third} Industrial Revolution pulled the previously somehow peripheral computing and information-based technologies toward a central position in contemporary societies and today's economy. The invention of the transistor at Bell Labs (1947) enabled the miniaturization of electronics, and put information technology on a trajectory of accelerated performance gains that continues today with the proliferation of AI accelerator chipsets. The emblem of the digital revolution is arguably the personal computer, which was introduced to households in the 1980s by firms like IBM and Apple. Independently, the first wide-area packet-switched network, known as ARPANET, became the foundation of yesterday's internet and today's Internet of Things (IoT). Building on the novel communication infrastructure, Tim Berners-Lee of CERN (by nature a political-scientific institution) introduced the World Wide Web in 1991. 

\qquad Computer Numerical Control (CNC) systems contributed to increased automation in industrial production, replacing human labor in routine manufacturing tasks and paved the way for more flexible (or ``lean'') paradigms in production management such as Toyota's system. Labor market outcomes were realigned too, but in a different way: the concept of \textit{skill-biased technological change} (SBTC) gained prominence, focusing attention on how the new information and communication technologies (ICT) favored  high-skilled labor and, by implication, college-educated workers \cite{katz_chapter_1999}.

\qquad Concurrently, the software sector became a new enabler of growth, to the point of commoditizing the underlying hardware platforms ---as best exemplified by Microsoft's non-exclusive license of its operating system to IBM. In the US, the National Science Foundation (NSF) contributed to the institutionalization of computing as a scientific discipline. Capital investment patterns also shifted: venture capital emerged as a financing mechanism for high-risk, high-reward innovation that continues to channel resources into digital infrastructure and platform-based business models. Economic theory caught up, and scholars like Paul Romer formalized an understanding of innovation based on the premise that ideas are non-rivalrous, partially non-excludable goods that can generate increasing returns \cite{romer_increasing_1986,arrow_economic_1972}. 

\qquad Policy remained a critical determinant of production organization and labor market outcomes. Extended deregulation, privatization, and labor market liberalization took place, prominently in the UK and the US under the Thatcher and Reagan administrations, and in most OECD economies. The reforms weakened union bargaining power, and expanded the demand of college educated workers to meet the increasing demands of service-oriented and knowledge-intensive sectors \cite{autor_skill_2003}. 

\section{Recent Developments}
As with previous stages of industrial transformation, the \textbf{Fourth} Industrial Revolution has also witnessed a wave of GPTs. Extending beyond mere automation, it incorporates even more elements of the digital revolution toward seamlessly blending physical and virtual components and processes.

\begin{figure}
    \centering
    \includegraphics[width=0.65\linewidth]{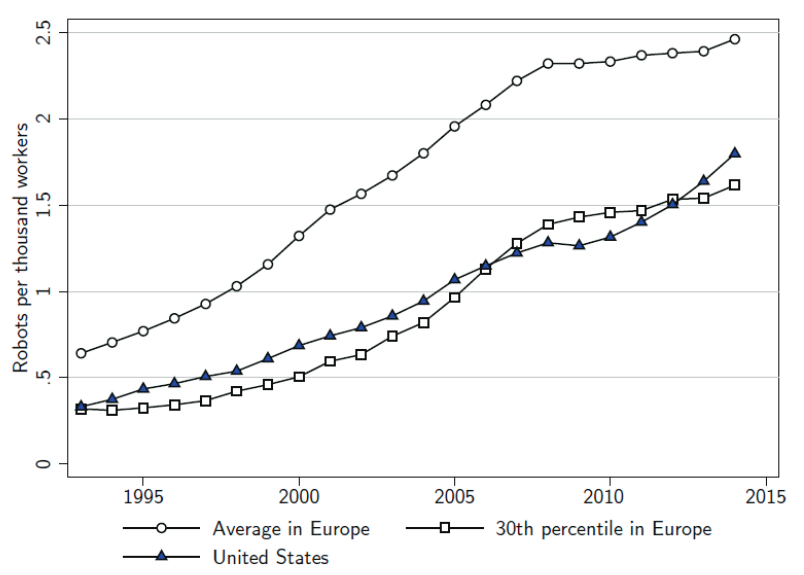}
    \caption{Industrial robots in the United States and Europe.}
    \label{fig:robots}
\end{figure}

\qquad For instance, interest in AI has been catalyzed by basic and applied research, the widespread availability of low-cost computing power, advanced semiconductor-based processing units (such as GPUs and TPUs), as well as exponential growth of digital data availability \cite{rumelhart_learning_1986,vaswani_attention_2017, oecd_blueprint_2023-1}. Applications of machine learning, currently the most prevalent paradigm of AI, expanded to include practical natural language processing, computer vision, and predictive analytics with the potential to automate complex processes beyond routine tasks \cite{mishra_ai_2023,wang_scientific_2023}. Robotics, increasingly powered by AI, have also been making significant strides in manufacturing: for example, according to data collected by the International Federation of Robotics (IFR), the number of robots per industrial worker has increased in the US almost by a factor of 7, while Europe has witness a similar increase (Figure~\ref{fig:robots}) \cite{acemoglu_robots_2020}. 

\qquad Labor markets are adjusting accordingly. The proliferation of gig economy, often mediated by non-transparent matching algorithms, obscure traditional definitions of employment, and further diminish access to labor protections. Employment insecurity in face of automation has reemerged, as automation displaces not only manual but also routine cognitive tasks \cite{frey_future_2017}. These \textit{task-specific effects} are fueling debates concerning the distributional consequences of technological change for capital and labor, and the future of work more broadly \cite{karabarbounis_global_2014}. Polarization is also evident at the level of industries as a handful of (super-)star firms lure top-tier talent away from competitors, who risk becoming unable to compete on innovation levels, with a detrimental aggregate effect on the economy \cite{firooz_automation_2025,koch_robots_2021,autor_fall_2020}.

\begin{figure}[H]
    \centering
    \includegraphics[width=0.75\linewidth]{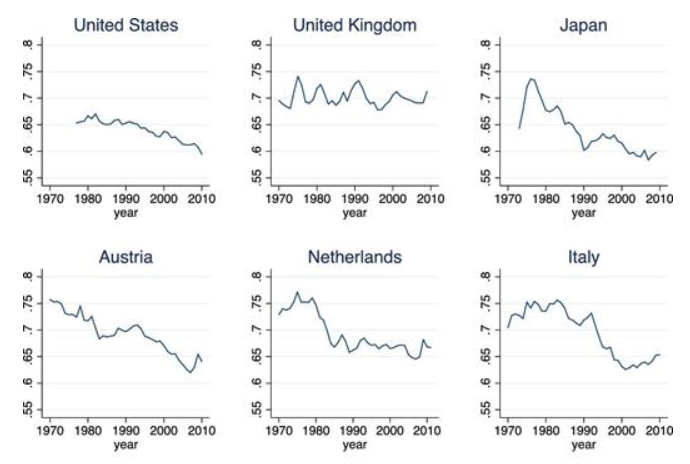}
    \caption{International comparison: labor share by country \cite{autor_fall_2020}.}
    \label{fig:laborsharedecline}
\end{figure}

\qquad The role of institutions in this phase of industrialization remains crucial yet underdeveloped, as persistent \textit{policy lags} prevent regulation from catching up in a timely manner with rapid technological change \cite{capraro_impact_2023}. An observed sharp decline in research productivity \cite{bloom_are_2020} may too necessitate targeted policy intervention to mitigate distortions. Crucially (and unsurprisingly), superstar technology firms are having a disproportionate influence over economic outcomes.

\qquad Much like in previous industrial revolutions, technological progress in the 21st century demands attention from regulators to its effects on production structures and labor markets. It has been argued that the knowledge foundation of the current wave of technologies is strictly linked to preceding technological paradigms (Table~\ref{tab:sectoral}) \cite{santarelli_automation_2023}. 

\begin{table}[H]
\small
\centering
\begin{tabular}{lll}
\toprule
\textbf{NAICS} & \textbf{Title} & \textbf{Overall (\%)} \\
\midrule
\multicolumn{3}{l}{\textit{Robotics}} \\
334 & Computer and Electronic Product Manufacturing & 21.64 \\
333 & Machinery Manufacturing & 14.28 \\
541 & Professional, Scientific, and Technical Services & 13.64 \\
325 & Chemical Manufacturing & 7.78 \\
336 & Transportation Equipment Manufacturing & 6.14 \\
522 & Credit Intermediation and Related Activities & 4.04 \\
339 & Miscellaneous Manufacturing & 3.81 \\
611 & Educational Services & 3.61 \\
335 & Electrical Equipment, Appliance, and Component Manufacturing & 3.12 \\
551 & Management of Companies and Enterprises & 2.31 \\
561 & Administrative and Support Services & 2.09 \\
423 & Merchant Wholesalers, Durable Goods & 2.05 \\
\midrule
\multicolumn{3}{l}{\textit{Artificial Intelligence}} \\
334 & Computer and Electronic Product Manufacturing & 25.66 \\
541 & Professional, Scientific, and Technical Services & 19.02 \\
511 & Publishing Industries (except Internet) & 7.34 \\
522 & Credit Intermediation and Related Activities & 5.99 \\
336 & Transportation Equipment Manufacturing & 5.52 \\
333 & Machinery Manufacturing & 4.27 \\
561 & Administrative and Support Services & 2.74 \\
335 & Electrical Equipment, Appliance, and Component Manufacturing & 2.36 \\
517 & Telecommunications & 2.30 \\
423 & Merchant Wholesalers, Durable Goods & 2.21 \\
551 & Management of Companies and Enterprises & 2.21 \\
611 & Educational Services & 2.02 \\
\bottomrule
\end{tabular}
\caption{Sectoral relevance to robotics and artificial intelligence \cite{santarelli_automation_2023}.}
\label{tab:sectoral}
\end{table}

Unique to the current phase however are the pronounced economic uncertainty, especially post-2008 and the \textit{geopolitical dimension} of frontier technologies. For instance, global competition over technological supremacy in areas such as AI, semiconductors, or quantum computing, has introduced strategic considerations into innovation policy as these technologies are viewed critical to national security and geopolitical influence.

\paragraph{Economic debate.} The current debate in economic literature concerning the effects of automation and AI on production and work is summarized below:
\begin{itemize}
\item \textit{Labor.} Task-based models have replaced traditional factor-augmenting frameworks to conceptualize automation as expansion of tasks performed by capital, complementing or displacing labor  \cite{acemoglu_simple_2025}. Autor and others have emphasized how computerization reduces demand for routine middle-skill job, leading to job polarization \cite{autor_skill_2003, acemoglu_skills_2010}. Aghion introduced ideas of \textit{directed technological change} in growth models, linking innovation with labor skills \cite{aghion_model_1992, acemoglu_why_1998}.

\item \textit{Productivity.} Bloom \textit{et al.} recently documented a long-run decline in research productivity despite rising R\&D expenditures \cite{bloom_are_2020}. The finding prompted a re-examination of innovation determinants and a discussion about mitigation measures (e.g., including higher levels of AI). Evidence at firm- and sector-level shows that adoption of robotics and AI is often concentrated in high-productivity, capital-rich firms, which promote their industries but they do so in a way that produces productivity dispersion and wage inequality even within them \cite{firooz_automation_2025,koch_robots_2021}. Nevertheless, measurement of the effects remains problematic, for example productivity metrics may understate intangible and organizational complementarities.

\item \textit{Institutions.} Economic models that account for the role of institutions, such as education and innovation systems, or collective bargaining, in shaping technological change are less frequent, despite its importance.
\end{itemize}

\begin{figure}[H]
\begin{minipage}{\linewidth}
    \includegraphics[width=0.6\linewidth]{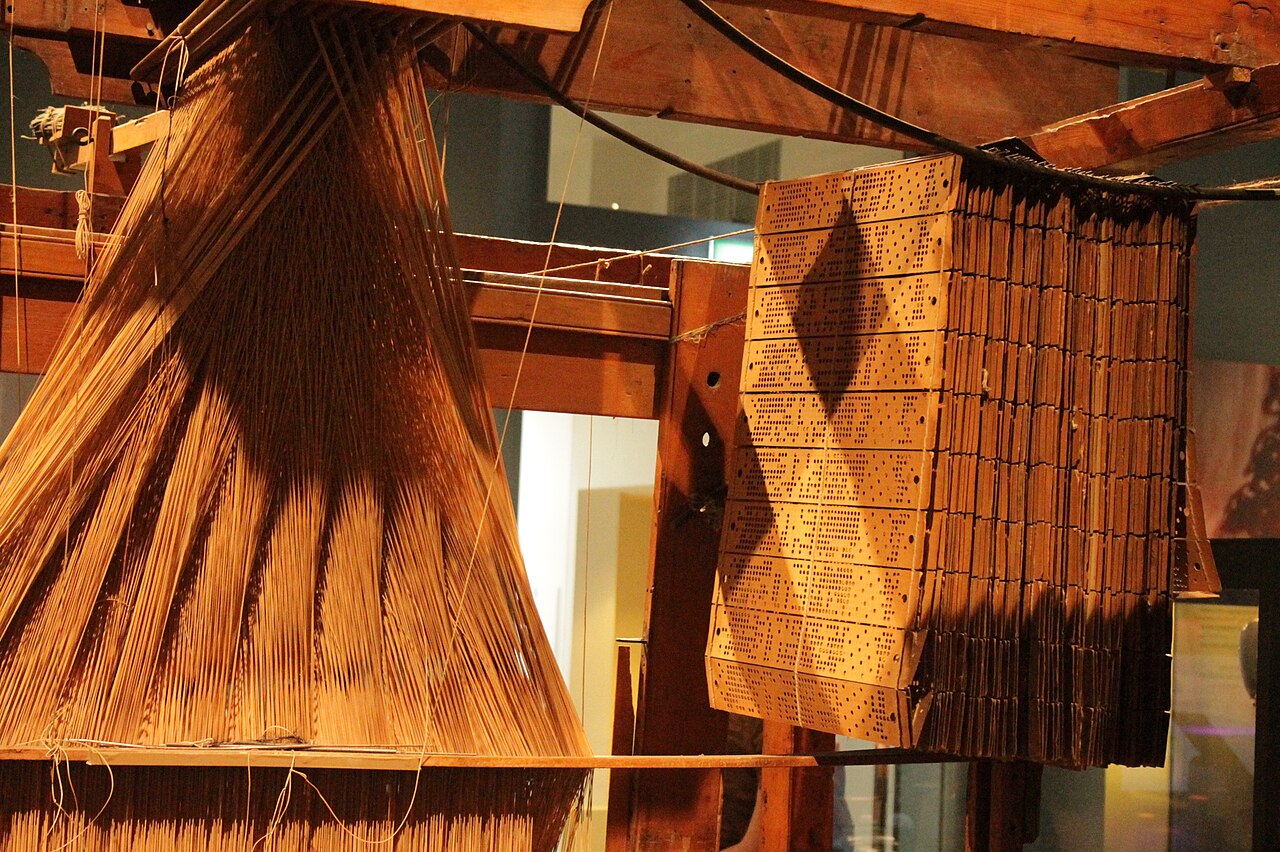}\\
    {\footnotesize Jacquard loom showing information punchcards.}
\end{minipage}

\vspace{2em}

\begin{minipage}{\linewidth}
    \begin{flushright}
    \includegraphics[width=0.55\linewidth]{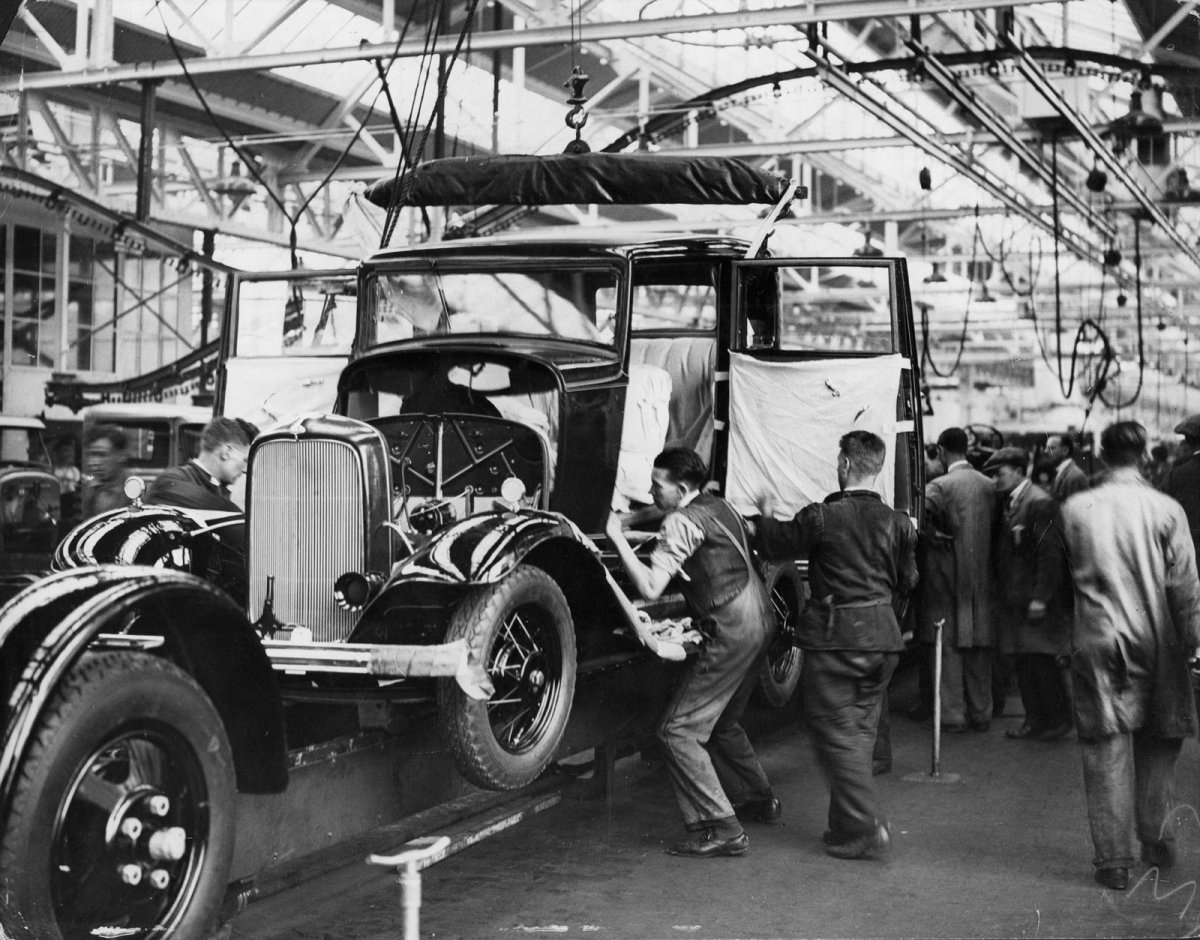}\\
    {\footnotesize Ford's moving assembly line.}
    \end{flushright}
\end{minipage}

\vspace{2em}

\begin{minipage}{\linewidth}    
    \includegraphics[width=0.65\linewidth]{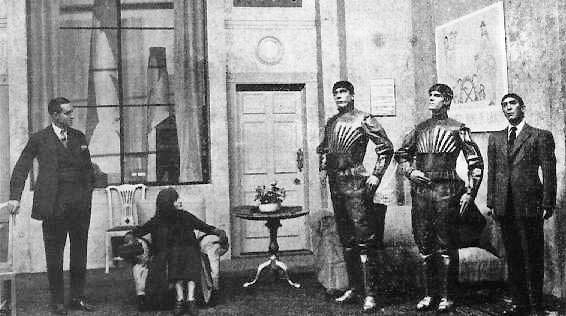}\\
    {\footnotesize Scene from Karel \v{C}apek's R.U.R. play, showing three robots.\footnote{Images taken from the public domain.}}
\end{minipage}
 
\end{figure}

\chapter{Theoretical Concepts}\label{ch:theory}
This chapter offers a brief introduction to seminal and current work in growth theory. The main motivation is to identify important determinants of productivity and economic growth across popular models, rather than to offer a detailed exposition of each \cite{acemoglu_introduction_2008}. 

\section{Concepts and Models}\label{sec:concepts}
Theory provides the framework for organizing underlying phenomena in an interpretable fashion. Specifically, growth theory forms the foundation for understanding long-run economic performance in terms of physical capital accumulation, human capital development and, crucially, technological progress. These continuously interact in non-trivial ways to shape production, the labor market, and the broader trajectory of the economy. 

\paragraph{Exogenous technology.} The \textbf{Solow-Swan} model is the classical representation of exogenous technological growth \cite{solow_contribution_1956, swan_economic_1956}. The model incorporates technology but considers it to be a fixed input to the neoclassical production function. Consider:
$$Y(t) = F(K(t), L(t), A(t)),$$
or, for example, simply:
$$Y(t) = A K(t), \quad A > 0,$$
where:
\begin{itemize}
    \item $Y(t)$ is the output at time $t$,
    \item $K(t)$ is the capital stock,
    \item $L(t)$ is the labor input, and
    \item $A(t)$ is a shifter of the production function, a broad notion of technology.
\end{itemize}
Capital evolves as:
$$\dot{K}(t) = s Y(t) - \delta K(t),$$
where:
\begin{itemize}
    \item $s$ is the fraction of output saved and invested, and
    \item $\delta$ is the depreciation rate.
\end{itemize}

The Solow-Swan model captures the core idea of capital accumulation as a critical driver of economic activity in a clear and analytically tractable form. This simplicity however is also the model's limitation: capital accumulation is determined by the savings rate, the depreciation rate, and the rate of population growth (via capital per worker), all of which are treated as exogenous. Thus, the model functions as a black box, limiting interpretability. Nevertheless, it offers a simple baseline against which more complex models can be evaluated.

\paragraph{Endogenous growth.} The theory of endogenous growth was developed as an attempt to address the disparity between measured economic performance and the theoretical predictions of earlier models. This set of economic growth theories and associated models maintain the core idea of technological growth as an important determinant of economic output, however, instead of treating it as exogenous, they endogenize it. Long-run economic growth is sustained through mechanisms such as human capital accumulation and knowledge spillovers.

\qquad \textbf{Romer's seminal} work offers a recipe for endogenizing technological progress by explicitly treating the positive externalities of knowledge accumulation  \cite{romer_increasing_1986}. It describes a mode of economic activity where economic agents decide how much to invest in developing human capital, and in doing so, generate unintentionally positive economic outcomes for others in the economy. The following version of the model using labor-augmenting technology $A(t)$ captures this important idea:
$$Y(t) = F(K(t),A(t)L(t)),$$
with:
$$A(t) = B K(t),$$
so that the knowledge stock of a firm depends on aggregate capital, and the production function of this economy exhibits increasing returns to scale. The model enables sustained per capita growth without relying on exogenous technology. In light of externalities however, policy intervention might be needed to ensure adequate investment.

\qquad Human capital accumulation via education and learning is a related albeit distinct mechanism that mediates economic growth when it is endogenized. For instance, the \textbf{Lucas model} \cite{lucas_mechanics_1988} describes human capital accumulation as:
$$ Y(t) = A h(t) u(t) L,$$
with 
$$ \dot{h}(t) = B(1-u(t))h(t),\quad B >0, $$
where:
\begin{itemize}
    \item h(t) is human capital per worker,
    \item u(t) is the fraction of time dedicated to production, and
    \item $B$ is the productivity of the education technology.
\end{itemize}
In this flavor of endogenous growth, households decide how much time to allocate to production, or education and skill accumulation. This is an individual decision, which does involve externalities as in the Romer model, and long-run economic growth is primarily driven by knowledge accumulation. The setup leaves a role for education policy, and offers a lens for rationalizing persistent differences in economic performance observed across countries.

\qquad As the idea of knowledge economy gained prominence, fueled by accelerated computerization of firms and the proliferation of software, attention was refocused more strongly on intangible assets such as information and knowledge. R\&D becomes an important driver of technological progress and ideas are conceptualized as non-rivalrous goods. Equally important are the mechanisms of \textit{knowledge diffusion} and \textit{recombination}, through which existing ideas are transmitted and combined to produce innovations.

\qquad An endogenous model by \textbf{Romer} treats new ideas (or technologies, or designs) as the product of the organized activity of a dedicated R\&D sector that feeds into production \cite{romer_endogenous_1990}. Suppose a final goods sector that is described by:
$$Y(t) = \left[ \int_0^{N(t)} x_i(t)^\alpha di \right]^{1/\alpha}, \quad 0 < \alpha < 1,$$
where:
\begin{itemize}
    \item $x_i(t)$ refers to intermediate input $i$, and
    \item $N(t)$ is the number of innovation inputs (or designs).
\end{itemize}
The intermediate goods sector comprises monopolistic firms, and the R\&D sector works to expand the set of available designs:
$$\dot{N}(t) = \delta N(t)^\phi L_A(t),$$
where:
\begin{itemize}
    \item $L_A(t)$ is the labor allocated to research,
    \item $\phi \in (0,1]$ quantifies the scale of intertemporal knowledge spillovers (``standing on shoulders'' effect).
\end{itemize}
Long-term growth then is determined by the volume of resources feeding into innovation activities (e.g., more researchers for faster growth), and how well the institutional framework prevents underinvestment in innovation (e.g., via R\&D subsidies and intellectual property protections).

\qquad \textbf{Jones} deviated from first-generation of endogenous growth theories, based on the real-world observation that larger economies grow unrealistically fast according to these theories \cite{jones_r_1995}. The core principle of Jones' proposed \textit{semi-endogenous} approach to growth is formalized as:
$$
\dot{A}(t) = \delta A(t)^\phi L_A(t)^{1 - \phi}. 
$$
A direct implication is that new ideas and designs contribute to the knowledge stock but with diminishing impact. To the extent that population increases and continually feeds into R\&D, growth can be sustained, and therefore economic behavior critically depends on population dynamics.

\qquad \textbf{Aghion and Howitt} conceptualized economic growth as the result of successive waves of technological replacement: entrepreneurs within either new or established firms disrupt existing technologies to make place for the development of new ideas \cite{aghion_model_1992}. In a basic version of the model:
$$
Y(t) = \int_0^1 y(i,t) \, di,
$$
with intermediate good produced using a unique input:
$$
y(i,t) = q(i,t) x(i,t),
$$
where:
\begin{itemize}
    \item $x(i,t)$ is the quantity of intermediate good $i$ (mass normalized to 1), and
    \item $q(i,t)$ is its quality level.
\end{itemize}
Innovation-inducing technology is modeled as a Poisson process:
$$
\lambda(i, t) = \kappa L_A(i,t),
$$
where:
\begin{itemize}
    \item $L_A(i,t)$ is the R\&D effort devoted to improving good $i$, and
    \item $\kappa$ is the productivity of R\&D.
\end{itemize}
The quality increases by a fixed factor when an innovation arrives. Here growth is rooted in \textit{sequential improvements }in quality, as opposed to expanding range of offerings. The model offers a microfoundation for industrial policy interventions as new entrants may overinvest (e.g., due to business-stealing externalities) and incumbents underinvest (e.g., due to cannibalization of existing products).

\qquad Building on these foundations, study of economic growth has evolved beyond mere endogenization to embrace a more refined treatment of production output, and conceptualize it as the aggregate product of a set (or batches) of smaller tasks. The approach, which has lately been championed by such scholars as \textbf{Acemoglu}, \textbf{Restrepo} and others, posits that technological change is directed by the relative allocation of labor and capital to tasks \cite{zeira_workers_1998,acemoglu_tasks_2021,acemoglu_simple_2025}. In Chapter~\ref{ch:modeling}, we develop analytic models using the task-based framework as the foundation.

\section{Model Classification}
Economic modeling pertains to the translation of theory into a formal, most often simplified, analytical representation. By solving the mathematical model accurately, theories can be tested. The models discussed here, depending on their methodological features, can be classified along the following structural dimensions:

\begin{itemize}
    \item \textit{Temporal structure.} Static models describe economic decisions within each time period, whereas dynamic models track how current decisions affect future states of the economy. Large-scale multi-region models such as the ECB's model for the Euro Area and Global Economy (EAGLE) or the IMF's Global Integrated Monetary and Fiscal model (GIMF) belong to the latter type \cite{gomes_eagle_2012,anderson_getting_2013}.
    \item \textit{Market closure.} Partial equilibrium assumes (some) prices to be fixed. On the other hand, general equilibrium models determine prices endogenously.
    \item \textit{Stochastic structure.} Stochastic models incorporate uncertainty through random shocks or probabilistic elements. In contrast, deterministic models assume no randomness so that it is the initial conditions and structural parameters that solely determine the system's dynamics.
    \item \textit{Agent representation.} Heterogeneous agent models account for difference in the characteristics within economic agent types. On the other hand, homogeneous agent models abstract from such differences assuming an average agent. 
    \item \textit{Technology specification.} Endogenous growth models assume technological progress is the outcome of economic decisions captured by the model (e.g., R\&D investments). In the case of exogenous growth models, technology is treated as externally given.    
\end{itemize}
Our analytical framework builds on a dynamic model with representative firm with internal heterogeneity (Chapter~\ref{ch:modeling}). Knowledge accumulates and automation levels evolve over long-run trajectories endogenously. For a given set of structural parameters the simulation treats the model as deterministic, although a large number of structural parameter sets are also used to assess the sensitivity of the model's long-run behavior (Chapter~\ref{ch:simulation}; see also Table~\ref{tab:model_dimensions}). Random shocks are used to visualize the effect of interventions.

\begin{table}[H]
\small
\centering
\begin{tabular}{@{}ll@{}}
\toprule
\textbf{Dimension}       & \textbf{Model Classification} \\ \midrule
Temporal structure           & Dynamic \\
Market closure      & Partial equilibrium \\
Stochastic structure              & Deterministic \\
Agent representation            & Representative \\
Technology specification              & Endogenous growth \\ \bottomrule
\end{tabular}
\caption{Classification of the full model.}
\label{tab:model_dimensions}
\end{table}

\section{Computational Strategy} 
After an economic problem has been specified analytically, it is often solved computationally to trace its evolution over time as faithfully as possible. By comparing simulation with model-based analytical results, the structural components of the model are verified (``Flow A'' in Figure~\ref{fig:data_driven}). Computer simulation of dynamic economic models relies on concepts, methods, and algorithms such as \cite{ljungqvist_recursive_2018}:
\begin{itemize}
    \item \textit{Dynamic programming.} The method relies on breaking down a multi-period optimization problem into simple subproblems, where economic agents act to maximize discounted rewards (e.g., household utility or firm profit). The recursive nature of the approach enables efficient computation. Central to the approach is the \textit{Bellman equation}, which expresses the value of a state variable (e.g., capital stock) as the maximum possible value from that point onwards. The canonical Bellman equation is:
$$
V(s) = \max_{a \in A(s)} \left\{ u(s, a) + \beta \mathbb{E} \left[ V(s') \mid s, a \right] \right\}, \quad \beta \in (0,1),
$$
where:
\begin{itemize}
    \item $V(s)$ is the value function, the maximal expected return attainable from state $s$,
    \item $a \in A(s)$ is a feasible action given state $s$,
    \item $u(s, a)$ describes instantaneous return,
    \item $\beta$ is the discount factor,
    \item $s'$ is the next-period state, and
    \item $\mathbb{E}[\cdot]$ is the expectation operator over future states.
\end{itemize}    

\item \textit{Value function iteration.} Another popular computational method for dynamic programming problems relies on making an initial guess for the value function and update it iteratively until convergence with:
$$
V_{n+1}(s) = \max_{a \in A(s)} \left\{ u(s,a) + \beta \mathbb{E}[V_n(s')|s,a] \right\}
$$
\item \textit{Forward simulation.} This approach is used to study the evolution of an economic system given a set of initial conditions and a set of governing equations. Laws of motions and decision rules are applied at each time step to move the system forward. It is particularly useful for complex systems or when closed-loop form solutions do not exist.
\end{itemize}

In Chapter~\ref{ch:simulation}, we use forward numerical simulation to analyze the evolution of our analytical model and the interplay between different mechanisms. 

\section{Data-Driven Analysis}
Observational data is used to guide the construction, calibration, and validation of theories, models, and simulations. Table~\ref{tab:calibration_sources} summarizes data repositories that are commonly used. Here, we do not perform model calibration for a specific country, instead we aim to extract general insights from statics and simulation. To that end, we use typical values directly from the literature for the structural parameters of our model. For example, the interpretation of a capital-labor ratio of 3 (model units) as structural input parameter is to be understood in relation to the fact that this ratio for a moderate innovator in Europe like Greece is ~60\% of the U.S. level at the global technological frontier.

\begin{table}[H]
\centering
\begin{adjustbox}{max width=\textwidth} 
\begin{tabular}{@{}ll@{}}
\toprule
\textbf{Variable} & \textbf{Common Sources} \\
\midrule
Gross Domestic Product & Eurostat; U.S. Bureau of Economic Analysis (BEA); OECD Nat. Acc. \\
Gross Fixed Capital Formation & Eurostat; BEA \\
Capital Stock & European Commission AMECO; Penn World Table \\
Capital Share & The Conference Board; OECD STAN \\
Hours Worked & The Conference Board \\
Population & World Bank Indicators; OECD Demographic Statistics \\
R\&D Labor Share & Eurostat (HRST); OECD MSTI \\
R\&D Expenditure & Eurostat (GBAORD); U.S. NSF NCSES; OECD ANBERD \\
AI Effectiveness & Eurostat (DSI); OECD (AI Indicators); Stanford AI index \\
Automation Levels & IFR (robot intensity) \\
Tasks \& Skills & Eurostat–ISCO; U.S. O*NET; OECD PIAAC \\
\bottomrule
\end{tabular}
\end{adjustbox}
\caption{Empirical data sources and proxies for calibration of growth models.}
\label{tab:calibration_sources}
\end{table}

\qquad Not to be confused with empirics, data-driven analysis describes methods that aim to discover trends and patterns in large datasets, and refine the understanding of the underlying generative economic model. These insights, in turn, support reliable inferences for economic forecasting or prescriptions for policy design. Machine learning techniques have been recently particularly successful in uncovering trends and latent structure in (high-dimensional) datasets \cite{bishop_pattern_2006,james_introduction_2013}.

\qquad In a supervised learning setting, the goal is to establish a causal relationship between inputs and outputs using existing data. Supervised machine learning models for regression include:
\begin{itemize}
    \item \textit{Regularized linear models.} Penalty terms are introduced in linear regression to manage overfitting and therefore improve generalization of the resulting calibrated model. Despite their simplicity, these models remain popular owing to their interpretability.
    \item \textit{Decision trees.} Also straightforward to interpret, these models rely on splitting data recursively into smaller subsets in a tree-like structure, where each node applies a binary test that partitions the data into two branches. Splitting stops at (leaf) nodes that make a prediction. Ensembles of decision trees can improve predictive accuracy and robustness by averaging over many different trees.
    \item \textit{Neural networks.} These approximate complex, nonlinear functions by means of successive non-linear transformations organized in layers. It is currently the most widely used class of machine learning models. However, they tend to be less interpretable than simpler models such as those above. 
\end{itemize}

\begin{figure}
    \centering
    \includegraphics[width=1\linewidth]{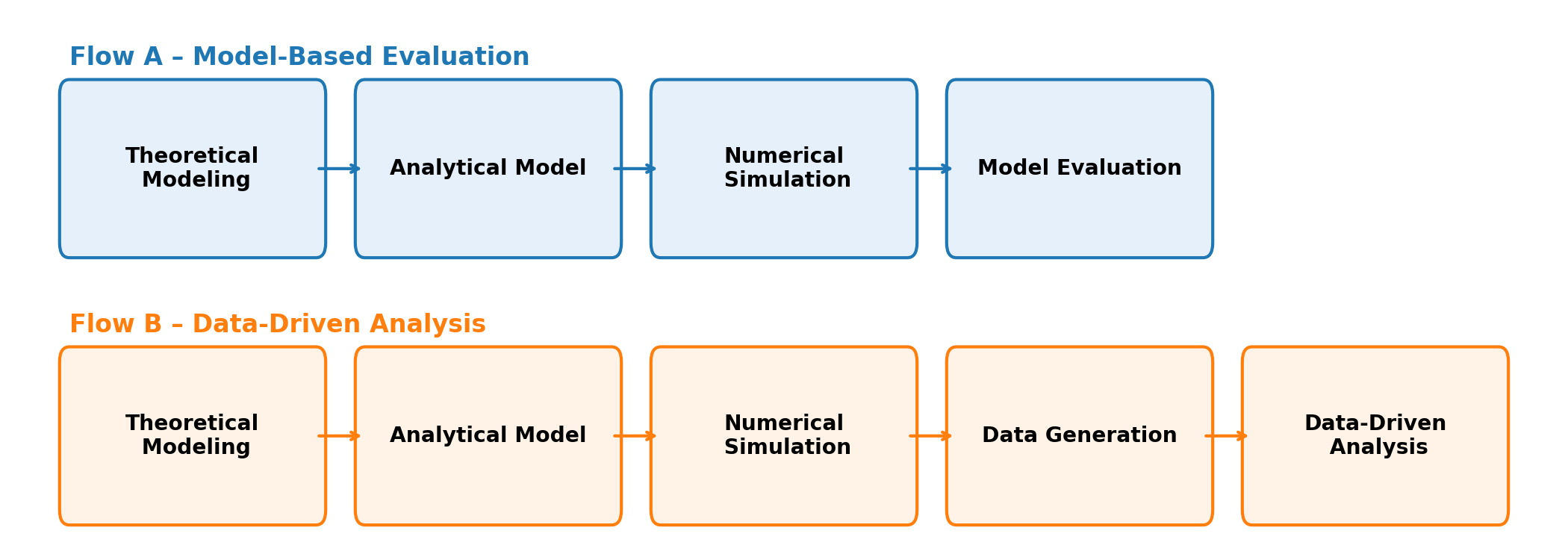}
    \caption{Computational workflows for evaluation and analysis.}
    \label{fig:data_driven}
\end{figure}

In Chapter~\ref{ch:simulation}, we rely on high-throughput calculations to establish a dataset of structural parameter inputs and key variable outputs. Then, we identify simple quantitative structure-output relationships by training and analysis of an ensemble of decision trees as \textit{surrogate} model of the dataset (``Flow B'' in Figure~\ref{fig:data_driven}).  

\paragraph{Software and tools.}
A variety of computational tools are used in the field for macroeconomic modeling, depending on the complexity of the model and the desirable accuracy. Popular solvers, libraries and platforms for the solution, calibration, simulation, and analysis of economic models include:
\begin{itemize}
    \item \textbf{Dynare.} Dynare is a software platform for handling a wide class of economic models, in particular dynamic stochastic general equilibrium (DSGE) and overlapping generations (OLG) models \cite{adjemian_dynare_2021}. It is a popular choice for policy analysis.
    \item \textbf{DSGE.jl}: A Julia-based implementation of the New York Fed DSGE model which provides general code to estimate user-specified models. 
    \item \textbf{MATLAB/Python/R}: Generic high-level languages with tailored toolboxes suitable for model simulation and advanced data operations. For example, \textit{scikit-learn} is a Python library for supervised and unsupervised machine learning tasks, which we use in Chapter~\ref{ch:simulation} to quantify the relative importance of model parameters in predicting key outputs.
\end{itemize}

\section{Task-based Framework}\label{sec:base}
We rely on the task-based formulation put forward by Acemoglu and co-authors as the foundational model \cite{acemoglu_race_2018,acemoglu_simple_2025}. Consider an economy in which output is produced by aggregating partial outputs from a set of tasks. The environment features exogenous capital and labor supply, and a social planner allocates these resources to maximize total economic output. Let the output, $Y$, be produced by aggregating a continuum of tasks indexed by $z \in [0,1]$ as:

$$
Y = \left( \int_0^1 y(z)^{\frac{\sigma - 1}{\sigma}} dz \right)^{\frac{\sigma}{\sigma - 1}}, \quad \sigma > 0, \sigma \ne 1,
$$
where:
\begin{itemize}
    \item $y(z)$ is output in task $z$, and
    \item $\sigma$ is the elasticity of substitution between tasks.
\end{itemize}

\paragraph{Task-based production.}
Each task can be performed using either labor or capital, but not both simultaneously. A discrete partition between labor tasks $L \subset [0,1]$ and capital tasks $K = [0,1] \setminus L$ therefore exists.

For labor-performed tasks $z \in L$:

$$
y(z) = a_L(z) \cdot l(z),
$$
with:
\begin{itemize}
    \item $a_L(z)$ the labor-specific productivity, and 
    \item $l(z)$ the labor input assigned to task $z$.
\end{itemize}
Likewise, for capital-performed tasks $z \in K$:
$$
y(z) = a_K(z) \cdot k(z),
$$
where:
\begin{itemize}
    \item $a_K(z)$ the capital-specific productivity, and 
    \item $k(z)$ the capital input assigned to task $z$.
\end{itemize}
The last expression can be rewritten to stress that capital and labor are perfect substitutes across tasks:
$$y(z) = a_L(z) \cdot l(z) + a_K(z) \cdot k(z)$$
To introduce heterogeneity in the tasks, we assume 
$a_L(z) / a_K(z)$ is strictly increasing with $z$. Tasks are allocated to capital or labor based on their relative efficiency based on the following decision rule:
$$
\text{assign task } z \text{ to capital if } \frac{w}{a_L(z)} \geq \frac{r}{a_K(z)}.
$$

The \textit{automation frontier} $z^*$ is defined such that:

$$
\frac{w}{a_L(z^*)} = \frac{r}{a_K(z^*)}.
$$

Tasks $z \leq z^*$ are automated, while $z > z^*$ are performed by labor (or, in other words, tasks produced with capital are in the range $[0,z^*]$ for wage, $w$, and capital rental, $r$ rates).

\paragraph{Optimization.}
The planner's objective is to maximize production output under total labor and capital supply constraints. Formally, the optimization problem for capital output is:
$$
\max_{k(z)} \int_{0}^{z*} \left[ a_K(z) \cdot k(z) \right]^{\frac{\sigma - 1}{\sigma}} dz
\quad \text{s.t.} \quad \int_{0}^{z*} k(z) \, dz = K,
$$
and the associated Lagrangian:
$$
\mathcal{L_K} = \int_{0}^{z*} \left[ a_K(z) \cdot k(z) \right]^{\frac{\sigma - 1}{\sigma}} dz - \lambda \left( \int_{0}^{z*} k(z) \, dz - K \right),
$$
conditional on the automation threshold $z^*$.
Taking the derivative with respect to $k(z)$ to derive the optimally allocated capital, the resulting \textit{effective} total capital output after algebraic manipulation is:
$$
y_K(z^*) = \left( \int_{0}^{z*} a_K(z)^{\sigma - 1} dz \right)^{\frac{1}{\sigma}} \cdot K^{\frac{\sigma - 1}{\sigma}},
$$
and analogously for labor tasks:
$$
y_L(z^*) = \left( \int_{z*}^{1} a_L(z)^{\sigma - 1} dz \right)^{\frac{1}{\sigma}} \cdot L^{\frac{\sigma - 1}{\sigma}}.
$$
The expression for the aggregated output $Y$ can be rewritten as:
$$
Y = \left(y_K + y_L\right)^{\frac{\sigma}{\sigma - 1}},
$$
leading to:
$$
Y(z^*) = \left[
\left( \int_{0}^{z*} a_K(z)^{\sigma - 1} dz \right)^{\frac{1}{\sigma}} K^{\frac{\sigma - 1}{\sigma}}
+
\left( \int_{z*}^{1} a_L(z)^{\sigma - 1} dz \right)^{\frac{1}{\sigma}} L^{\frac{\sigma - 1}{\sigma}}
\right]^{\frac{\sigma}{\sigma - 1}}.
$$

\paragraph{Key variables.}
We assume labor markets are perfectly competitive and wages are determined solely by the marginal product in labor-performed production tasks. By differentiating $Y$ with respect to $L$ we obtain for the wage:
$$
w = \frac{\partial Y}{\partial L} = \left( \frac{\partial Y}{\partial y_L} \cdot \frac{\partial y_L}{\partial L} \right).
$$

This gives:
$$
w = \left( y_K + y_L \right)^{1/(\sigma - 1)} \cdot \frac{y_L}{L} 
$$
To evaluate the \textit{wage effect} of automation, we take:
$$
\frac{d w}{d z^*}
= \frac{\partial w}{\partial y_K} \cdot \frac{d y_K}{d z^*}
+ \frac{\partial w}{\partial y_L} \cdot \frac{d y_L}{d z^*}
$$

For $\sigma > 1$, it is straightforward to show that all the partial derivatives in the RHS are positive, except for 
$$\frac{dy_L}{dz^*}<0,$$
since increased automation $z^*$ decreases labor output $y_L$. Thus, the first term of the RHS is positive, demonstrating a \textit{productivity effect} whereas the second term is negative, quantifying the \textit{displacement effect}. The net effect depends on the relative magnitudes of the two terms. 
For $\sigma<1$, the first term in the RHS becomes negative instead. The labor share is defined as:
$$
s_L = \frac{w \cdot L}{Y}.
$$
Plugging in the expressions for equilibrium wage and outputs, we obtain after simplification the following expression:
$$
s_L = \frac{y_L}{y_K + y_L}.
$$
To evaluate the \textit{labor share effect} of automation, we take:
$$
\frac{d s_L}{d z^*}
= \frac{\partial s_L}{\partial y_K} \cdot \frac{d y_K}{d z^*}
+ \frac{\partial s_L}{\partial y_L} \cdot \frac{d y_L}{d z^*}.
$$

In this case both terms in the RHS are negative and automation unambiguously reduces the labor share.

\begin{figure}
    \centering
    \includegraphics[width=1.\linewidth]{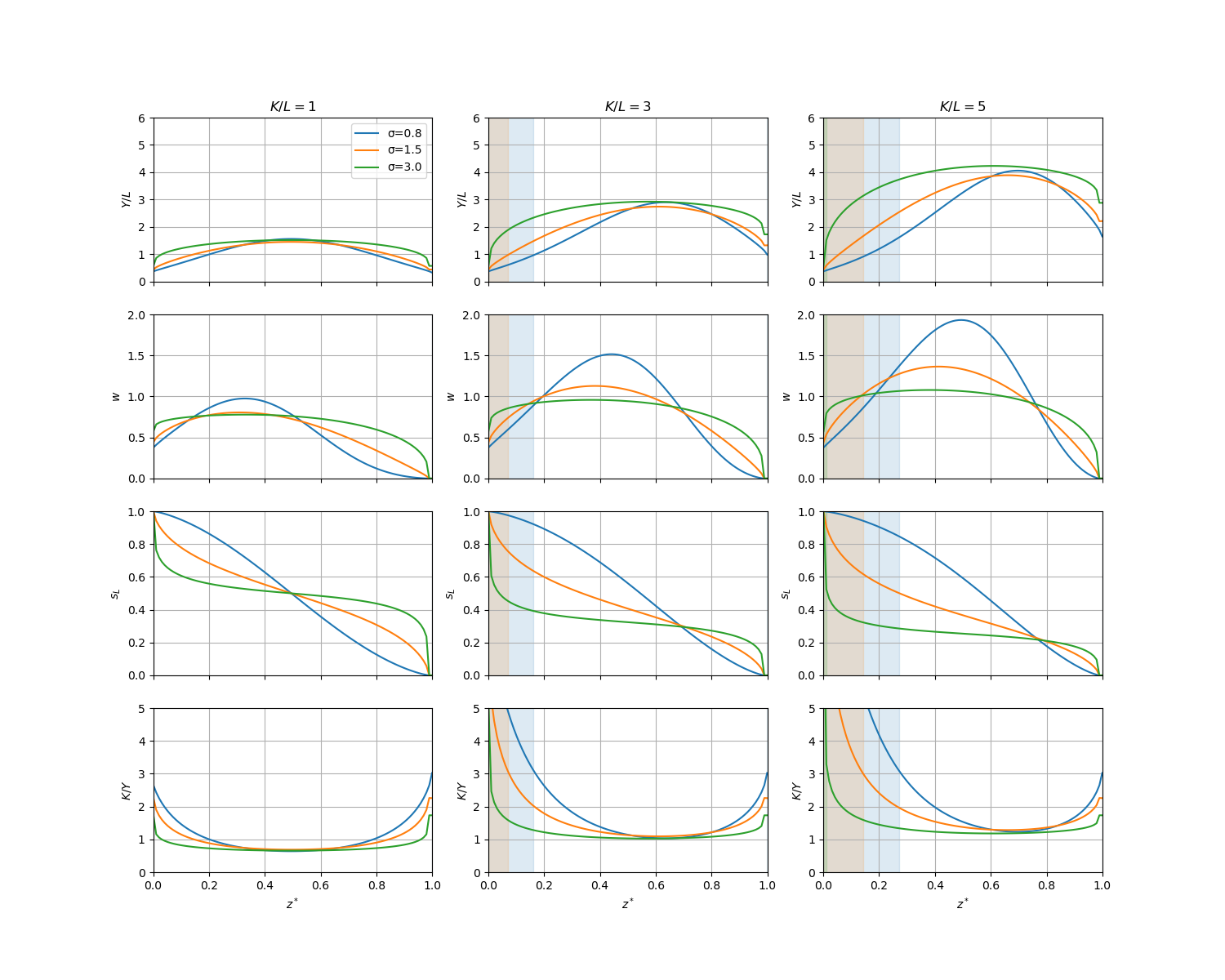}
    \caption{Numerical calculation of key variables of the baseline model.}
    \label{fig:1statics_1}
\end{figure}

\qquad Figures~\ref{fig:1statics_1} and~\ref{fig:1statics_2} present numerical results from the model for how four key variables depend on $z^*$:
\begin{enumerate}
    \item \textit{Output $Y$ per production labor $L$}. As more tasks are performed with capital, output increases. Output also increases with increasing capital-to-labor ratios $K/L$, and with $\sigma$ (more efficient allocation of tasks to factors). It is non-zero even if wages or labor share collapse.
    \item \textit{Wage $w$}. It increases with increasing automation up to the point where productivity and substitution effects balance out. Then, it decreases to reach zero (0).
    \item \textit{Labor share $s_L$}. It monotonously decreases with increasing automation. At the limit of complete automation $z^*=1$, labor collapses.
    \item \textit{Capital to output ratio $K/Y$}. Shaded areas are regimes where $K/Y>3$, used here as a proxy of ineffective capital allocation and therefore automation levels.
\end{enumerate}
    
\section{Open Questions} 
A number of open questions remain in the economic literature:
\begin{itemize}
    \item What are the key factors that determine whether technological change displaces or augments labor?
    \item How to best describe the interaction of automation, AI, and R\&D as co-determinants of economic output?
    \item Are exponential growth trajectories possible in light of self-improving AI?
    \item What is the impact of institutional lags, limits, and failures on current scientific and technological progress?
    \item What are effective policy levers for shaping the impact of automation and AI on production and labor market dynamics?
\end{itemize}
In the remainder of the work we attempt to give answers to these questions. We avoid discussing any strand of growth models in depth, instead we focus on productive combination of elements.

\chapter{Analytical Framework}\label{ch:modeling}

In this chapter we develop a series of extensions to the baseline model presented in Section~\ref{sec:base}. These include the introduction of capital frictions, knowledge accumulation, and direct coupling between production and growth, reflecting core principles of the canonical models in growth theory presented in Section~\ref{sec:concepts}.

\section{Physical Capital Frictions}
We first extend the baseline model to capture the idea that the more tasks are automated, the system-level cost of maintaining flexibility increases in the face of technological lock-in. This reflects ideas incorporated in semi-endogenous models, but tailored for the task-based framework.

\paragraph{Friction.}
We denote the friction cost associated with technological lock-in of an automated task $z \leq z^*$ at time $t$ as $\phi(t, z)$. Assuming time-invariant friction, the aggregate cost $\Phi(z^*)$ for a level of automation $z^*$ then is:
$$
\Phi(z^*) = \gamma \cdot \int_0^{z^*} \phi(z) \, dz, \quad \gamma \geq 0,
$$
where $\gamma$ is a system-level scaling parameter that captures institutional susceptibility to lock-in. The limit of no frictions, i.e., the baseline model, is restored for $\gamma = 0$.

A simple functional form for the per-task friction cost that allows for convexity is:
$$\phi(z) = z_0 + z^\eta, \quad \eta > 0,\ z_0 \geq 0,$$
where $\eta$ reflects how quickly lock-in costs escalate.
The total cost $\Phi(z^*)$ is: 
$$\Phi(z^*; z_0 = 0) = \gamma \cdot \int_0^{z^*} z^\eta dz = \gamma \cdot \frac{(z^*)^{\eta+1}}{\eta + 1}.
$$

\paragraph{Optimization.}
Here, the social planner maximizes total \textit{net} output, additionally considering technological lock-in costs:
$$
\max_{z^*} \left[ Y(z^*) - \Phi(z^*) \right] = \max_{z^*} \left[Y(z^*) - \gamma \cdot \frac{(z^*)^{\eta+1}}{\eta + 1}\right].
$$
The first-order condition yields:
$$
\frac{\partial Y}{\partial z^*} = \gamma \cdot (z^*)^\eta.
$$
The LHS is the marginal output gain from automating one more task, and the RHS is the marginal cost associated with reallocating, updating, or scrapping that task (as in the case of non-fixed tasks mass as discussed later). 

\paragraph{Effect on key variables.} 
We use Taylor expansion around the frictionless equilibrium to explore the effect of the friction-adjusted automation frontier on the key variables of interest. Let $z^*_0$ be the automation frontier in the case of no frictions ($\gamma = 0$). For small strength of frictions the production output is described by:
$$
Y(z^*; \gamma) \approx Y(z^*_0) + \frac{\partial Y}{\partial z^*}\bigg|_{z^*_0} \cdot \frac{dz^*}{d\gamma}\bigg|_{\gamma=0} \cdot \gamma,
$$
with
$$
\frac{\partial Y}{\partial z^*} = \frac{\sigma}{\sigma - 1} \left[ y_K + y_L \right]^{\frac{1}{\sigma - 1}} \cdot \left( \frac{dy_K}{dz^*} + \frac{dy_L}{dz^*} \right).
$$
By differentiating both sides of the first-order condition:
\begin{align*}
\frac{d^2Y}{d{z^*}^2} \cdot \frac{dz^*}{d\gamma}
 &= (z^*)^\eta + \gamma \cdot \eta \cdot (z^*)^{\eta - 1} \cdot \frac{dz^*}{d\gamma} \Rightarrow\ \\
\frac{dz^*}{d\gamma} &= \frac{(z^*)^\eta}{\frac{d^2Y}{d{z^*}^2} - \gamma \cdot \eta \cdot (z^*)^{\eta - 1}}.
\end{align*}
If $Y''(z^*)$ is sufficiently negative, then $\frac{dz^*}{d\gamma} < 0$, that is, increasing friction reduces automation. Conversely, this corresponds to the real-world intuition that automating early tasks (routine, low-skill) adds lots of value but later tasks increasingly require more careful planning and implementation. Using the first-order Taylor expansion around $\gamma = 0$ we obtain for the wage:
\begin{align*}
w(z^*; \gamma) &\approx w(z^*_0) + \frac{\partial w}{\partial z^*}\bigg|_{z^*_0} \cdot \frac{dz^*}{d\gamma}\bigg|_{\gamma=0} \cdot \gamma.
\end{align*}
The effect of parameter $\gamma$ on the wage is ambiguous and it depends on the relative balance between the productivity and displacement effect. Similarly, for the labor share:
$$
s_L(z^*; \gamma) \approx s_L(z^*_0) + \frac{\partial s_L}{\partial z^*}\bigg|_{z^*_0} \cdot \frac{dz^*}{d\gamma}\bigg|_{\gamma=0} \cdot \gamma.
$$
In this case, increasing friction results in increased labor share. 

\begin{figure}[H]
    \centering
    \includegraphics[width=1\linewidth]{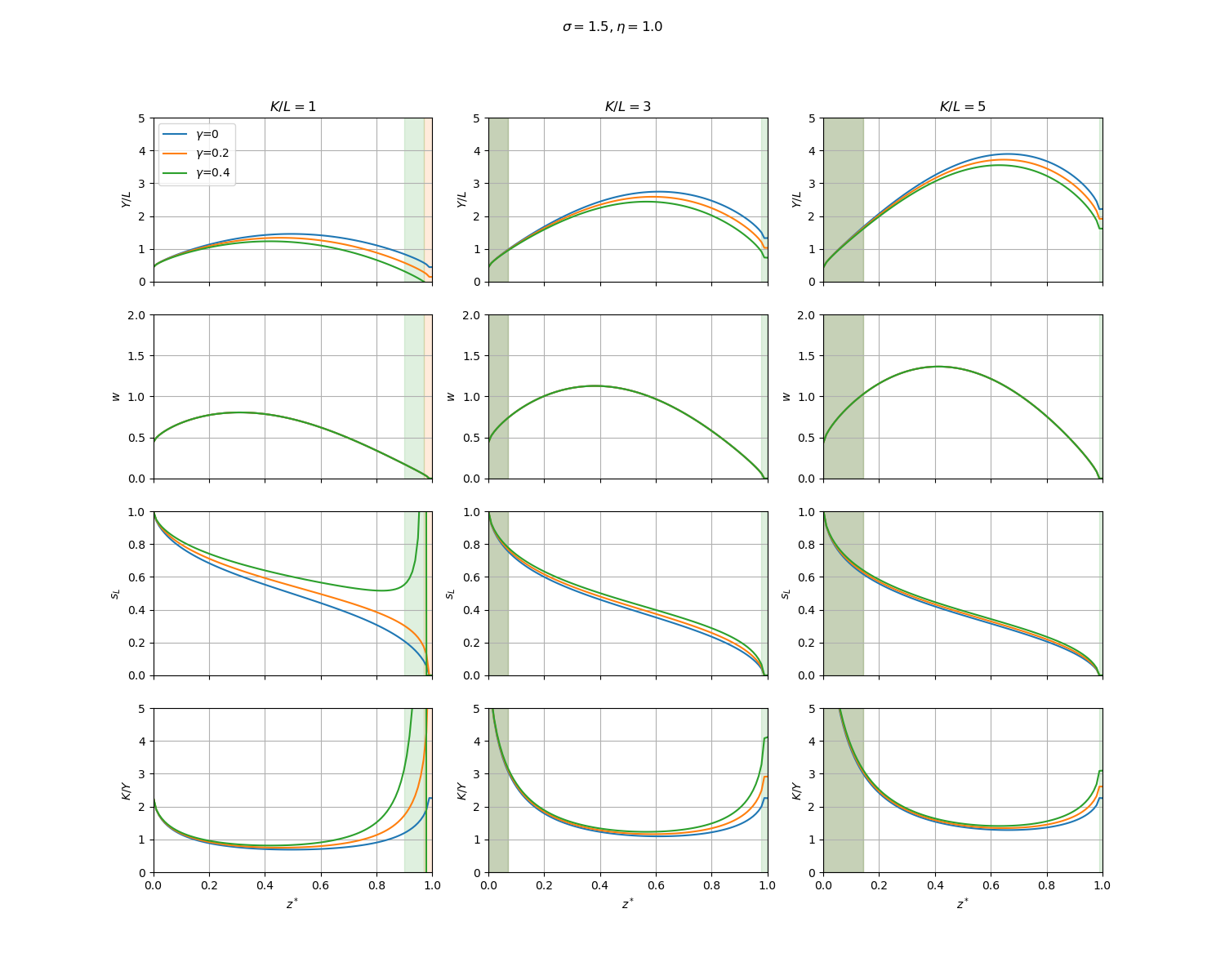}
    \caption{Numerical calculation of key variables of the model with frictions.}
    \label{fig:2statics_1}
\end{figure}

\qquad Figures~\ref{fig:2statics_1} and~\ref{fig:2statics_2} present numerical results for the model. The output reduces and the labor share increases with increasing $\gamma$. Higher capital-labor ratios mute however the effect on the latter.

\section{Knowledge-Driven Productivity}\label{sec:knowdriven}
We proceed to introduce an endogenous growth component that relies on a stock of knowledge $\mathcal{K}(t)$. This is a logical next step in generalization, as it allows technological progress to arise from purposeful investment in knowledge generation and use in a manner consistent with canonical (Section~\ref{sec:concepts}) and contemporary growth literature \cite{almeida_artificial_2024, agrawal_artificial_2023}.

\paragraph{Knowledge accumulation.} We assume that knowledge is generated by dedicated R\&D, which enhances productivity uniformly across all tasks (labor or capital) as follows:
$$
a_L(z,t) = \bar{a}_L(z) \cdot \mathcal{K}(t)^\beta, \quad a_K(z,t) = \bar{a}_K(z) \cdot \mathcal{K}(t)^\beta, \quad \beta \geq 0,
$$
with:
\begin{itemize}
    \item $\bar{a}_{K,L}(z, t)$ the constant component of task-specific productivities $a_{K,L}$, and
    \item $\beta$ the output elasticity with respect to the knowledge stock. In the limiting case of $\beta = 0$, access to the knowledge stock has no impact on production.
\end{itemize}
The evolution of the knowledge stock is described by a \textit{knowledge production function} such as:
$$
\dot{\mathcal{K}}(t) = \zeta \cdot R(t)^\alpha \cdot \mathcal{K}(t)^\phi, \quad \zeta \geq 0,\ \alpha \in (0,1],\ \phi \in [0,1],
$$
where:
\begin{itemize}
    \item $R(t)$ is the labor allocated to research,
    \item $\alpha$ describes the elasticity of knowledge production with respect to R\&D labor,
    \item $\phi$ quantifies strength of \textit{intertemporal }knowledge spillovers, i.e., how much past ideas matter for new ideas (``standing on shoulders'' effect), and
    \item $\zeta$ is a R\&D productivity parameter. The limiting case of $\zeta=0$ treats the knowledge stock as exogenous.
\end{itemize}

Even though the function has a simple form, it leads to rich behavior that depends on elected parameters (for instance, $\alpha+\phi=1$ corresponds to scale-invariant knowledge production). Table \ref{tab:knowledge_regimes} summarizes parameter regimes.

\begin{table}[H]
\centering
\begin{adjustbox}{max width=\textwidth} 
\begin{tabular}{@{}lll@{}}
\toprule
\textbf{Parameter Regime} & \textbf{Condition} & \textbf{Growth implication} \\ \midrule
\textit{No endogenous growth} & \( \zeta = 0 \) & No endogenous knowledge accumulation \\
\textit{Constant returns to scale} & \( \alpha + \phi = 1 \) & Scale-invariant knowledge production \\
\textit{Increasing returns} & \( \alpha + \phi > 1 \) & Superlinear accumulation; explosive or accelerating growth possible \\
\textit{Decreasing returns} & \( \alpha + \phi < 1 \) & Growth slows unless offset by rising inputs \\
\textit{No intertemporal spillovers} & \( \phi = 0 \) & Knowledge accumulation depends solely on current R\&D \\ \bottomrule
\end{tabular}
\end{adjustbox}
\caption{Parameter regimes in knowledge production.}
\label{tab:knowledge_regimes}
\end{table}

\paragraph{Production.}
Let total labor supply be fixed (exogenous) at $\bar{L}$, allocated dynamically between the production sector $L(t)$ and the R\&D sector $R(t)$:
$$
L(t) + R(t) = \bar{L}.
$$
The total labor supply constraint can be rewritten as:
$$
(1 - S_R(t)) \bar{L} + S_R(t) \bar{L} = \bar{L}, \quad S_R(t) \in [0,1],
$$
with $S_R(t)$ the share of R\&D labor force. This introduces a production-growth trade-off: production labor contributes to task performance, while R\&D labor generates knowledge that improves productivity. The task-aggregate output $Y$ then becomes:
\begin{align*}
Y(t) &= \mathcal{K}(t)^\beta \cdot \left[ \left( \int_0^{z^*(t)} \bar{a}_K(z)^{\sigma - 1} dz \right)^{\frac{1}{\sigma}} K^{\frac{\sigma - 1}{\sigma}} + \left( \int_{z^*(t)}^1 \bar{a}_L(z)^{\sigma - 1} dz \right)^{\frac{1}{\sigma}} L(t)^{\frac{\sigma - 1}{\sigma}} \right]^{\frac{\sigma}{\sigma - 1}} =\\
&= \mathcal{K}(t)^\beta \cdot \tilde{Y},
\end{align*}
where $\mathcal{K}(t)^\beta$ has been factored out of the task productivities. It becomes readily apparent that output can grow over time solely via an increase in $\mathcal{K}(t)$.

\paragraph{Long-run growth.} Taking logs and differentiating with respect to time $t$, we obtain the growth rate $g_Y$ of production:
\begin{align*}
g_Y(t) &= \frac{\dot{Y}(t)}{Y(t)} = \beta \cdot \frac{\dot{\mathcal{K}}(t)}{\mathcal{K}(t)} \Rightarrow\\
g_Y(t) &= \beta \cdot g_\mathcal{K}(t),
\end{align*}
where $g_\mathcal{K}(t)$ denotes the growth rate of the knowledge stock.
Plugging in the knowledge production function gives:
$$
g_\mathcal{K}(t) = \frac{\dot{\mathcal{K}}(t)}{\mathcal{K}(t)} = \zeta \cdot R(t)^\alpha \cdot \mathcal{K}(t)^{\phi - 1}.
$$

For $\phi = 1$ and assuming a fixed R\&D labor share, the \textit{long-term} growth rates become:
\begin{align*}
g_\mathcal{K} &= \zeta \cdot (S_R \bar{L})^\alpha, \\
g_Y &= \beta \cdot \zeta \cdot (S_R \bar{L})^\alpha.
\end{align*}

In contrast, when $\phi < 1$:
$$g_Y(t) \propto g_{\mathcal{K}}(t) \propto \frac{1}{t} \quad \text{as } t \to \infty.$$
Growth converges to zero unless offset by growth in the R\&D labor force or R\&D productivity rises. Table \ref{tab:baseline_growth} summarizes the behavior of the growth rates and knowledge stock in the long-run.

\begin{table}[H]
\small
\centering
\begin{adjustbox}{max width=\textwidth} 
\begin{tabular}{@{}lll@{}}
\toprule
\textbf{Variable} & \textbf{Growth Expression} & \textbf{Asymptotic Behavior} \\ \midrule
\multicolumn{3}{l}{$\phi = 1$} \\
$g_{\mathcal{K}}$ & $\zeta \cdot R^\alpha$ & Constant \\
$g_Y$ & $\beta \cdot \zeta \cdot R^\alpha$ & Constant \\
$\mathcal{K}(t)$ & $\propto e^{\zeta R^\alpha t}$ & Exponential \\[1ex]
\multicolumn{3}{l}{$\phi < 1$} \\
$g_{\mathcal{K}}$ & $\zeta \cdot R^\alpha \cdot \mathcal{K}(t)^{\phi - 1}$ & $\sim \frac{1}{t}$ \\
$g_Y$ & $\beta \cdot \zeta \cdot R^\alpha \cdot \mathcal{K}(t)^{\phi - 1}$ & $\sim \frac{1}{t}$ \\
$\mathcal{K}(t)$ & $\propto t^{\frac{1}{1 - \phi}}$ & Sub-exponential \\ \bottomrule
\end{tabular}
\end{adjustbox}
\caption{Long-run growth outcomes in the endogenous knowledge model.}
\label{tab:baseline_growth}
\end{table}

\paragraph{Effect on key variables.}
For a static analysis, the knowledge stock is assumed to be exogenous. Since knowledge scales task productivity multiplicatively and uniformly, output increases monotonically with the knowledge stock:

$$
\frac{\partial Y}{\partial \mathcal{K}} = \beta \cdot \mathcal{K}^{\beta - 1} \cdot \tilde{Y} > 0
$$

Higher values of $\beta$ lead to stronger amplification effects. Thus, both $\mathcal{K}$ and $\beta$ positively affect output.

\qquad Given that wages derive from the marginal product of production labor $L(t)$, they inherit the same scaling behavior:
\begin{align*}
w &= \frac{\partial Y}{\partial L} = \mathcal{K}^\beta \cdot \frac{\partial \tilde{Y}}{\partial L} \Rightarrow\\
w &= \mathcal{K}^\beta \tilde{w} \Rightarrow \\
w &\propto \mathcal{K}^\beta.
\end{align*}
Here, we have abstracted from explicitly modeling compensation to R\&D labor, which is treated as a resource input to idea generation (and not an input used in the production sector), that is subsidized, e.g., using research grants. For a given level of automation, by differentiation:
$$
\frac{\partial w}{\partial \mathcal{K}} = \beta \cdot \mathcal{K}^{\beta - 1} \cdot \tilde{w} > 0.
$$
Wages rise proportionally with knowledge intensity, and more so when $\beta$ is large. For the production labor share of total output:
\begin{align*}
s_L &= \frac{w \cdot L}{Y} = \frac{\mathcal{K}^\beta \cdot \tilde{w} \cdot L}{\mathcal{K}^\beta \cdot \tilde{Y}} \Rightarrow\\
s_L &= \frac{\tilde{w} \cdot L}{\tilde{Y}} \Rightarrow \\
\frac{\partial s_L}{\partial \mathcal{K}} &= 0.
\end{align*}

\begin{figure}
    \centering
    \includegraphics[width=1\linewidth]{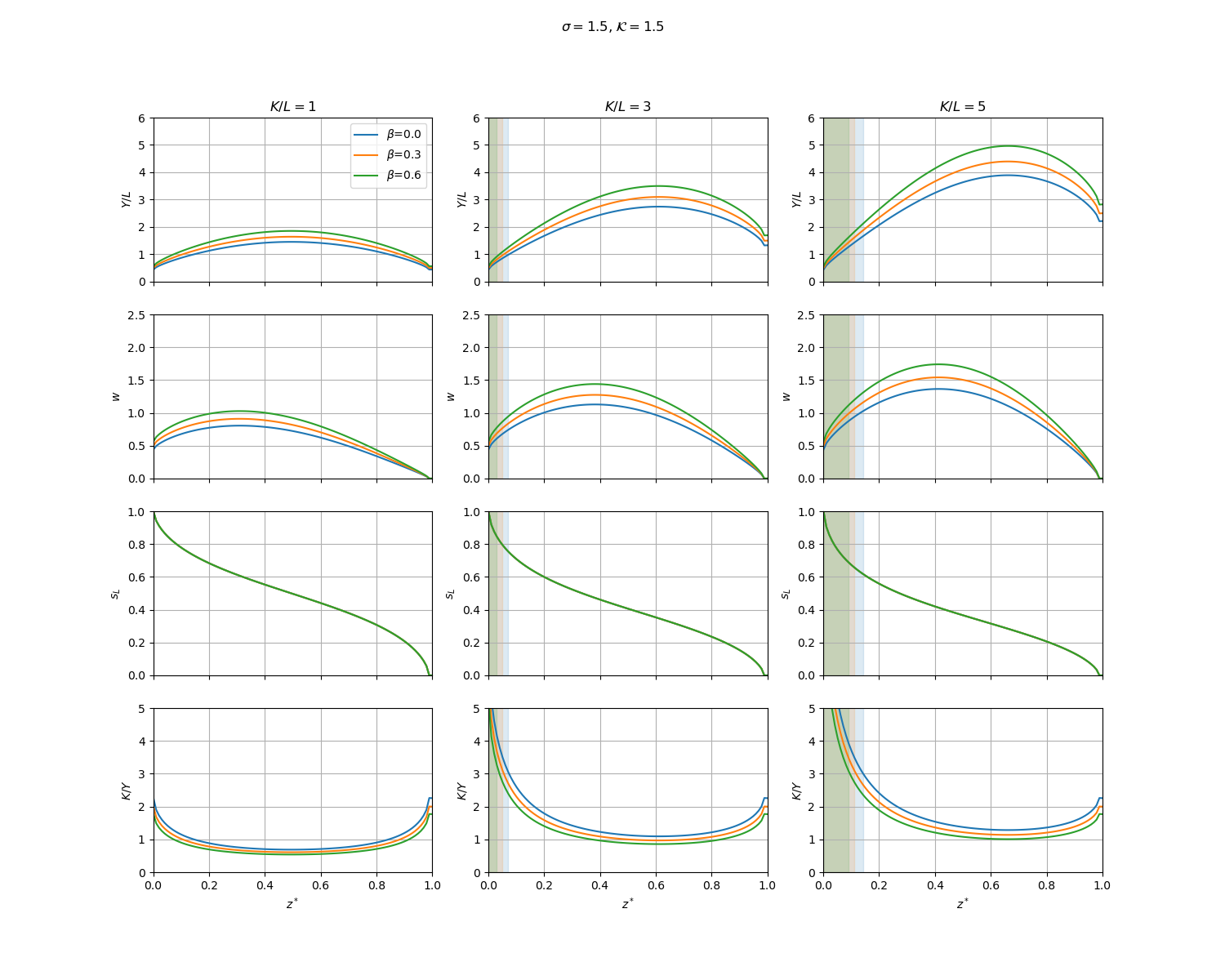}
    \caption{Numerical calculation of key variables of the model with knowledge accumulation.}
    \label{fig:3statics_1}
\end{figure}

\qquad Thus, holding all else constant, the labor share is unaffected by the knowledge stock. Figures~\ref{fig:3statics_1} and~\ref{fig:3statics_2} present numerical results for the model. Output and wages increase with increasing $\beta$, while the capital-labor ratio affects the scale of the effect for a given level of the knowledge stock.

\section{Autonomous Knowledge Generation}\label{sec:aiknow}
We refine the model presented in Section~\ref{sec:knowdriven} by introducing a GPT that autonomously raises R\&D productivity \cite{damioli_impact_2021, brynjolfsson_generative_2023}. For example, AI can enhance the efficiency of idea generation and validation in R\&D by improving search, prediction, and knowledge recombination with reduced or no need for human supervision, possibly super-linearly \cite{weitzman_recombinant_1998,agrawal_finding_2018} (see the work of Wang \textit{et al.} for an account of the use of AI in scientific research \cite{wang_scientific_2023}). The OECD defines an AI system as:
\begin{center}
\textit{...a machine-based system that, for explicit or implicit objectives, infers, from the input it receives, how to generate outputs such as predictions, content, recommendations, or decisions that can influence physical or virtual environments. Different AI systems vary in their levels of autonomy and adaptiveness after deployment.}
\end{center}

In this section we examine the attribute of \textit{autonomy}, and in Section~\ref{sec:adaptive} the attribute of \textit{adaptiveness}. Importantly, both these elements necessitate a different modeling approach than for (mechanistic) automation.

\paragraph{Knowledge accumulation.} We augment the knowledge production function as:
$$
\dot{\mathcal{K}}(t) = \zeta \cdot A(t)^\xi \cdot R(t)^\alpha \cdot \mathcal{K}(t)^\phi,$$
$$\zeta \geq 0,\ \xi \geq 0,\ \alpha \in (0,1],\ \phi \in [0,1],
$$
where:
\begin{itemize}
    \item $A(t)$ denotes the effectiveness of the GPT in R\&D efforts, and
    \item $\xi$ is the elasticity of knowledge production with respect to the GPT's contribution. Elasticity of $\xi > 1$ captures the idea of superlinear returns. The limiting case of no GPT-driven augmentation is restored for $\xi = 0$. 
\end{itemize}
In the case of AI, the term $A(t)$ can be thought of as a distinct class of workers, namely \textit{intelligent autonomous agents}. Consider, for example, the following relationship between the physical (or human), virtual (or non-human), and total R\&D labor $\bar{R}$:
\begin{align*}
R(t) + A(t) &= \bar{R} \Rightarrow \\
(1 - s_A(t)) \bar{R} + s_A(t) \bar{R} &= \bar{R}, \quad s_A \in [0,1],
\end{align*}
where $s_A(t)$ is the ratio of virtual R\&D workers. In the following however we do not use such constraint.

\paragraph{Long-run growth.}
Since technologies such as AI have not yet produced widespread effects on production and economies, it is reasonable to assume that their levels are determined exogenously as $A(t) = A$. We also assume that R\&D labor share remains constant over time with $S_R(t) = S_R$. Then, the growth rate of the knowledge stock is:
\begin{align*}
g_\mathcal{K}(t) &= \zeta \cdot A(t)^\xi \cdot R(t)^\alpha \cdot \mathcal{K}(t)^{\phi - 1} =\\
&= \zeta \cdot A^\xi \cdot (S_R \bar{L})^\alpha \cdot \mathcal{K}(t)^{\phi - 1}.
\end{align*}

On a balanced growth path, $g_\mathcal{K}(t)$ is constant. If $\phi=1$ \cite{romer_endogenous_1990}, the knowledge stock grows at a constant rate:
$$
g_\mathcal{K} = \zeta \cdot A^\xi \cdot (S_R \bar{L})^\alpha.
$$

The corresponding growth rate of production output is:
$$
g_Y = \beta \cdot g_{\mathcal{K}} = \beta \cdot \zeta \cdot A^\xi \cdot (S_R \bar{L})^\alpha,
$$
implying that long-run output growth is sustained, and increasing in AI effectiveness. For $\phi < 1$, the long-run behavior is:
$$
g_Y(t) = \beta \cdot g_\mathcal{K}(t) = \beta \cdot \zeta \cdot A^\xi \cdot (S_R \bar{L})^\alpha \cdot \mathcal{K}(t)^{\phi - 1}.
$$

\begin{figure}
    \centering
    \includegraphics[width=1\linewidth]{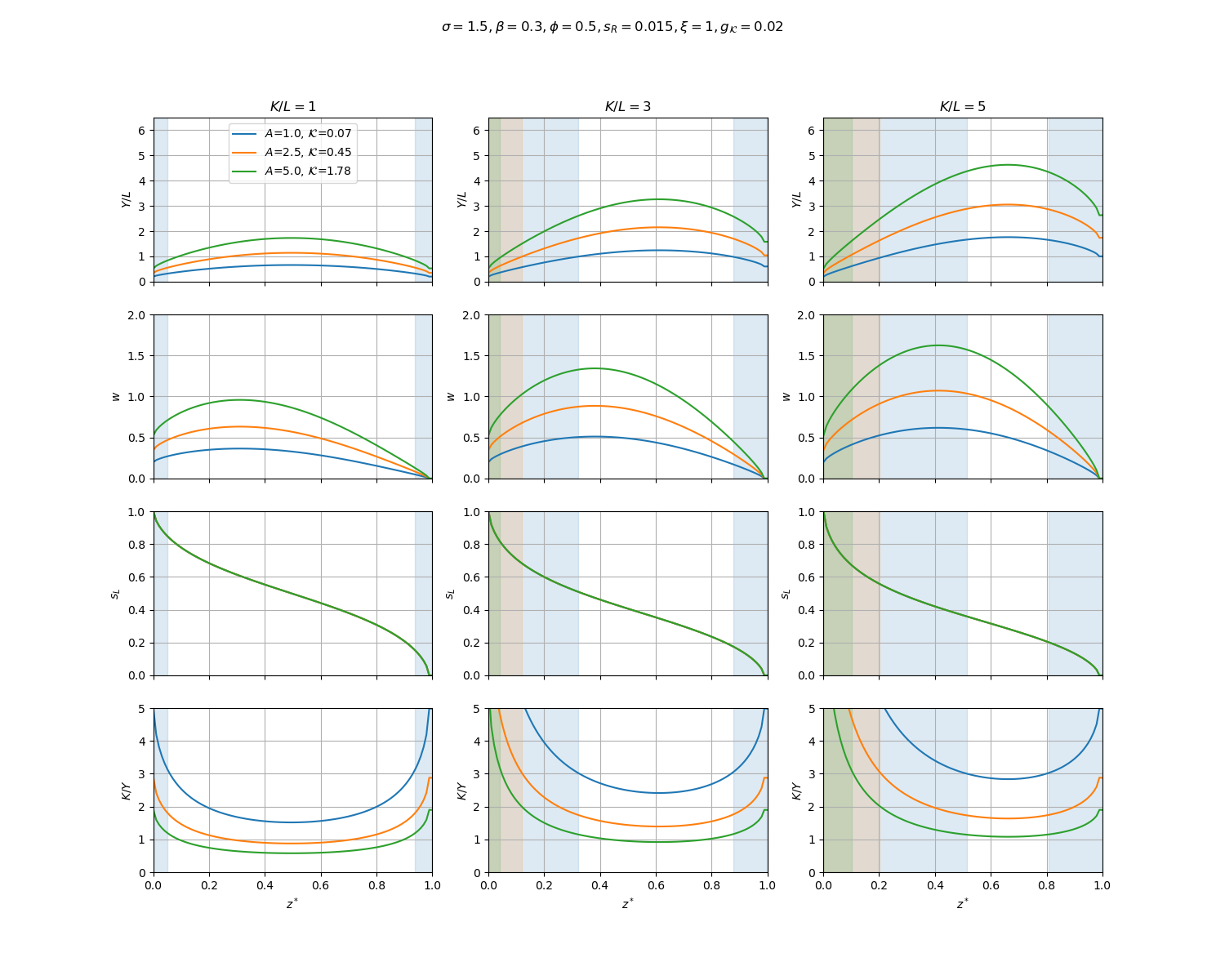}
    \caption{Numerical calculation of key variables of the model with autonomous knowledge generation.}
    \label{fig:4statics_1}
\end{figure}

In this case, sustained growth is feasible only if either $A(t)$ or $R(t)$ increase over time. Table \ref{tab:aiauto_growth} summarizes the behavior of the growth rates and knowledge stock in the long-run. Figures~\ref{fig:4statics_1} and~\ref{fig:4statics_2} present numerical results for the model: increasing effectiveness of GPT contributions increases output and wages, but also decreases significantly the range of $z^*$ where capital remains effective.

\begin{table}[H]
\small
\centering
\begin{adjustbox}{max width=\textwidth} 
\begin{tabular}{@{}lll@{}}
\toprule
\textbf{Variable} & \textbf{Growth Expression} & \textbf{Asymptotic Behavior} \\ \midrule
\multicolumn{3}{l}{$\phi = 1$ (linear spillovers)} \\
$g_{\mathcal{K}}$ & $\zeta \cdot A^\xi \cdot (S_R \bar{L})^\alpha$ & Constant \\
$g_Y$ & $\beta \cdot \zeta \cdot A^\xi \cdot (S_R \bar{L})^\alpha$ & Constant \\
$\mathcal{K}(t)$ & $\propto e^{\zeta A^\xi (S_R \bar{L})^\alpha t}$ & Exponential \\[1ex]

\multicolumn{3}{l}{$\phi < 1$ (sublinear spillovers)} \\
$g_{\mathcal{K}}$ & $\zeta \cdot A^\xi \cdot (S_R \bar{L})^\alpha \cdot \mathcal{K}(t)^{\phi - 1}$ & $\sim \frac{1}{t}$ \\
$g_Y$ & $\beta \cdot \zeta \cdot A^\xi \cdot (S_R \bar{L})^\alpha \cdot \mathcal{K}(t)^{\phi - 1}$ & $\sim \frac{1}{t}$ \\
$\mathcal{K}(t)$ & $\propto t^{\frac{1}{1 - \phi}}$ & Sub-exponential \\[1ex]

\end{tabular}
\end{adjustbox}
\caption{Long-run growth outcomes with autonomous knowledge generation.}
\label{tab:aiauto_growth}
\end{table}

\section{Knowledge Accumulation Costs}
Next, we extend the model presented in Section~\ref{sec:aiknow} by incorporating knowledge generation and validation costs that increase with the proliferation of ideas. The main idea is to realistically limit combinatorial explosion in the knowledge stock. Such frictions include: 
\begin{itemize}
    \item screening funnels and idea prioritization,
    \item peer review for quality control,
    \item construction of costly prototypes,
    \item cognitive or institutional filters,
    \item compliance with regulatory requirements, and
    \item fundamental thermodynamic or energy bounds.
\end{itemize}
For instance, rapid increase in energy demands of state-of-the-art AI models may constrain their ability to (self-)improve, casting doubt on ideas about emergent ``superintelligence'' \cite{nordhaus_are_2015,aghion_artificial_2017}. In fact, as of 2025, this bottleneck has contributed to a renewed interest in nuclear power supply. Simply put, physical constraints substitute for cognitive bottlenecks.  

\paragraph{Knowledge accumulation.} Dampening is achieved by using a \textit{net} knowledge production rate that incorporates increasing generation and validation costs as:
$$
\dot{\mathcal{K}}_{\text{net}}(t) = \dot{\mathcal{K}}(t) - \Psi(\mathcal{K}(t))
$$
where $\Psi(\mathcal{K}(t))$ is a cost function representing the resource or efficiency loss in selecting good ideas from the growing pool.
Without loss of generality, we adopt the simple functional form:
$$
\Psi(\mathcal{K}(t)) = \kappa \cdot \mathcal{K}(t)^\theta, \quad \kappa \geq 0,\ \theta > 0,
$$
where:
\begin{itemize}
    \item $\kappa$ controls the burden of validation (the limit of no frictions is restored for $\kappa = 0$), and
    \item $\theta$ determines the rate at which friction costs escalate with knowledge stock. Convexity is ensured when $\theta > 1$. In fact, net knowledge growth becomes negative if the validation cost term dominates raw idea generation and validation, which implies the possibility of knowledge stagnation or decline. 
\end{itemize}

\paragraph{Long-run growth.}
Assuming fixed GPT level $A(t) = A$ and fixed R\&D labor share $S_R(t) = S_R$ (so that $R = S_R \bar{L}$), the net growth rate of the knowledge stock is:

$$
g_{\mathcal{K},\text{net}}(t) = \zeta \cdot A^\xi \cdot R^\alpha \cdot \mathcal{K}(t)^{\phi - 1} - \kappa \cdot \mathcal{K}(t)^{\theta - 1}.
$$

The system's asymptotic behavior depends on the $(\phi, \theta)$ regime:
\begin{itemize}
    \item $\phi < \theta$. Friction costs rise faster than ideas are generated. 

We define the steady-state $\mathcal{K}^*$ as the solution to:
\begin{align*}
\dot{\mathcal{K}}_{\text{net}} = 0 \quad \Rightarrow\\
\zeta A^\xi R^\alpha \cdot (\mathcal{K}^*)^{\phi} = \kappa \cdot (\mathcal{K}^*)^\theta.
\end{align*}
This implies:
\begin{align*}
(\mathcal{K}^*)^{\theta - \phi} = \frac{\zeta A^\xi R^\alpha}{\kappa} \quad \Rightarrow \\
\mathcal{K}^* = \left( \frac{\zeta A^\xi R^\alpha}{\kappa} \right)^{\frac{1}{\theta - \phi}}.
\end{align*}

Since $\phi < \theta$, it follows that $\mathcal{K}^* < \infty$. It is straightforward to show local stability near $K^*$. Thus, as $t \to \infty$ the terms in the expression of growth rates cancel out: 
$$\lim_{t \to \infty} g_Y(t) = 0, \lim_{t \to \infty} g_\mathcal{K}(t) = 0.$$ 

\item $\phi = \theta$. Both terms scale identically with $\mathcal{K}$ in the long run. The system sits on a knife-edge: growth is constant if and only if $\zeta A^\xi R^\alpha = \kappa$. Any deviation leads to divergence or stagnation.

\item $\phi > \theta$. Idea generation outpaces friction costs asymptotically. The first term dominates and net growth accelerates without bound. A finite-time blow-up in knowledge stock occurs at critical time $t_c$:
$$
\mathcal{K}(t) \sim (t_c - t)^{1/(1 - \phi)} \to \infty.
$$
\end{itemize}

In short, $\theta$ alone does not determine long-run dynamics. It is the relative values of $\phi$ and $\theta$ that determine the trade-off between knowledge accumulation benefits and burdens, enabling an expanded set of intervention entry points. 

\begin{figure}
    \centering
    \includegraphics[width=1\linewidth]{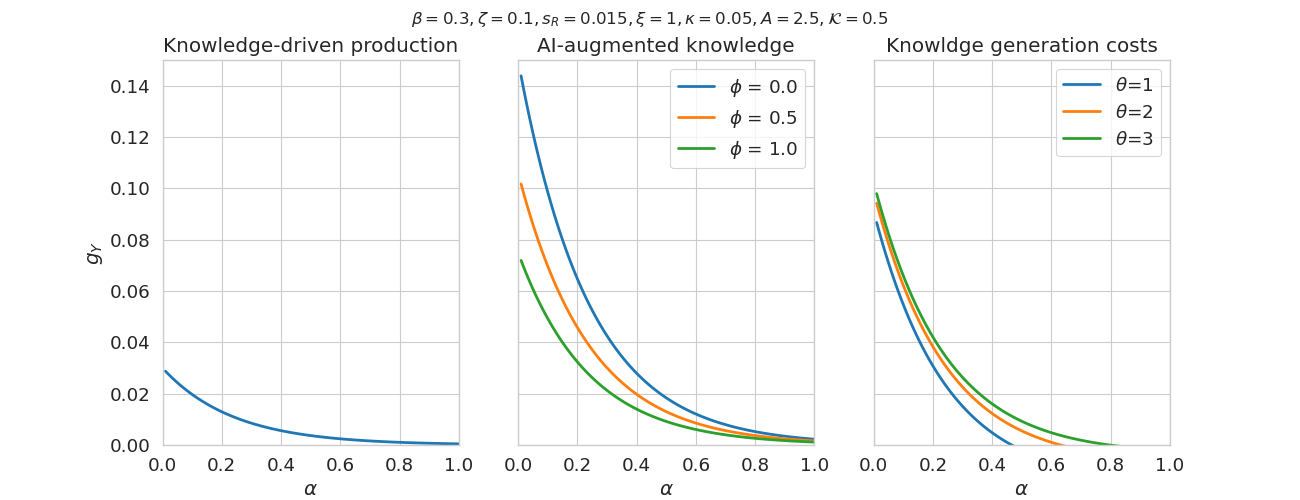}
    \caption{Production growth for three different environments.}
    \label{fig:345g_Y}
\end{figure}

\begin{figure}
    \centering
    \includegraphics[width=1\linewidth]{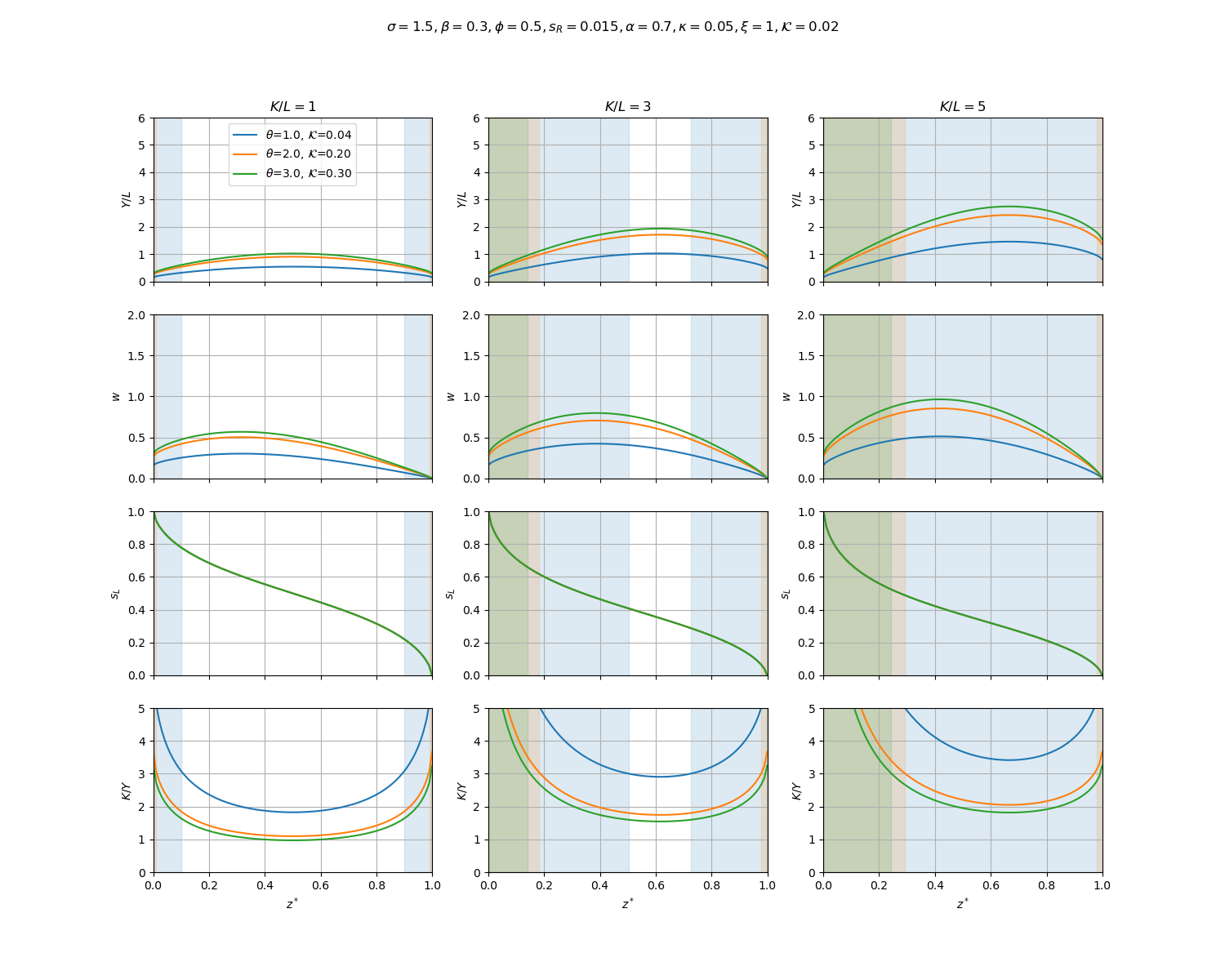}
    \caption{Numerical calculation of key variables of the model with GPT and knowledge accumulation costs.}
    \label{fig:5statics_1}
\end{figure}

\qquad Figure~\ref{fig:345g_Y} shows how production growth changes with parameter $\alpha$, juxtaposed with corresponding trends from Sections~\ref{sec:knowdriven} and~\ref{sec:aiknow}. Fixed knowledge stock and research labor (with total labor normalized to 1) are assumed for comparative static evaluation across different sets of the key parameters $\phi$ and $\theta$. Increasing $\phi$ or decreasing $\theta$ depresses growth, thus the resulting rate depends on their relative values (left panel). Figures~\ref{fig:5statics_1} and~\ref{fig:5statics_2} present numerical results for the model.

\section{Adaptive Knowledge Generation}\label{sec:adaptive}
Finally, we proceed to explicitly define a feedback loop between production and growth (driven by R\&D): automation not only reallocates tasks to capital but it can also enhance the effectiveness of AI-accelerated R\&D. For instance, increased automation is often associated with increased digitalization and improved data governance, which contribute to improvements in AI models and algorithms. A case in point is the deployment of industrial robots which are manipulated by and feed training data back to online AI models. Digital integration of sensors and self-improving AI models is another common theme in advanced manufacturing. A firm operating in this domain has the incentive to strategically invest in capital-intensive projects biasing long-term trajectory of growth.

\qquad We continue to treat the GPT as partially decoupled from broader R\&D effort for analytical clarity. Nevertheless, the model can be readily extended to endogenize it, for instance by allowing $\bar{A}(t)$ to depend explicitly on $\mathcal{K}(t)$. Given that presently much of AI purposefully improving and learning from production systems is concentrated within private firms, while frontier AI R\&D remains primarily with the academic sector, we consider such distinction a grounded abstraction.

\paragraph{Knowledge accumulation.}
To illustrate the effect of production-growth (or automation-AI for that matter) complementarities, we continue with a simple extension to the model. 
Let the level $A(t)$ be increasing in the automation frontier $z^*(t)$. Without loss of generality, we formalize the complementarity with an effective $A(t)$ of the simple form:
\begin{align*}
A(t) &= \bar{A}(t) \cdot \Lambda(z^*(t)), \\
\Lambda (z^*(t)) &= 1 + \lambda z^*(t), \quad \lambda \geq 0,
\end{align*}
where:
\begin{itemize}
    \item $\bar{A}$ is the GPT level absent any contribution from automation, and
    \item $\lambda$ governs the degree to which automation improves GPT effectiveness. Decoupling is restored in the limiting case of $\lambda = 0$.
\end{itemize}
The knowledge production equation is then modified as:
$$
\dot{\mathcal{K}}(t) = \zeta \cdot \bar{A}(t)^\xi \cdot R(t)^\alpha \cdot \mathcal{K}(t)^\phi,$$
with:
$$\zeta = \bar{\zeta} \cdot \Lambda(z^*(t)) = \bar{\zeta} \cdot (1 + \lambda z^*(t)).$$
Consequently, the growth rate of knowledge becomes:
\begin{align*}
    g_\mathcal{K}(t) &=  \bar{\zeta} \cdot A(t)^\xi \cdot R(t)^\alpha \cdot \mathcal{K}(t)^{\phi-1}\\
    &= \zeta \cdot \bar{A}(t)^\xi \cdot R(t)^\alpha \cdot \mathcal{K}(t)^{\phi - 1},
\end{align*}
and the output growth evolves as:
\begin{align*}
g_Y(t) &= \beta \cdot g_\mathcal{K}(t) \\
&= \beta \cdot \zeta \cdot \bar{A}(t)^\xi \cdot R(t)^\alpha \cdot \mathcal{K}(t)^{\phi - 1}.
\end{align*}

The complementarity therefore affects the knowledge production function and rates multiplicatively.

\paragraph{Task creation.}
To further tighten the coupling between production and growth, we allow the set of economically meaningful tasks to expand \textit{endogenously}. For example, the advent of AI created new tasks such as prompt engineering, synthetic data generation, etc.

\qquad Let $M(t)$ denote the size of the task domain, where the economy performs tasks $z \in [0, M(t)]$, with initial value $M(0) = 1$. We adopt the form of the knowledge production function to task creation:
$$
\dot{M}(t) = \chi \cdot A(t)^\xi \cdot R(t)^\alpha \cdot M(t)^\phi, \quad \chi \geq 0.
$$
Thus, GPT-accelerated R\&D not only enhances productivity, but also expands the set of tasks the organization can perform. For $\chi = 0$, the task set remains immutable. The fact that $A(t)$ affects $\dot{M}(t)$, which in turn affects $z^*$ and consequently $A(t)$ via $\Lambda(t)$, creates a closed feedback loop between production and growth. This reflects the idea that even with automation in production, without commensurate innovation efforts growth cannot be sustained in the long run. Conversely, advancing the knowledge frontier yields no sustained output increase unless new knowledge is applied to effectively improve production processes \cite{mokyr_gifts_2005}. We refer to this reinforcing dynamic as \textit{adaptive} growth. Without loss of generality, we assume that the knowledge production function shares the same spillover parameter $\phi$. This simplifies the models, while maintain realism: for example, we expect a tight coupling between knowledge and task creation in AI-native firms (high $\phi$), while for less AI- and automation-intensive firms (low $\phi$) factors other than such spillovers become relevant.

\paragraph{Effect on key variables.}
Output increases due to:
\begin{itemize}
    \item scaling with knowledge accumulation ($Y \propto \mathcal{K}(t)^\beta$), and
    \item integration over a growing task domain $[0, M(t)]$.
\end{itemize}
Differentiating the production function with respect to $M(t)$ and using the Leibniz rule yields:
$$\frac{dY}{dM} > 0,$$
that is, increasing task mass allows more tasks to be performed, thereby raising $Y(t)$ (provided that any added task at the margin has positive output, $y(z,t) > 0$).

\qquad Crucially, the effect of $M(t)$ on the wage or labor share hinges on whether the new task set becomes more or less labor (or capital) intensive. For wages:

$$
w = \mathcal{K}^\beta \cdot \frac{\partial \tilde{Y}}{\partial L}.
$$

\begin{figure}
    \centering
    \includegraphics[width=1\linewidth]{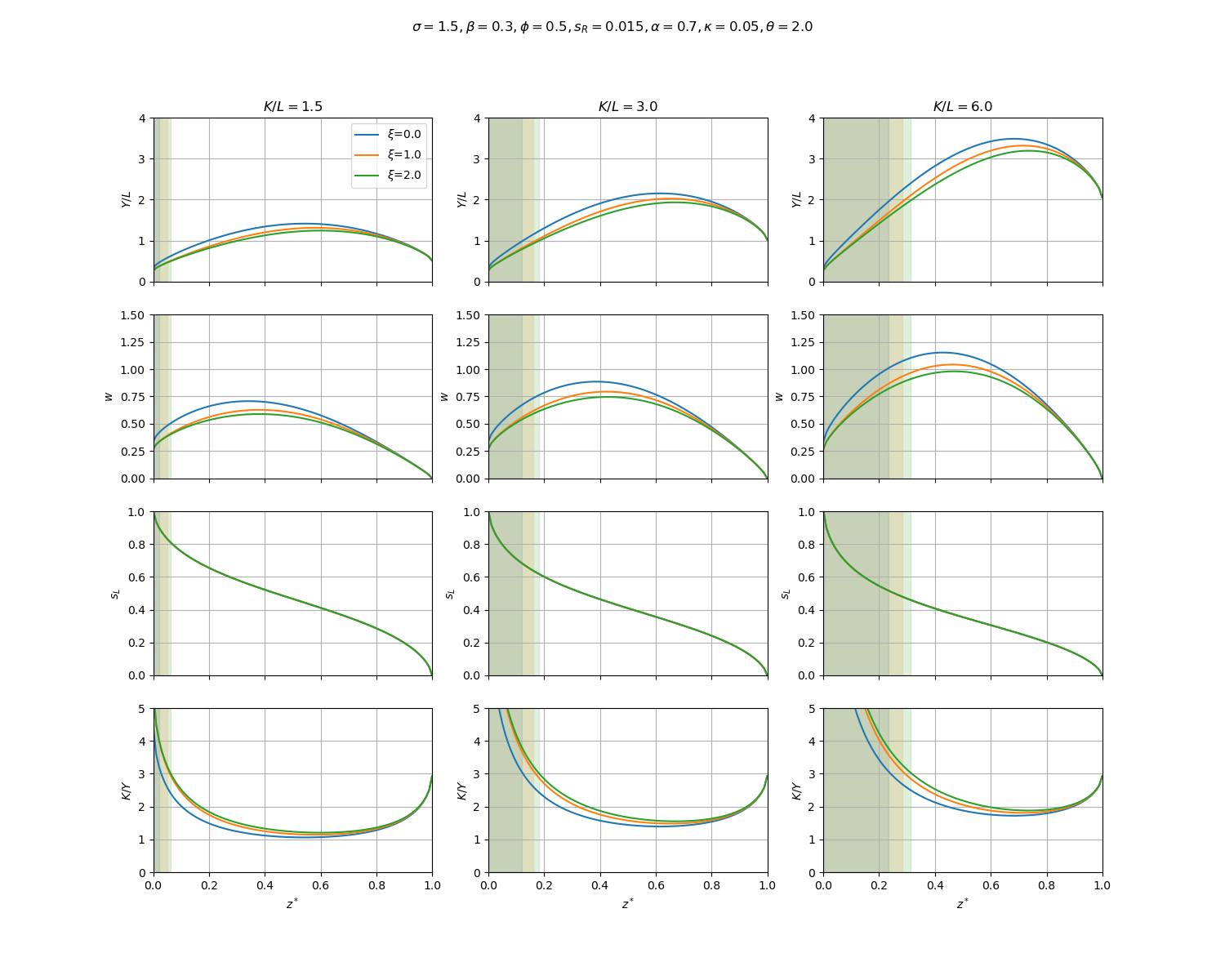}
    \caption{Numerical calculation of key variables of the model with GPT, knowledge accumulation costs, and adaptive knowledge generation.}
    \label{fig:6statics_1}
\end{figure}

If new tasks are labor-intensive, $\partial \tilde{Y} / \partial L(t)$ increases, thus $w$ also increases.
For the labor share:
$$
s_L = \frac{w L}{Y} = \frac{\tilde{w} L}{\tilde{Y}}
$$
Added tasks beyond $M(t)=1$ are increasingly labor-intensive, however if $z^*(t)$ concurrently increases disproportionately, there will eventually be displacement of labor. Figures~\ref{fig:6statics_1} and~\ref{fig:6statics_2} present numerical results for the model. Increasing coupling between automation and growth depresses output and wages, given fixed task mass.

\section{Refinements}
In this section, we briefly discuss straightforward refinements to treating capital and labor, for completeness.

\paragraph{Capital accumulation.} The framework can be readily extended to include capital accumulation in the planner's problem:
$$\dot{K}(t)= s Y(t)-\delta K(t), \quad \delta \in [0,1),$$
where:
\begin{itemize}
    \item $s$ is the savings rate, and
    \item $\delta$ the depreciation rate (for $\delta = 0$ capital becomes permanent). 
\end{itemize}

The extension allows for modeling dynamic capital deepening effects and capital-automation complementarity more accurately. Still, in the following chapters we continue to treat capital as a key structural parameter.

\qquad Similarly, labor can be an explicit state variable, and the share between production and R\&D labor dynamically determined to further improve the model's realism. Finally, the total population has also been held fixed, a non-trivial assumption: although analytically convenient, it obscures an important determinant of future growth (or deceleration, in light of current demographic trends).

\paragraph{Labor heterogeneity.} Another realistic and relevant extension of the model is to introduce heterogeneity in labor inputs beyond the single labor aggregate. For example, the labor input $l(z)$ could be differentiated by skill level (e.g., low- versus high-skill workers). This modification enables the model to account for distributional impacts of automation and AI across heterogeneous workers.

\chapter{Quantitative Analysis}\label{ch:simulation}
Building on the baseline model presented in Section~\ref{sec:base}, Chapter~\ref{ch:modeling} presented a logical sequence of extensions. In this chapter, we proceed to combine them into a unified model, which we will study numerically. The aim is to establish quantitative cause-effect relationships between the model's structural parameters and calculated outputs. 

\section{Full Model}\label{sec:full} We begin by outlining the complete model to be simulated. Definitions from earlier chapters are repeated here for clarity and convenience. 

\paragraph{Structure and dynamics.} 
Let the output at time $t$ be:
\begin{equation}
Y(t) = \mathcal{K}(t)^\beta \cdot \left( \int_0^{M(t)} y(z,t)^{\frac{\sigma - 1}{\sigma}} dz \right)^{\frac{\sigma}{\sigma - 1}} = \mathcal{K}(t)^\beta \cdot \tilde{Y},
\end{equation}
with $y(z,t)$ the output of task $z$ at time $t$, and $\sigma \ne 1, \sigma > 0$ the elasticity of substitution across tasks. Tasks are assigned to capital or labor:
\begin{equation}
y(z,t) =
\begin{cases}
  a_K(z,t) \cdot k(z,t), & \text{if } z \in [0, z^*(t)] \\
  a_L(z,t) \cdot l(z,t), & \text{if } z \in (z^*(t), M(t)]
\end{cases}
\end{equation}
Productivities are scaled by knowledge:
\begin{equation}
a_K(z,t) = \bar{a}_K(z) \mathcal{K}(t)^\beta, \quad a_L(z,t) = \bar{a}_L(z) \mathcal{K}(t)^\beta, \quad \beta \geq 0.
\end{equation}
The expression for the aggregated output can then be rewritten as:
\begin{equation}
\tilde{Y}(z^*) = \left[
\left( \int_{0}^{z^*(t)} \bar{a}_K(z)^{\sigma - 1} dz \right)^{\frac{1}{\sigma}} K^{\frac{\sigma - 1}{\sigma}}
+
\left( \int_{z^*(t)}^{M(t)} \bar{a}_L(z)^{\sigma - 1} dz \right)^{\frac{1}{\sigma}} L^{\frac{\sigma - 1}{\sigma}}
\right]^{\frac{\sigma}{\sigma - 1}},
\end{equation}
given capital and labor resource constraints:
$$
\int_0^{z^*(t)} k(z,t) dz = K(t), \quad \int_{z^*(t)}^{M(t)} l(z,t) dz = L(t).
$$

For the simulation, fixed total labor supply is assumed:
$$
L(t) + R(t) = \bar{L}, \quad L(t) = (1 - S_R(t)) \bar{L}, \quad S_R \in [0,1]
$$

Tasks are assigned to capital up to $z^*(t)$ such that:
\begin{equation}
\frac{w(t)}{a_L(z^*,t)} = \frac{r(t)}{a_K(z^*,t)}.
\end{equation}

Accelerated knowledge accumulation with automation complementarity in light of knowledge generation and validation costs, is described as:
\begin{equation} \label{eq:knowledge}
\dot{\mathcal{K}}(t) = \zeta \cdot \bar{A}(t)^\xi \cdot (1 + \lambda z^*(t))^\xi \cdot R(t)^\alpha \cdot \mathcal{K}(t)^\phi - \Psi(\mathcal{K}(t)), 
\end{equation}
$$
\zeta \geq 0,\ \lambda \geq 0,\ \xi \geq 0,\ \alpha \in (0,1],\ \phi \in [0,1],
$$
with:
\begin{equation}
\Psi(\mathcal{K}(t)) = \kappa \cdot \mathcal{K}(t)^\theta, \quad \kappa \geq 0, \theta \geq 0.
\end{equation}
The two-way feedback between automation and growth is captured by:
\begin{equation}
\dot{M}(t) = \chi \cdot A(t)^\xi \cdot R(t)^\alpha \cdot M(t)^\phi, \quad \chi \geq 0, \xi \geq 0,
\end{equation}
or, for $\xi = \alpha$,
$$
\dot{M}(t) = \chi \cdot \left( A(t) \cdot R(t) \right)^\alpha \cdot M(t)^\phi.
$$
Finally, friction costs associated with increasing automation reduce production output by:
\begin{equation}
\Phi(z^*(t)) = \gamma \cdot \frac{(z^*(t))^{\eta + 1}}{\eta + 1}, \quad \gamma \geq 0, \eta > 0.
\end{equation}

\paragraph{Key variables.} 
Based on this formulation, key variables include the wage and labor share of output. Wage is calculated as:
\begin{align*}
w(t) = \frac{\partial Y(t)}{\partial L(t)} &= \mathcal{K}(t)^\beta \cdot \frac{\partial \tilde{Y}(t)}{\partial L(t)}\\
&= \mathcal{K}(t)^\beta \cdot \tilde{w}(t),
\end{align*}

and the labor share as:
$$
s_L(t) = \frac{w(t) \cdot L(t)}{Y(t)} = \frac{\tilde{w}(t) \cdot L(t)}{\tilde{Y}(t)}.
$$

Growth rates for output and knowledge are calculated from:
\begin{align*}
g_Y(t) &= \beta \cdot g_\mathcal{K}(t) \\
 &= \beta \cdot \frac{\dot{\mathcal{K}}(t)}{\mathcal{K}(t)} \\
 &= \beta \cdot \left[ \zeta \cdot \bar{A}(t)^\xi \cdot (1 + \lambda z^*(t))^\xi \cdot R(t)^\alpha \cdot \mathcal{K}(t)^{\phi - 1} - \kappa \cdot \mathcal{K}(t)^{\theta - 1} \right].
\end{align*}

\paragraph{Optimization.}
For numerical implementation and analysis, the optimization can be decomposed into two layers:
\begin{enumerate}
\item An \textit{inner} problem of solving:
$$
\max_{\{k(z), l(z)\}} \left( \int_0^{M(t)} y(z,t)^{\frac{\sigma - 1}{\sigma}} dz \right)^{\frac{\sigma}{\sigma - 1}},
$$
at each time step $t$. 

\item An \textit{outer} problem of consistently maximizing net output:
$$
\max_{\{z^*_t, s_{R,t}\}_{t=0}^T} \sum_{t=0}^T b^t \cdot \left[ Y_t - \Phi_t \right],
$$
given discount factor $b$, subject to the model's transition equations.
\end{enumerate}

\begin{figure}[ht]
    \centering
    \includegraphics[width=1\linewidth]{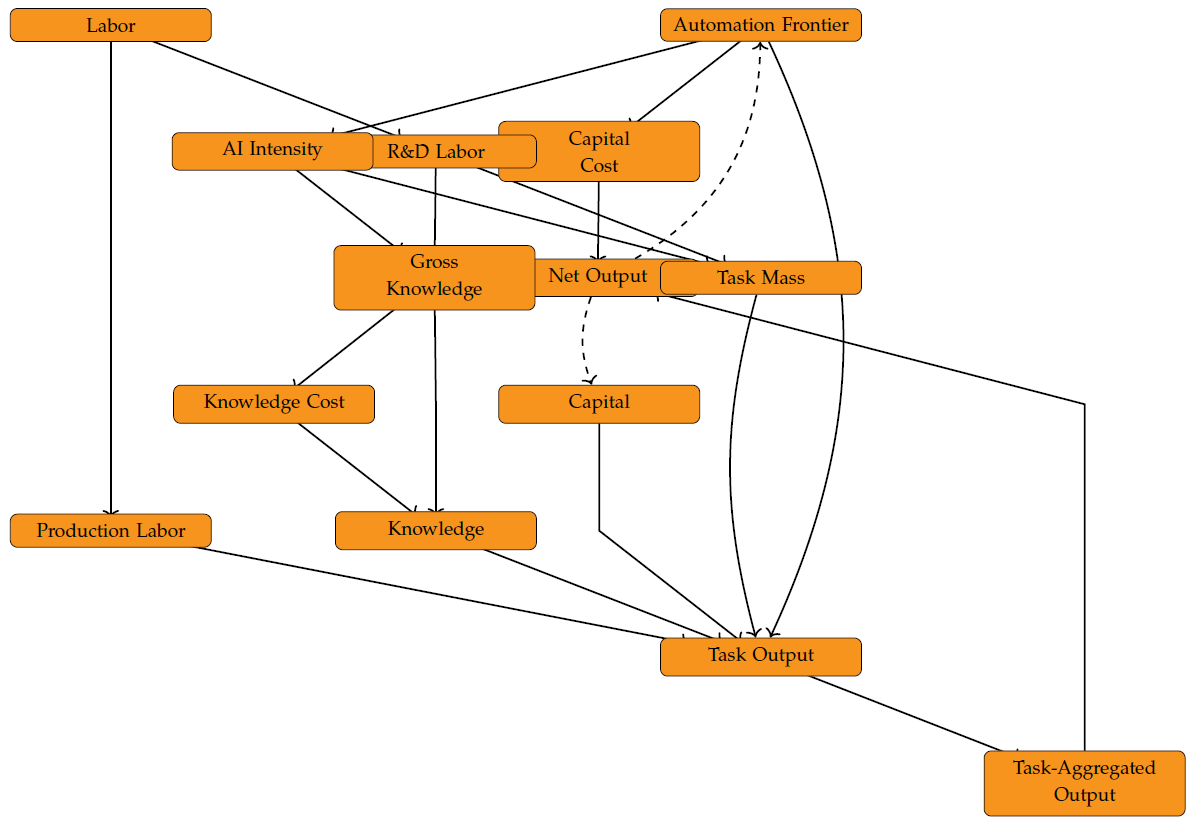}
\caption{Schematic representation of the full model.}
\label{model_schema}
\end{figure}

\qquad While it is possible to simplify the model by absorbing capital frictions and knowledge accumulation costs into variables or exponentials, we elect to describe them explicitly to allow for increased interpretability of the system's dynamics, and to avoid obscuring important feedback mechanisms or possible policy levers. A schematic representation of the full model is shown in Figure~\ref{model_schema}.

\section{Numerical Simulation}\label{sec:numerical}
To evaluate the macroeconomic outcomes of the full model, we numerically simulate its forward dynamics. In light of feedback loops, it is possible depending on the details of the parametrization for different mechanisms to either magnify each other's effect or dampen it.

\qquad We start from numerical simulation of four representative scenarios, summarized in Table~\ref{tab:simulation_scenarios}. Task mass $M(t)$ remains variable across all parametrizations but the scenario with $\zeta,\lambda,\gamma=0$ (labeled ``0''). Based on typical values reported in the literature (Chapter~\ref{ch:theory}) and model statics (Chapter~\ref{ch:modeling}), we explore the parameter space presented in Table~\ref{tab:parameters}. \footnote{To improve stability and accuracy, the predictor-corrector method is used for estimating the change in $\mathcal{K}(t)$ and $M(t)$ (i.e., a ``predictor'' step is combined with an implicit ``corrector'' refinement at the time midpoint).}

\begin{table}[H]
\centering
\small
\begin{tabular}{@{}clccc@{}}
\toprule
\textbf{Parameters} & \textbf{Knowledge Acc.} & \textbf{Task Mass} & \textbf{Automation-AI Integration} & \textbf{Capital Frictions} \\ \midrule
$\zeta, \lambda, \gamma=0$   & No  & Fixed     & No & No \\
$\zeta > 0; \lambda, \gamma=0$ & Yes & Fixed  & No & No \\
$\zeta, \lambda > 0; \gamma=0$ & Yes & Variable  & Yes & No \\
$\zeta, \lambda, \gamma > 0$ & Yes & Variable  & Yes & Yes \\
\bottomrule
\end{tabular}
\caption{Numerical simulation under different parametrizations.}
\label{tab:simulation_scenarios}
\end{table}

\begin{table}[H]
\centering
\small
\begin{tabular}{@{}llll@{}}
\toprule
\textbf{Parameter} & \textbf{Value} & \textbf{Range} & \textbf{Description} \\ \midrule
$\alpha$   & $0.4$  & (0.4, 0.7)                      & Elasticity of knowledge production w.r.t. R\&D labor \\
$\beta$    & $0.3$ & ($0.2$, $0.6$)     & Elasticity of output w.r.t. knowledge stock \\
$\gamma$    & $0.3$ & ($0.0$, $1.0)$  & Frictional capital costs \\
$\zeta$    & $0.1$ & ($0.0$, $0.4$)  & Baseline R\&D productivity parameter \\
$\eta$  & $2.0$ & $(1.0, 3.0)$ & Curvature for capital costs \\
$\theta$   & $2.0$ & ($1.0$, $3.0$)     & Curvature for knowledge generation and validation \\
$\kappa$   & $0.1$ & ($0.0$, $0.3)$   & Scale of knowledge generation and validation costs \\
$\lambda$   & $2.0$  & ($0.0$, $3.0$)    & Prefactor of AI-automation complementarity \\
$\xi$    & $0.4$ &  ($0.2$, $0.6$)             & Exponent on AI productivity in knowledge production \\
$\sigma$   & $2.0$ & ($0.8$, $3.0$)   & Elasticity of substitution across tasks \\
$S_R$      & $0.015$& $(0.01,0.03)$  & Share of labor allocated to R\&D \\ 
$\phi$     & $0.5$ & ($0.25$, $1.00$)   & Spillover effect strength \\
$\chi$  & $0.003$ & $(0.0, 0.01)$ & Task domain expansion rate \\
 \bottomrule
\end{tabular}
\caption{Model parameters, baseline values, and exploration values.}
\label{tab:parameters}
\end{table}

\begin{figure}[H]
    \centering
    \includegraphics[width=1\linewidth]{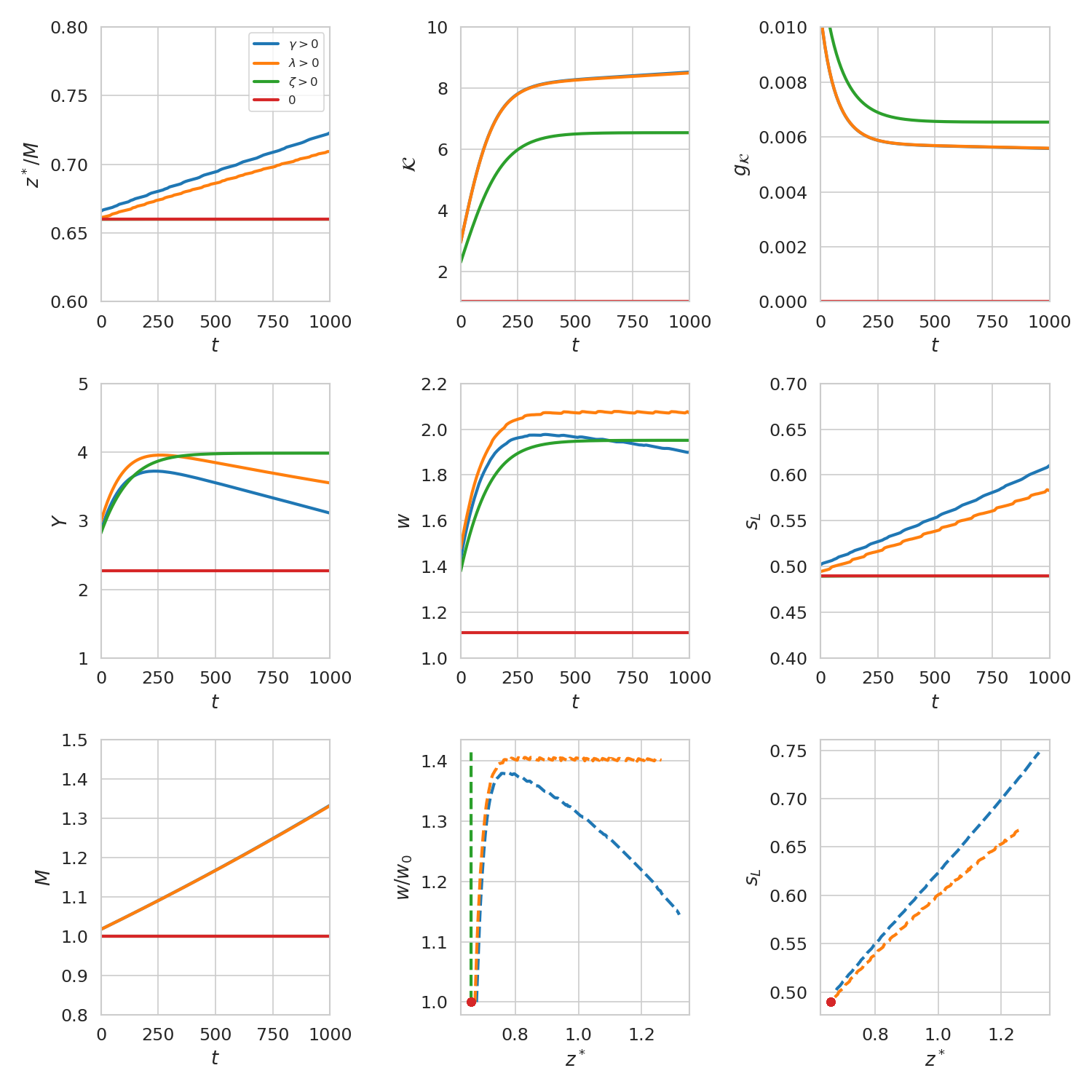}
    \caption{Numerical simulation of the full model for the representative scenarios of Table~\ref{tab:simulation_scenarios} (label ``0'' for $\zeta, \lambda, \gamma=0$).}
    \label{fig:simulation}
\end{figure}

\qquad Figure~\ref{fig:simulation} presents time series of key variables, the knowledge accumulation path, and phase diagrams with respect to $z^*$ for each of the four numerical simulations. Across all scenarios, except the benchmark $\zeta, \lambda, \gamma=0$, $z^*$ increases steadily, confirming an expanding automation frontier. Moreover, the knowledge stock initially undergoes rapid expansion and then slows down, while the direct production-growth coupling significantly amplifies growth. Expansion of the automation frontier and knowledge stock drive an initial growth phase in production output.

\qquad However, capital frictions and an expanding set of tasks under conditions of fixed capital and labor supply $\bar{K}/(1-S_R)\bar{L}=3$, have a detrimental effect on the output. Increases in wages are coupled to expanding knowledge stock, but soon decouple from output to subsequently either saturate or decrease slowly. This is an important implication of the task-based framework. Knowledge costs exacerbate the effect. Interestingly, the labor share increases with task expansion. This is a minor effect nevertheless and $s_L$ is primarily controlled by the capital-labor factor ratio as we show below. The main takeaway is that wages, labor share, and output, although correlated, are not directly linked.

\qquad The phase diagram of scaled wage $w/w_0$ in Figure~\ref{fig:simulation} shows a threshold of automation above which marginal wage gains can reverse. In the realistic scenario of capital frictions, the finding suggests that an optimal $z^*$ might exist from the perspective of social welfare, necessitating public policy intervention to achieve it. 

\section{Policy Design}
The scenarios examined in Section~\ref{sec:numerical} provide a first insight into the compounding effects arising from the interaction between the model's otherwise distinct structural components. To devise policy instruments that effectively tip the balance towards desirable outcomes, it is desirable to understand how the structural parameters affect production and labor outcomes in the long-run. To that end, we go beyond selected parametrizations to explore in an \textit{automated, high-throughput} fashion the parameter space in the neighborhood of a set of baseline parameters (Table~\ref{tab:parameters}) and attempt to uncover trends in the model's behavior. This is a reasonable approach in absence of closed-form solutions and in light of many interacting mechanisms. 

\begin{figure}
    \centering
    \includegraphics[width=1\linewidth]{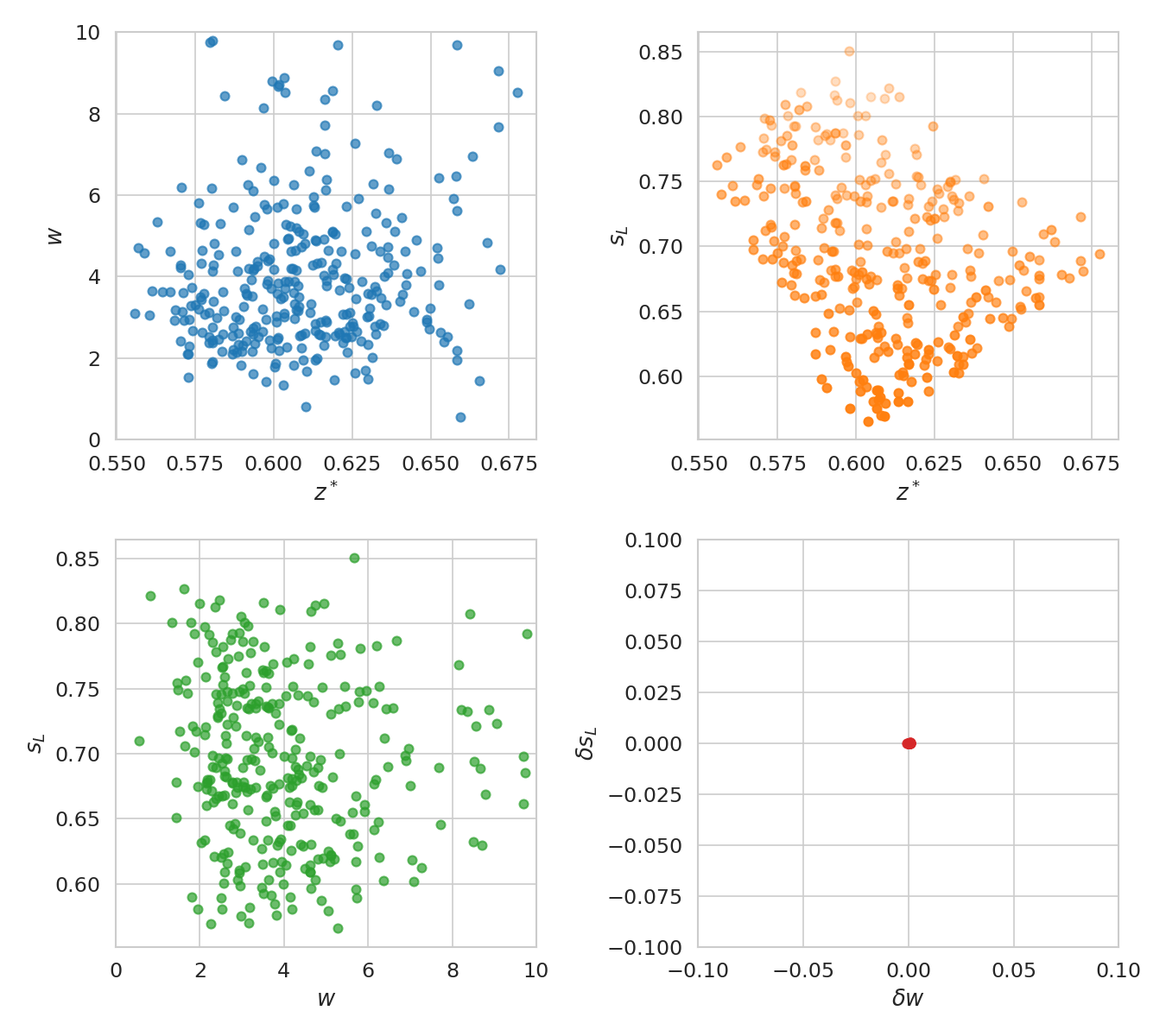}
    \caption{Wages $w$, labor share $s_L$, and automation $z^*$ from high-throughput calculations across the structural parameter space.}
    \label{fig:screening}
\end{figure}

\qquad First, we generate a reliable dataset that associates input structural parameters with the calculated (or output) variables $w$ and $s_L$. Five hundred simulations were performed, from which only those that have converged ($\delta s_L \approx 0, \delta w \approx 0$) are selected for further analysis. Figure~\ref{fig:screening} summarizes the results. 

\qquad There is a negative correlation between the labor share $s_L$ and automation frontier $z^*$ as higher automation levels are associated with lower labor shares, on average, consistent with what the analytical model predicts. A similar pattern holds between $s_L$ and the capital-labor ratio: higher ratios (shown as more opaque points) drive the simulated system into lower $s_L$. This can be understood as the effect of \textit{capital deepening}, where access to more capital effectively displaces labor \cite{ray_growth_2020}. 

\qquad A clear correlation between $w$ and $z^*$ or $s_L$ is harder to establish. The high dispersion observed in $(w,z^*)$ and $(w,s_L)$ is an indication of the increased sensitivity of wages to multiple underlying structural parameters. Depending on the parameter regime, automation and access to capital can either raise or depress wages at difference pace, in agreement with prior findings \cite{acemoglu_capital_2024}. One possible outcome is \textit{complete collapse} of the labor share at very high capital availability. On the other hand, for any given capital-output (or labor) ratio, there are a host of possible scenarios where labor share may stabilized, if not increase. 

\qquad To identify simple rules in the model's behavior, we use a \textit{random forest regressor}, as implemented in the machine learning Python library \textit{scikit-learn}. With this ensemble supervised approach a meta estimator fits a number of decision tree regressors on various sub-samples of the high-throughput calculations, and uses averaging to improve predictive accuracy.\footnote{Model selection was conducted using a 80/20 train/validation split for optimal depth of the tree, and number of samples at an internal nodes.} Decision trees were selected for their computational tractability. The ensemble functions effectively as a \textit{surrogate} model, i.e., an approximation of the underlying system that is agnostic to its internal structural mechanisms. This reduction of complexity however comes at a cost of \textit{interpretability}, which we restore next.

\begin{figure}
    \centering
    \includegraphics[width=1\linewidth]{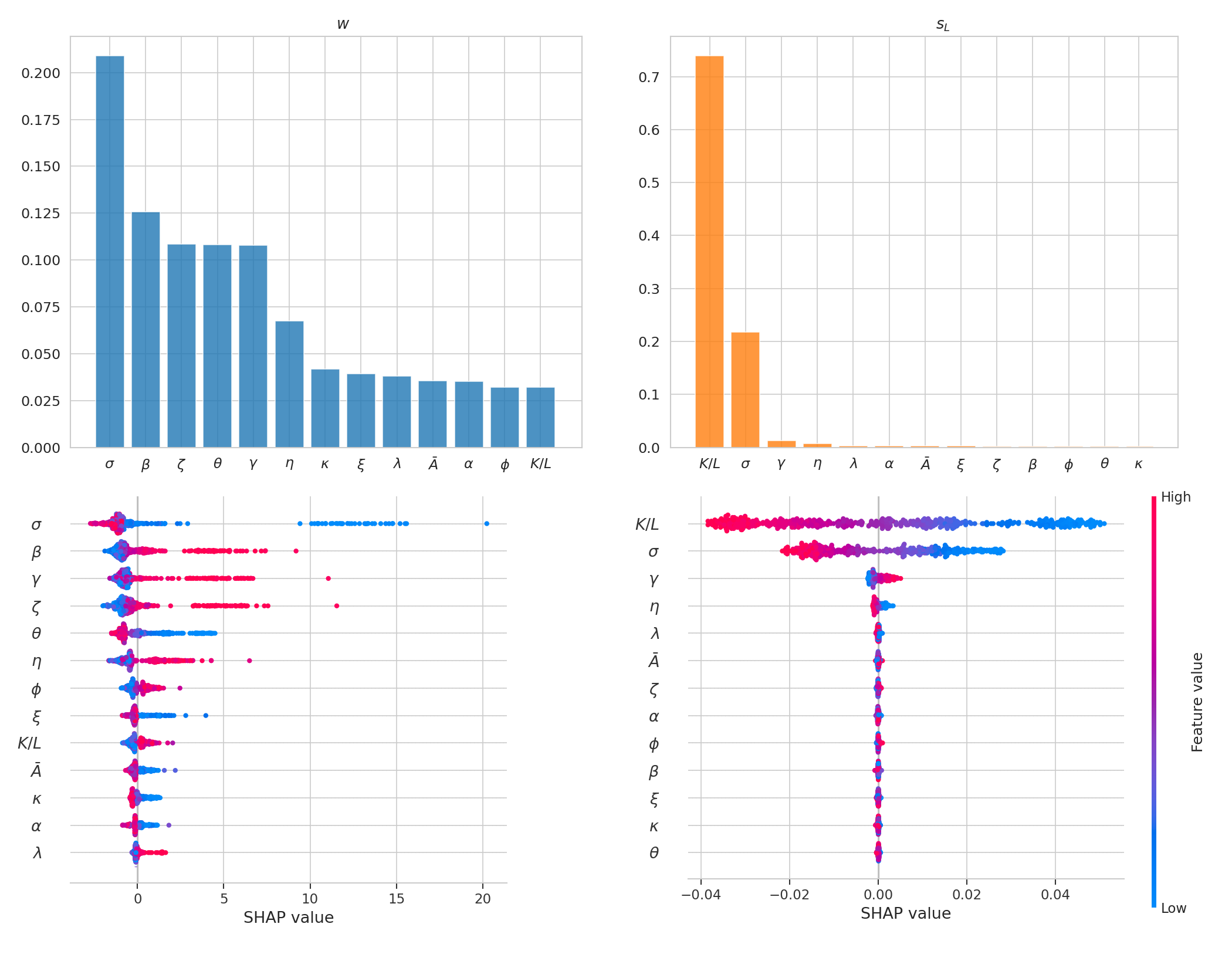}
    \caption{Influence of the full model's structural parameters on its outputs using (top) impurity-based feature importance, and (bottom) Shapley additive explanations.}
    \label{fig:features}
\end{figure}

\qquad Figure~\ref{fig:features} ranks parameters based on how much they influence the surrogate's model predictive performance for $w$ and $s_L$. Two different approaches are used for assessing the influence of structural parameters:
\begin{itemize}
    \item \textit{Impurity-based feature importance}. It is calculated as reduction in the mean squared error brought by a given input across the tree-based model. The higher the value, the more important the feature.
    \item \textit{Shapley additive explanations; SHAP}. A game-theoretic method that assigns each feature a value that describes the average marginal contribution of that feature to the prediction, across all possible feature coalitions (or orderings). Let a model $f(x)$ take a vector of $N$ features. The Shapley value for feature $i$ with respect to input $x$ is:
    $$
    \phi_i(x) = \sum_{S \subseteq N \setminus \{i\}} \frac{|S|! \, (n - |S| - 1)!}{n!} \left[ f_{S \cup \{i\}}(x_{S \cup \{i\}}) - f_S(x_S) \right],
    $$
    where:
    \begin{itemize}
        \item $f_S(x_S)$ is the model prediction when only features in $S$ are known,
        \item $f_{S \cup \{i\}}(x_{S \cup \{i\}}) - f_S(x_S)$ describes the marginal contribution of feature $i$ to coalition $S$. 
    \end{itemize}
\end{itemize}
The weight accounts for the number of possible permutations in which $S$ precedes $i$. 

\qquad Figure~\ref{fig:features} shows results for the two methods. Clearly, wages are sensitive to many structural parameters, mostly related to knowledge accumulation. On the other hand, $s_L$ strongly depends on the capital-labor ratio. Each dot in the bottom panels in Figure~\ref{fig:features} represents a calculation run. Positive (negative) SHAP values correspond to calculations where a structural parameter increased (reduced) the prediction of the variable. Red (blue) color is mapped to high (low) value for a structural parameter on a percentile scale.

\begin{figure}
    \centering
    \includegraphics[width=1\linewidth]{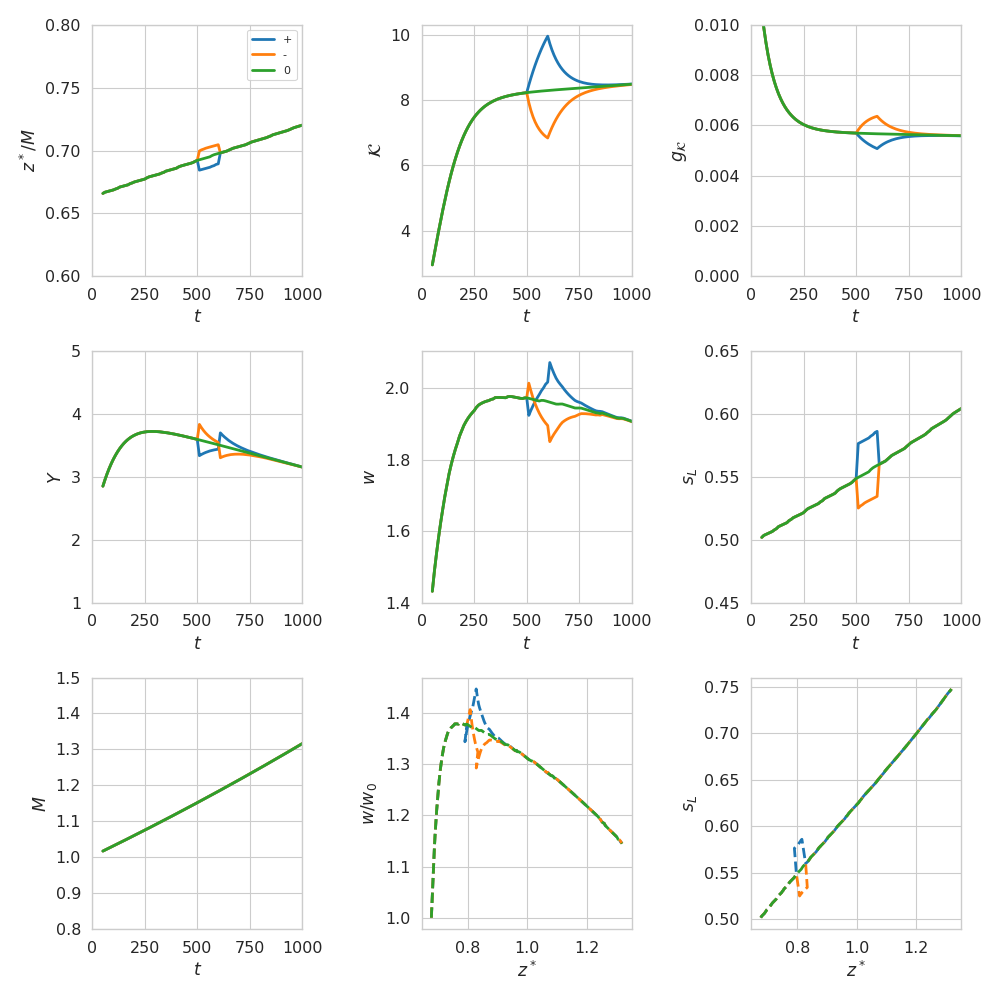}
    \caption{Simulated trajectories under two stylized policy intervention scenarios (blue and orange). The unperturbed model simulation (green) is shown for reference.}
    \label{fig:policy}
\end{figure}

\qquad Crucially, with the exception of $\sigma$, $s_L$ is less sensitive to the same parameters that influence more $w$, such as $\theta$ (costs associated with knowledge accumulation) and $\zeta$ (R\&D productivity). Effectively, the two parameters can be controlled independently. Based on this insight, we devise two stylized scenarios that we test by introducing a perturbation (or \textit{shock}) to the simulation in Section~\ref{sec:numerical} ($\zeta, \lambda, \gamma > 0$):
\begin{enumerate}
    \item increasing temporarily $K/L$ and $\theta$ by 10\% (label ``-'' in Figure~\ref{fig:policy}),
    \item decreasing temporarily $K/L$ and $\theta$ by 10\% (``+'').
\end{enumerate}
The scenarios vary the parameters in opposite directions symmetrically, which functions as a simple robustness check of the model's mechanism. The calculated response of important variables like $\mathcal{K}, w,$ and $s_L$ is likewise symmetric. It is straightforward to extrapolate these results to regimes characterized by low growth and high capital stocks, and vice versa.

\paragraph{Policy levers.} For the reference $K/L=3$ and $\theta=2$, a 10\% perturbation results in $\sim 4\%$ shift in $w$ and $s_L$ (Figure~\ref{fig:policy}), a non-trivial trivial change. The capital-labor ratio can be realistically affected by such policy instruments as targeted public procurement or occupational regulation. Knowledge generation, validation and related costs can be affected by, for example, promotion of open innovation and open standards, establishment of regulatory sandboxes, and improved linkages between public R\&D centers and private firms. Table~\ref{tab:policy-mapping} offers an overview of interventions that map to key mechanisms of the model. 

\begin{table}
\small
\centering
\begin{tabular}{p{5cm} p{9cm}}
\toprule
\textbf{Mechanism} & \textbf{Policy Instrument(s)} \\
\midrule
Capital efficiency & Employment-linked investment allowances; progressive tax on capaital; strategic public procurement \\
Labor efficiency & Education reform; digital skills; occupational regulation; labor standards  \\
Knowledge accumulation & R\&D tax credits; regulatory sandboxes; open innovation platforms; public-private partnerships \\
Technological lock-in & Innovation tax credits; interoperability mandates; infrastructure transition grants \\
\bottomrule
\end{tabular}
\caption{Mapping interventions to model mechanisms.}
\label{tab:policy-mapping}
\end{table}

\chapter{Economic Development}\label{ch:development}
Economic development goes beyond mere economic growth and productivity improvement to bring in issues of structural transformation of production, human development, and institutional reform, to name a few \cite{herrendorf_chapter_2014, todaro_economic_2020}. In this chapter, we focus on few selected elements of development that directly relate to earlier discussion in long-run productivity trends, technological change, and shifts in labor and capital allocation. 

\section{Theories and Models}
\begin{figure}
    \centering
    \includegraphics[width=0.75\linewidth]{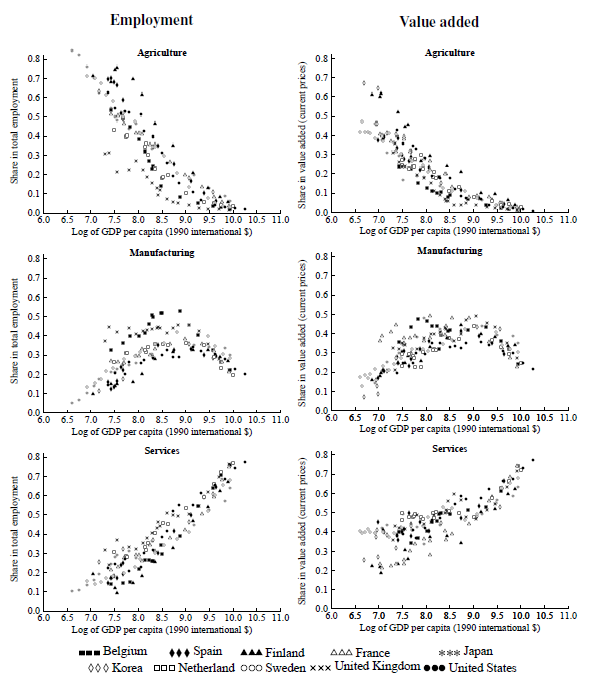}
    \caption{Shares in (left) employment and (right) value added of main economic sectors \cite{herrendorf_chapter_2014}.}
    \label{fig:dev_industries}
\end{figure}

As discussed in Section~\ref{sec:concepts}, early growth models such as the \textbf{Solow-Swan} model imply that economies should conditionally converge in the long run \cite{solow_contribution_1956, swan_economic_1956}. Empirical evidence suggests differently: economic trajectories globally have diverged since the industrial revolution (Figure~\ref{fig:dev_industries}). The very notion of a balanced growth trajectory is not immediately relevant to the more volatile economic and political environment of developing countries. 

\qquad As Kuznets had already noted early on, GDP is not a complete measure of a nation's well-being or societal progress in terms of health, education, and other dimensions \cite{kuznets_national_1941}. Developing countries are often less endowed with human resources, have higher person-to-land ratios, and are in a disadvantaged position with regards to scientific and technological capabilities \textit{vis-\`a-vis} developed countries. 

\qquad Naively, it can be claimed that by increasing savings and investments the transition through the \textbf{stages of development} will be accelerated. This line of reasoning (theoretically supported by the Rostow and Harrod-Domar models \cite{rostow_stages_1991,harrod_essay_1972,domar_capital_1946}), has been used to justify large transfers of capital and technical assistance from developed to less developed nations in the past, without uniform success nevertheless. Higher savings and investments are a necessary, but not a sufficient condition for accelerated progress if economic development is stimulated within a deficient institutional framework or in absence of competent labor that prevent the efficient allocation of capital. 

\qquad The early model by \textbf{Lewis} provided a framework for describing \textbf{structural transformation} from traditional, low-productivity sectors (e.g., agriculture) to modern, high-productivity ones (e.g., manufacturing) \cite{lewis_economic_1954}. This dual-sector approach became the main theory of the development process in the late 1960s. 

\qquad Suppose that the low-productivity sector is associated with surplus labor, $L_1$, which is the only input to production, and that the high-productivity sector is associated with deficiency in labor, $L_2$. The total labor supply is:
$$L = L_1 + L_2.$$
The surplus in the traditional sector implies:
$$
\frac{\partial Y_1}{\partial L_1} = 0,
$$
for large $L_1$.
In contrast, the modern sector operates under the neoclassical production function:
$$
Y_2 = F(K_2, L_2), \quad \frac{\partial F}{\partial L_2} > 0, \quad \frac{\partial F}{\partial K_2} > 0.
$$
Production in this sector increases as a result of reinvestment. With the objective of profit maximization, firms offer in the competitive labor market a marginal wage $w_2 > w_1$. This incentivizes migration of workers from the traditional to the modern sector until all surplus labor has been absorbed (known as the \textit{Lewis turning point}).

\qquad The model's success lies in conceptualizing economic development as a staged process of structural transformation that roughly captures historical trends in industrialization, while highlighting the importance of initial conditions such as market structure or capital-labor endowments. Of course, migration is not instantaneous in practice (e.g., due to differences in educational and community enforcements costs), while profits can be increasingly invested in labor-saving capital equipment with an adverse effect on employment shares. 

\qquad Between the 1970s and the 1990s, versions of the \textbf{international-dependence theory} became popular. The theory viewed institutional, economic, and political rigidities in developing countries as the result of their being caught in a dependence and dominance relationship with the industrialized countries. Dependence theories however mostly lacked practical prescriptions for initiating and sustaining development. In fact, developing countries that pursued a policy of autarky such as China have experienced mixed results. In the 1980s, \textbf{economic growth theory} gained renewed momentum (see also Section~\ref{sec:concepts}) \cite{barro_convergence_1991}, while more recently interest in empirically grounded microeconomic analysis also grew \cite{banerjee_poor_2011}.

\section{Trade and Global Integration}
Historically, industrialization has enabled accelerated economic growth, while international trade offered a path to economies of scale in manufactured goods based on a country's comparative advantage. The work of Rodrick documents a significant de-industrialization trend in recent decades however, and provides evidence that both globalization and labor-saving technological progress have been behind these developments \cite{rodrik_premature_2016}. The challenge for developing countries lies in making the transition to a service-based economy prematurely, in the absence of a solid industrial base to sustain it. Within the manufacturing sector too, countries tend to be more successful when diversifying into nearby and related products that require similar knowhow to build on existing capabilities \cite{hidalgo_product_2007}. For example, countries with comparative advantage in automotives are more likely to also have competency in electronics. In contrast, the persistent \textit{digital divide}, that is, cross-country asymmetries in effective absorption of technical knowledge and access to computational resources critical to robotics, AI, and related frontier technology, exacerbates developmental divergence.

\begin{figure}
    \centering
    \includegraphics[width=0.85\linewidth]{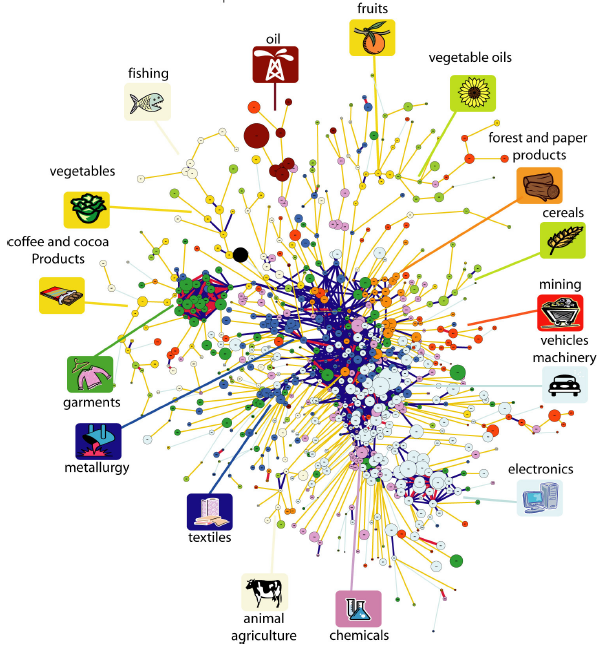}
    \caption{Relatedness of over 800 products using international trade data \cite{hidalgo_product_2007}.}
    \label{fig:product_space}
\end{figure}

\begin{figure}
    \centering
    \includegraphics[width=0.60\linewidth]{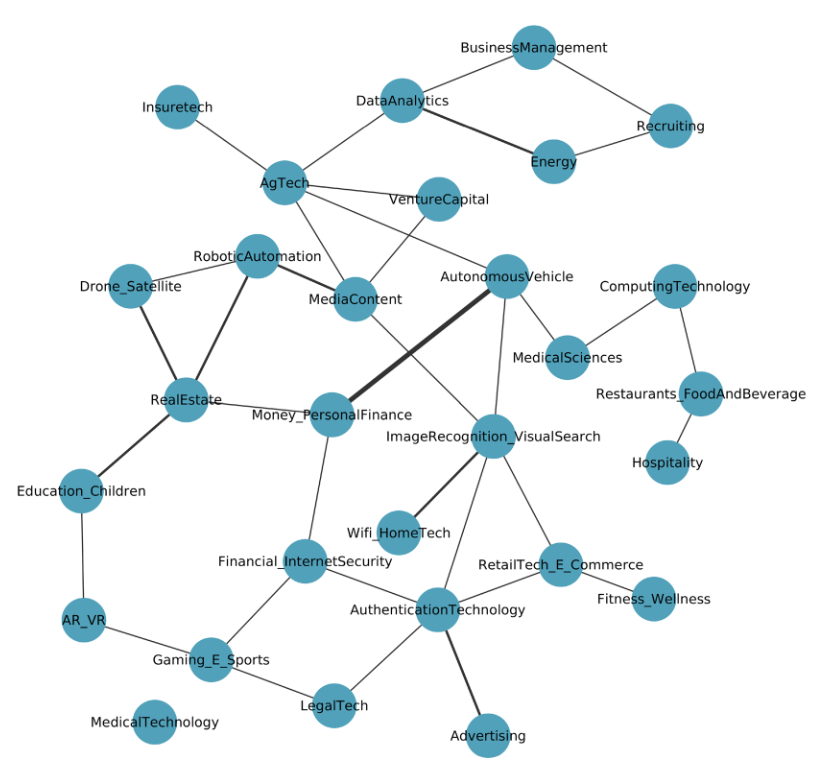}
    \caption{Relatedness of AI specializations \cite{mishra_ai_2023}}
    \label{fig:ai_specialization}
\end{figure}

\qquad Figure~\ref{fig:product_space} shows the network of relatedness of over 800 products using international trade data. Each node is a product, and two products are connected if they are frequently co-exported by countries with revealed comparative advantage (RCA) in both. Developed economies export more of the products that are found near the core of the network, while less developed economies produce more products in the periphery of the network. Mishra \textit{et al.} used international trade and private market data (e.g., UN COMTRADE and Crunchbase) to uncover similar connections specifically across AI specializations (Figure~\ref{fig:ai_specialization}) \cite{mishra_ai_2023}. Countries may attempt to increase the complexity and value-added of their productive structures by jumping empirically ``infrequent'' distances, for example from applications of AI in hospitality services or energy to autonomous vehicles. They are then likely to face implementation challenges, unless they have commensurate state capacity to support targeted industrial, infrastructure, and investment policies, and tolerate accompanying social transformations. 

\qquad The picture becomes more complicated as the world continues to undergo a shift towards an international order shaped by heightened trade and security concerns. Foreign direct investment (FDI) and infrastructure finance are becoming less instruments of open markets and multilateral cooperation, and more of geopolitical influence \cite{farrell_weaponized_2019}. For example, in contrast to the multilateral lending and investment frameworks championed by global financial institutions such as the IMF and the World Bank, or by regional actors such as the EU with its emphasis on policy conditionality, transparency, and institutional reform, China's infrastructure-first development model (exemplified by its Belt and Road Initiative), offers flexible financing with fewer political conditions, which dilutes multilateral norms \cite{imf_world_2025}. 

\section{Sustainable Development}
This most widely accepted definition of sustainable development was given by the UN Brundtland Commission in 1987 as ``meeting the needs of the present without compromising the ability of future generations to meet their own needs.'' In most applied frameworks, this principle integrates in some form three dimensions:
\begin{itemize}
    \item \textit{Economic.} Sustained growth, technological progress, innovation.
    \item \textit{Social.} Access to education, health, and opportunities.
    \item \textit{Environmental.} Respect for the global commons, biodiversity, and planetary bounds.
\end{itemize}

A common theme is the global interdependence of nations: actions of one country can directly affect economic growth and development in others, often in asymmetric ways. Consider for example the population-energy-climate nexus. Rising population in a developing economy can stimulate economic growth but a concurrent surge in energy consumption may in addition generate climate externalities, no less consequential than the emissions of high performance computing centers in an advanced economy at work to develop proprietary state-of-the-art AI technology. 

\begin{figure}
    \centering
    \includegraphics[width=0.8\linewidth]{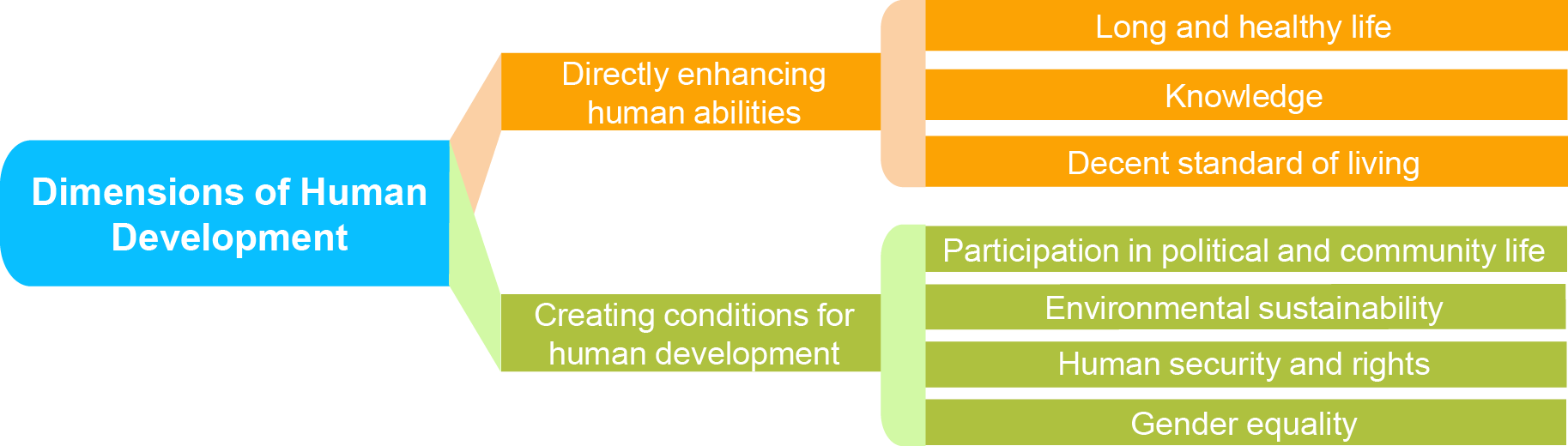}
    \caption{Dimensions of human development according to the UNDP.}
    \label{fig:human_dimensions}
\end{figure}

\qquad A related concept is that of human development (Figure~\ref{fig:human_dimensions}). According to the UNDP, the human development lens focuses on:
\begin{itemize}
    \item \textit{People}. Improving the lives people lead rather than assuming that economic growth will lead, automatically, to greater opportunities for all.
    \item \textit{Opportunities.} Developing people's abilities and giving them a chance to use them.
    \item \textit{Choices.} Providing people with opportunities, not insisting that they make use of them. 
\end{itemize}

Simple metrics such as the \textit{Human Development Index} (HDI) allow for a human-centered view of progress. It is based on three distinct dimensions, namely:
\begin{itemize}
    \item \textit{Health.} Life expectancy at birth.
    \item \textit{Education.} Mean years of schooling, and expected years of schooling.
    \item \textit{Living Standards.} Gross national income per capita (PPP-adjusted).
\end{itemize}

\begin{figure}
\centering
    \includegraphics[width=\textwidth]{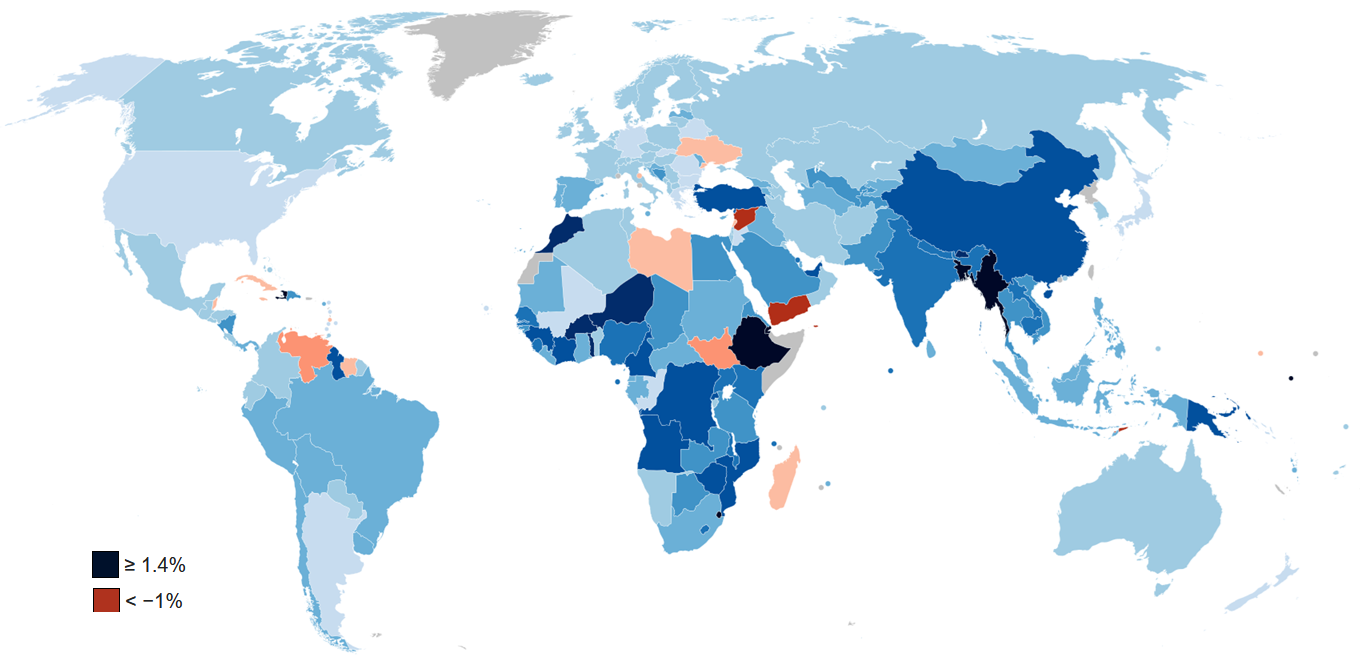}
  \hfill
    \includegraphics[width=\textwidth]{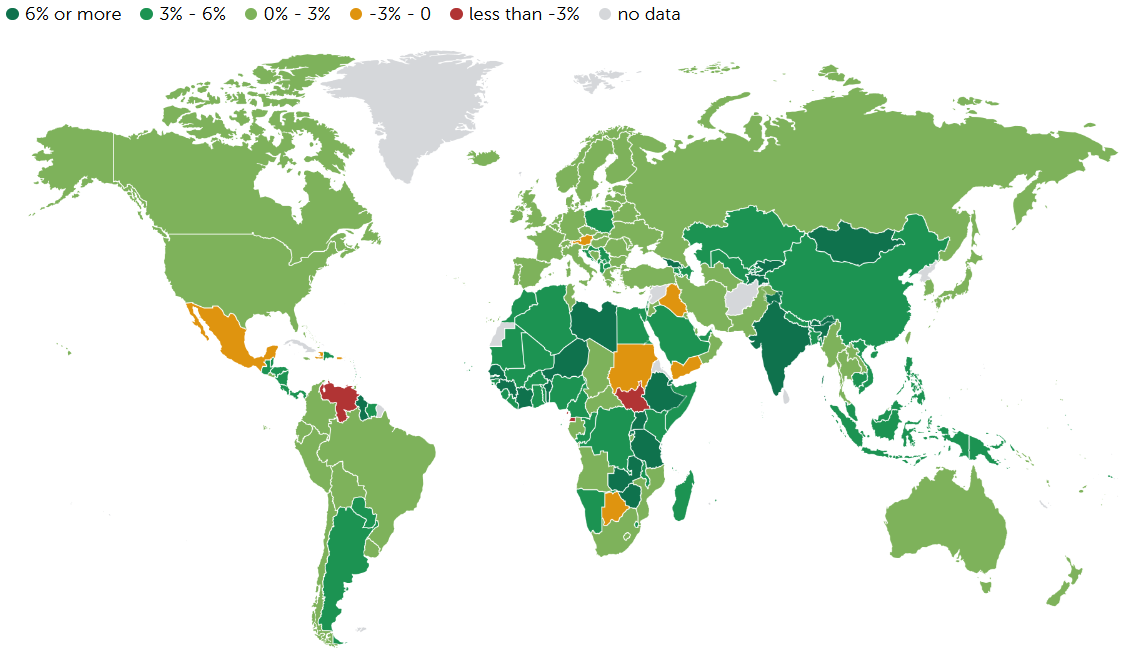}
  \caption{Annual growth rate of (top) human development index and (bottom) real GDP. Data by UNDP and IMF (published 2025).}
  \label{fig:dev_metrics}
\end{figure}

Figure~\ref{fig:dev_metrics} shows average annual HDI growth and real GDP growth. Comparison between the two reveals that although economic growth and human development are generally correlated, they diverge in various regions. For instance, countries such as Turkey and South Africa perform more poorly on the HDI than would be predicted from their income level, while the reverse is true of Cuba and Kenya \cite{todaro_economic_2020}. Lower levels of human development can impose additional constraints on how capital and labor are mobilized. 

\section{Knowledge Diffusion}
Although it is beyond the scope of this work to address in depth the effect of cross-national interdependencies, our analytical model can be readily modified to reflect the resulting asymmetries in the development and adoption of frontier technology across countries \cite{stiglitz_markets_1989,comin_exploration_2010}. For example, in a country that lacks the capacity (or will) to develop national computing and AI infrastructure, firms may instead import AI technology (e.g., by using a non-domestic, proprietary large language models via a web interface). To describe this scenario, the GPT effectiveness in Equation \ref{eq:knowledge} is modeled more realistically as following a convergence process:
$$
\dot{\bar{A}}(t) = \rho \cdot T(t) \cdot \left( \tilde{A} - \bar{A}(t) \right), \quad \rho \geq 0,
$$
where:
\begin{itemize}
    \item $T(t)$ represents trade openness, FDI inflows, or absorptive capacity (e.g., due to incompatible industrial structures or workforce skills),
    \item $\tilde{A}$ is the global technology frontier, exogenous to the domestic economy, and
    \item $\rho$ is the speed of technology diffusion ($\rho = 0$ eliminates the effect).
\end{itemize}

This introduces a lag in benefiting fully and timely from developments at the technology frontier. Time delays can also be used to model how economic outputs and structural changes lag the development of frontier or general-purpose technologies even within an economy \cite{griliches_hybrid_1957,brynjolfsson_artificial_2017,bloom_diffusion_2021}. As a case in point, wages increased only a lifetime after the onset of the industrial revolution, during which total factor productivity changed little.

\chapter{Conclusions and Outlook}\label{ch:conclusions}
The Fourth Industrial Revolution, currently driven by such general-purpose technologies as robotics and artificial intelligence, has been marked by an apparent divergence in the role of labor and capital. Crucially, for many exposed to these developments, whether on the factory floor or in the office, what is at stake is a redefinition of human relevance and agency.

\qquad To interrogate the impact of automation and artificial intelligence on production and labor structures, I relied on a conceptualization of occupations as bundles of tasks that can be allocated either to capital or labor. The main contribution of the work is the combination of a task-based formulation of production with endogenous knowledge accumulation and frictions associated with technological lock-in and the burden of generating and validating new knowledge. These have been treated mostly separately in prior work, which obscures how feedback loops between production and growth shape the system's trajectory. 

\qquad Numerical simulation is used to maintain tractability in spite of the increased complexity of the model. To discover underlying patterns in the model's response to varying structural parameters, I used data-driven techniques based on supervised machine learning such as random forest regression. For example, in a stylized simulation scenario where the capital-labor ratio and knowledge accumulation costs are simultaneously increased, both wages and labor share decline. A key finding is that wages and labor share are not directly linked. Instead, they can be raised or lowered independently. Generally, labor share depends sensitively on capital-labor ratio, while wages respond positively to larger knowledge stocks.

\qquad The model does not consider elements of otherwise practical importance such as collective bargaining by labor unions and employment protection mechanisms. Instead, it aims to offer a simple baseline for empirical comparisons. Treating explicitly labor as heterogeneous, and the diffusion of technology across economic sectors or countries as a staged rather than instantaneous process are natural refinements towards increased realism. 

\qquad The production's sensitivity primarily to the knowledge and capital stocks is not particularly surprising. More significantly, the results underscore the potential need for policy intervention. The discussion includes possible instruments for targeted interventions for mitigating pressures on labor that arise from increased capital deployment for automation. In catch-up economies, for example, raising capital share through use of non-native automation technology may have an adverse long-run economic effect in the absence of policies that aim to simultaneously develop the broader research and innovation ecosystem and expand workforce capabilities \cite{nelson_technical_1993,tritsaris_interdisciplinary_2019}. 

\qquad Future work could extend the framework to incorporate political economy considerations: ultimately, it is institutions that dictate which policy instruments are activated. The greater risk associated with current emerging technologies may well prove to be the slow diffusion of impactful applications due to insufficient complementary investments and regulatory frameworks, notwithstanding the risk of regulatory capture by frontier firms in technology.

{
\bibliography{references}

@misc{almeida_artificial_2024,
	address = {Rochester, NY},
	type = {{SSRN} {Scholarly} {Paper}},
	title = {Artificial {Intelligence} and the {Discovery} of {New} {Ideas}: {Is} an {Economic} {Growth} {Explosion} {Imminent}?},
	shorttitle = {Artificial {Intelligence} and the {Discovery} of {New} {Ideas}},
	url = {https://papers.ssrn.com/abstract=4709699},
	doi = {10.2139/ssrn.4709699},
	language = {en},
	urldate = {2024-02-18},
	author = {Almeida, Derick and Naudé, Wim and Sequeira, Tiago},
	month = feb,
	year = {2024},
}

@misc{capraro_impact_2023,
	title = {The impact of generative artificial intelligence on socioeconomic inequalities and policy making},
	url = {http://arxiv.org/abs/2401.05377},
	doi = {10.48550/arXiv.2401.05377},
	urldate = {2024-02-18},
	publisher = {arXiv},
	author = {Capraro, Valerio and Lentsch, Austin and Acemoglu, Daron and Akgun, Selin and Akhmedova, Aisel and Bilancini, Ennio and Bonnefon, Jean-François and Brañas-Garza, Pablo and Butera, Luigi and Douglas, Karen M. and Everett, Jim A. C. and Gigerenzer, Gerd and Greenhow, Christine and Hashimoto, Daniel A. and Holt-Lunstad, Julianne and Jetten, Jolanda and Johnson, Simon and Longoni, Chiara and Lunn, Pete and Natale, Simone and Rahwan, Iyad and Selwyn, Neil and Singh, Vivek and Suri, Siddharth and Sutcliffe, Jennifer and Tomlinson, Joe and van der Linden, Sander and Van Lange, Paul A. M. and Wall, Friederike and Van Bavel, Jay J. and Viale, Riccardo},
	month = dec,
	year = {2023},
	note = {arXiv:2401.05377 [cs]},
}

@article{acemoglu_robots_2020,
	title = {Robots and {Jobs}: {Evidence} from {US} {Labor} {Markets}},
	volume = {128},
	issn = {0022-3808},
	shorttitle = {Robots and {Jobs}},
	url = {https://www.journals.uchicago.edu/doi/abs/10.1086/705716},
	doi = {10.1086/705716},
	number = {6},
	urldate = {2024-02-18},
	journal = {Journal of Political Economy},
	author = {Acemoglu, Daron and Restrepo, Pascual},
	month = jun,
	year = {2020},
	pages = {2188--2244},
}

@article{acemoglu_colonial_2001,
	title = {The {Colonial} {Origins} of {Comparative} {Development}: {An} {Empirical} {Investigation}},
	volume = {91},
	issn = {0002-8282},
	shorttitle = {The {Colonial} {Origins} of {Comparative} {Development}},
	url = {https://www.aeaweb.org/articles?id=10.1257/aer.91.5.1369},
	doi = {10.1257/aer.91.5.1369},
	language = {en},
	number = {5},
	urldate = {2024-02-26},
	journal = {American Economic Review},
	author = {Acemoglu, Daron and Johnson, Simon and Robinson, James A.},
	month = dec,
	year = {2001},
	pages = {1369--1401},
}

@article{acemoglu_reversal_2002,
	title = {Reversal of {Fortune}: {Geography} and {Institutions} in the {Making} of the {Modern} {World} {Income} {Distribution}},
	volume = {117},
	issn = {0033-5533},
	shorttitle = {Reversal of {Fortune}},
	url = {https://doi.org/10.1162/003355302320935025},
	doi = {10.1162/003355302320935025},
	number = {4},
	urldate = {2024-02-26},
	journal = {The Quarterly Journal of Economics},
	author = {Acemoglu, Daron and Johnson, Simon and Robinson, James A.},
	month = nov,
	year = {2002},
	pages = {1231--1294},
}

@misc{acemoglu_institutions_2004,
	address = {Rochester, NY},
	type = {{SSRN} {Scholarly} {Paper}},
	title = {Institutions as the {Fundamental} {Cause} of {Long}-{Run} {Growth}},
	url = {https://papers.ssrn.com/abstract=541706},
	language = {en},
	urldate = {2024-02-26},
	author = {Acemoglu, Daron and Johnson, Simon and Robinson, James A.},
	month = may,
	year = {2004},
}

@article{olsson_biogeography_2005,
	title = {Biogeography and long-run economic development},
	volume = {49},
	issn = {0014-2921},
	url = {https://www.sciencedirect.com/science/article/pii/S0014292103001016},
	doi = {10.1016/j.euroecorev.2003.08.010},
	number = {4},
	urldate = {2024-02-26},
	journal = {European Economic Review},
	author = {Olsson, Ola and Hibbs, Douglas A.},
	month = may,
	year = {2005},
	pages = {909--938},
}

@article{acemoglu_capital_2024,
	title = {Capital and {Wages}},
	issn = {1556-5068},
	url = {https://www.ssrn.com/abstract=4746801},
	doi = {10.2139/ssrn.4746801},
	language = {en},
	urldate = {2024-03-07},
	journal = {NBER Working Papers},
	author = {Acemoglu, Daron},
	month = mar,
	year = {2024},
}

@article{autor_skill_2003,
	title = {The {Skill} {Content} of {Recent} {Technological} {Change}: {An} {Empirical} {Exploration}},
	volume = {118},
	issn = {0033-5533},
	shorttitle = {The {Skill} {Content} of {Recent} {Technological} {Change}},
	url = {https://www.jstor.org/stable/25053940},
	number = {4},
	urldate = {2024-03-12},
	journal = {The Quarterly Journal of Economics},
	author = {Autor, David H. and Levy, Frank and Murnane, Richard J.},
	year = {2003},
	note = {Publisher: Oxford University Press},
	pages = {1279--1333},
}

@article{alesina_culture_2015,
	title = {Culture and {Institutions}},
	volume = {53},
	issn = {0022-0515, 2328-8175},
	language = {English},
	number = {4},
	urldate = {2024-03-29},
	journal = {Journal of Economic Literature},
	author = {Alesina, Alberto and Giuliano, Paola},
	month = dec,
	year = {2015},
	pages = {898--944},
}

@article{glaeser_institutions_2004,
	title = {Do institutions cause growth?},
	volume = {9},
	issn = {1381-4338, 1573-7020},
	language = {English},
	number = {3},
	urldate = {2024-03-29},
	journal = {Journal of Economic Growth},
	author = {Glaeser, E. L. and La Porta, R. and Lopez-de-Silanes, F. and Shleifer, A.},
	month = sep,
	year = {2004},
	pages = {271--303},
}

@article{koch_robots_2021,
	title = {Robots and {Firms}},
	volume = {131},
	issn = {0013-0133},
	url = {https://doi.org/10.1093/ej/ueab009},
	doi = {10.1093/ej/ueab009},
	number = {638},
	urldate = {2024-04-13},
	journal = {The Economic Journal},
	author = {Koch, Michael and Manuylov, Ilya and Smolka, Marcel},
	month = aug,
	year = {2021},
	pages = {2553--2584},
}

@article{acemoglu_institutions_2014,
	title = {Institutions, {Human} {Capital} and {Development}},
	issn = {1556-5068},
	url = {http://www.ssrn.com/abstract=2392106},
	doi = {10.2139/ssrn.2392106},
	language = {en},
	urldate = {2024-05-09},
	journal = {SSRN Electronic Journal},
	author = {Acemoglu, Daron and Gallego, Francisco A. and Robinson, James A.},
	year = {2014},
}

@misc{aghion_artificial_2017,
	type = {Working {Paper}},
	series = {Working {Paper} {Series}},
	title = {Artificial {Intelligence} and {Economic} {Growth}},
	url = {https://www.nber.org/papers/w23928},
	doi = {10.3386/w23928},
	urldate = {2024-06-24},
	publisher = {National Bureau of Economic Research},
	author = {Aghion, Philippe and Jones, Benjamin F. and Jones, Charles I.},
	month = oct,
	year = {2017},
	doi = {10.3386/w23928},
}

@article{acemoglu_race_2018,
	title = {The {Race} between {Man} and {Machine}: {Implications} of {Technology} for {Growth}, {Factor} {Shares}, and {Employment}},
	volume = {108},
	issn = {0002-8282},
	shorttitle = {The {Race} between {Man} and {Machine}},
	url = {https://www.aeaweb.org/articles?id=10.1257/aer.20160696},
	doi = {10.1257/aer.20160696},
	language = {en},
	number = {6},
	urldate = {2024-06-24},
	journal = {American Economic Review},
	author = {Acemoglu, Daron and Restrepo, Pascual},
	month = jun,
	year = {2018},
	pages = {1488--1542},
}

@misc{agrawal_finding_2018,
	type = {Working {Paper}},
	series = {Working {Paper} {Series}},
	title = {Finding {Needles} in {Haystacks}: {Artificial} {Intelligence} and {Recombinant} {Growth}},
	shorttitle = {Finding {Needles} in {Haystacks}},
	url = {https://www.nber.org/papers/w24541},
	doi = {10.3386/w24541},
	urldate = {2024-07-01},
	publisher = {National Bureau of Economic Research},
	author = {Agrawal, Ajay and McHale, John and Oettl, Alex},
	month = apr,
	year = {2018},
	doi = {10.3386/w24541},
}

@misc{agrawal_artificial_2023,
	type = {Working {Paper}},
	series = {Working {Paper} {Series}},
	title = {Artificial {Intelligence} and {Scientific} {Discovery}: {A} {Model} of {Prioritized} {Search}},
	shorttitle = {Artificial {Intelligence} and {Scientific} {Discovery}},
	url = {https://www.nber.org/papers/w31558},
	doi = {10.3386/w31558},
	urldate = {2024-07-01},
	publisher = {National Bureau of Economic Research},
	author = {Agrawal, Ajay K. and McHale, John and Oettl, Alexander},
	month = aug,
	year = {2023},
	doi = {10.3386/w31558},
}

@misc{ray_growth_2020,
	type = {Working {Paper}},
	series = {Working {Paper} {Series}},
	title = {Growth, {Automation} and the {Long} {Run} {Share} of {Labor}},
	url = {https://www.nber.org/papers/w26658},
	doi = {10.3386/w26658},
	urldate = {2024-07-01},
	publisher = {National Bureau of Economic Research},
	author = {Ray, Debraj and Mookherjee, Dilip},
	month = jan,
	year = {2020},
	doi = {10.3386/w26658},
}

@article{mishra_ai_2023,
	title = {{AI} specialization for pathways of economic diversification},
	volume = {13},
	copyright = {2023 The Author(s)},
	issn = {2045-2322},
	url = {https://www.nature.com/articles/s41598-023-45723-x},
	doi = {10.1038/s41598-023-45723-x},
	language = {en},
	number = {1},
	urldate = {2024-07-02},
	journal = {Scientific Reports},
	author = {Mishra, Saurabh and Koopman, Robert and De Prato, Giuditta and Rao, Anand and Osorio-Rodarte, Israel and Kim, Julie and Spatafora, Nikola and Strier, Keith and Zaccaria, Andrea},
	month = nov,
	year = {2023},
	pages = {19475},
}

@article{frey_future_2017,
	title = {The future of employment: {How} susceptible are jobs to computerisation?},
	volume = {114},
	issn = {0040-1625},
	shorttitle = {The future of employment},
	url = {https://www.sciencedirect.com/science/article/pii/S0040162516302244},
	doi = {10.1016/j.techfore.2016.08.019},
	urldate = {2024-07-02},
	journal = {Technological Forecasting and Social Change},
	author = {Frey, Carl Benedikt and Osborne, Michael A.},
	month = jan,
	year = {2017},
	pages = {254--280},
}

@article{santarelli_automation_2023,
	title = {Automation and related technologies: a mapping of the new knowledge base},
	volume = {48},
	issn = {1573-7047},
	shorttitle = {Automation and related technologies},
	url = {https://doi.org/10.1007/s10961-021-09914-w},
	doi = {10.1007/s10961-021-09914-w},
	language = {en},
	number = {2},
	urldate = {2024-07-05},
	journal = {The Journal of Technology Transfer},
	author = {Santarelli, Enrico and Staccioli, Jacopo and Vivarelli, Marco},
	month = apr,
	year = {2023},
	pages = {779--813},
}

@misc{acemoglu_skills_2010,
	type = {Working {Paper}},
	series = {Working {Paper} {Series}},
	title = {Skills, {Tasks} and {Technologies}: {Implications} for {Employment} and {Earnings}},
	shorttitle = {Skills, {Tasks} and {Technologies}},
	url = {https://www.nber.org/papers/w16082},
	doi = {10.3386/w16082},
	urldate = {2024-07-07},
	publisher = {National Bureau of Economic Research},
	author = {Acemoglu, Daron and Autor, David},
	month = jun,
	year = {2010},
	doi = {10.3386/w16082},
}

@article{damioli_impact_2021,
	title = {The impact of artificial intelligence on labor productivity},
	volume = {11},
	issn = {2147-4281},
	url = {https://doi.org/10.1007/s40821-020-00172-8},
	doi = {10.1007/s40821-020-00172-8},
	language = {en},
	number = {1},
	urldate = {2024-07-08},
	journal = {Eurasian Business Review},
	author = {Damioli, Giacomo and Van Roy, Vincent and Vertesy, Daniel},
	month = mar,
	year = {2021},
	pages = {1--25},
}

@incollection{katz_chapter_1999,
	title = {Chapter 26 - {Changes} in the {Wage} {Structure} and {Earnings} {Inequality}},
	volume = {3},
	url = {https://www.sciencedirect.com/science/article/pii/S1573446399030072},
	urldate = {2024-07-08},
	booktitle = {Handbook of {Labor} {Economics}},
	publisher = {Elsevier},
	author = {Katz, Lawrence F. and Autor, David H.},
	editor = {Ashenfelter, Orley C. and Card, David},
	month = jan,
	year = {1999},
	doi = {10.1016/S1573-4463(99)03007-2},
	pages = {1463--1555},
}

@book{north_institutions_1990,
	title = {Institutions, {Institutional} {Change} and {Economic} {Performance}},
	isbn = {978-0-521-39734-6},
	language = {en},
	publisher = {Cambridge University Press},
	author = {North, Douglass C.},
	month = oct,
	year = {1990},
}

@misc{bloom_diffusion_2021,
	type = {Working {Paper}},
	series = {Working {Paper} {Series}},
	title = {The {Diffusion} of {New} {Technologies}},
	url = {https://www.nber.org/papers/w28999},
	doi = {10.3386/w28999},
	urldate = {2024-07-15},
	publisher = {National Bureau of Economic Research},
	author = {Bloom, Nicholas and Hassan, Tarek Alexander and Kalyani, Aakash and Lerner, Josh and Tahoun, Ahmed},
	month = jul,
	year = {2021},
	doi = {10.3386/w28999},
}

@misc{brynjolfsson_generative_2023,
	type = {Working {Paper}},
	series = {Working {Paper} {Series}},
	title = {Generative {AI} at {Work}},
	url = {https://www.nber.org/papers/w31161},
	doi = {10.3386/w31161},
	urldate = {2024-08-17},
	publisher = {National Bureau of Economic Research},
	author = {Brynjolfsson, Erik and Li, Danielle and Raymond, Lindsey R.},
	month = apr,
	year = {2023},
	doi = {10.3386/w31161},
}

@article{bresnahan_general_1995,
	title = {General purpose technologies ‘{Engines} of growth’?},
	volume = {65},
	issn = {0304-4076},
	url = {https://www.sciencedirect.com/science/article/pii/030440769401598T},
	doi = {10.1016/0304-4076(94)01598-T},
	number = {1},
	urldate = {2024-08-19},
	journal = {Journal of Econometrics},
	author = {Bresnahan, Timothy F. and Trajtenberg, M.},
	month = jan,
	year = {1995},
	pages = {83--108},
}

@article{acemoglu_simple_2025,
	title = {The simple macroeconomics of {AI}},
	volume = {40},
	issn = {0266-4658},
	url = {https://doi.org/10.1093/epolic/eiae042},
	doi = {10.1093/epolic/eiae042},
	number = {121},
	urldate = {2025-06-03},
	journal = {Economic Policy},
	author = {Acemoglu, Daron},
	month = jan,
	year = {2025},
	pages = {13--58},
}

@article{zeira_workers_1998,
	title = {Workers, {Machines}, and {Economic} {Growth}*},
	volume = {113},
	issn = {0033-5533},
	url = {https://doi.org/10.1162/003355398555847},
	doi = {10.1162/003355398555847},
	number = {4},
	urldate = {2025-06-04},
	journal = {The Quarterly Journal of Economics},
	author = {Zeira, Joseph},
	month = nov,
	year = {1998},
	pages = {1091--1117},
}

@misc{acemoglu_tasks_2021,
	type = {Working {Paper}},
	series = {Working {Paper} {Series}},
	title = {Tasks, {Automation}, and the {Rise} in {US} {Wage} {Inequality}},
	url = {https://www.nber.org/papers/w28920},
	doi = {10.3386/w28920},
	urldate = {2025-06-15},
	publisher = {National Bureau of Economic Research},
	author = {Acemoglu, Daron and Restrepo, Pascual},
	month = jun,
	year = {2021},
	doi = {10.3386/w28920},
}

@book{mokyr_gifts_2005,
	address = {Princeton N.J},
	title = {The {Gifts} of {Athena}: {Historical} {Origins} of the {Knowledge} {Economy}},
	isbn = {978-0-691-09483-0},
	shorttitle = {The {Gifts} of {Athena}},
	language = {English},
	publisher = {Princeton University Press},
	author = {Mokyr, Joel},
	year = {2005},
}

@article{lucas_mechanics_1988,
	title = {On the mechanics of economic development},
	volume = {22},
	issn = {0304-3932},
	url = {https://www.sciencedirect.com/science/article/pii/0304393288901687},
	doi = {10.1016/0304-3932(88)90168-7},
	number = {1},
	urldate = {2025-06-19},
	journal = {Journal of Monetary Economics},
	author = {Lucas, Robert E.},
	month = jul,
	year = {1988},
	pages = {3--42},
}

@article{swan_economic_1956,
	title = {Economic growth and capital accumulation},
	volume = {32},
	issn = {1475-4932},
	language = {en},
	number = {2},
	urldate = {2025-06-19},
	journal = {Economic Record},
	author = {Swan, T. W.},
	year = {1956},
	pages = {334--361},
}

@article{romer_increasing_1986,
	title = {Increasing {Returns} and {Long}-{Run} {Growth}},
	volume = {94},
	issn = {0022-3808},
	url = {https://www.journals.uchicago.edu/doi/10.1086/261420},
	doi = {10.1086/261420},
	number = {5},
	urldate = {2025-06-19},
	journal = {Journal of Political Economy},
	author = {Romer, Paul M.},
	month = oct,
	year = {1986},
	pages = {1002--1037},
}

@article{jones_r_1995,
	title = {R \& {D}-{Based} {Models} of {Economic} {Growth}},
	volume = {103},
	issn = {0022-3808},
	number = {4},
	urldate = {2025-06-19},
	journal = {Journal of Political Economy},
	author = {Jones, Charles I.},
	year = {1995},
	pages = {759--784},
}

@book{clark_farewell_2010,
	address = {Princeton, N.J.},
	title = {A {Farewell} to {Alms}: {A} {Brief} {Economic} {History} of the {World}},
	isbn = {978-0-691-14128-2},
	shorttitle = {A {Farewell} to {Alms}},
	language = {English},
	publisher = {Princeton University Press},
	author = {Clark, Gregory},
	year = {2010},
}

@book{warsh_knowledge_2007,
	address = {New York, NY},
	title = {Knowledge and the {Wealth} of {Nations}: {A} {Story} of {Economic} {Discovery}},
	isbn = {978-0-393-32988-9},
	shorttitle = {Knowledge and the {Wealth} of {Nations}},
	language = {English},
	publisher = {W. W. Norton \& Company},
	author = {Warsh, David},
	year = {2007},
}

@book{acemoglu_power_2024,
	address = {New York},
	title = {Power and {Progress}},
	isbn = {978-1-5417-0254-7},
	language = {English},
	publisher = {PublicAffairs},
	author = {Acemoglu, Daron and Johnson, Simon},
	year = {2024},
}

@inproceedings{vaswani_attention_2017,
	title = {Attention is {All} you {Need}},
	volume = {30},
	url = {https://proceedings.neurips.cc/paper/2017/hash/3f5ee243547dee91fbd053c1c4a845aa-Abstract.html},
	urldate = {2025-06-20},
	booktitle = {Advances in {Neural} {Information} {Processing} {Systems}},
	publisher = {Curran Associates, Inc.},
	author = {Vaswani, Ashish and Shazeer, Noam and Parmar, Niki and Uszkoreit, Jakob and Jones, Llion and Gomez, Aidan N and Kaiser, Lukasz and Polosukhin, Illia},
	year = {2017},
}

@article{bloom_are_2020,
	title = {Are {Ideas} {Getting} {Harder} to {Find}?},
	volume = {110},
	issn = {0002-8282},
	url = {https://www.aeaweb.org/articles?id=10.1257/aer.20180338},
	doi = {10.1257/aer.20180338},
	language = {en},
	number = {4},
	urldate = {2025-06-20},
	journal = {American Economic Review},
	author = {Bloom, Nicholas and Jones, Charles I. and Van Reenen, John and Webb, Michael},
	month = apr,
	year = {2020},
	pages = {1104--1144},
}

@article{firooz_automation_2025,
	title = {Automation and the rise of superstar firms},
	volume = {151},
	issn = {0304-3932},
	url = {https://www.sciencedirect.com/science/article/pii/S0304393225000042},
	doi = {10.1016/j.jmoneco.2025.103733},
	urldate = {2025-06-21},
	journal = {Journal of Monetary Economics},
	author = {Firooz, Hamid and Liu, Zheng and Wang, Yajie},
	month = apr,
	year = {2025},
	pages = {103733},
}

@article{acemoglu_why_1998,
	title = {Why {Do} {New} {Technologies} {Complement} {Skills}? {Directed} {Technical} {Change} and {Wage} {Inequality}},
	volume = {113},
	issn = {0033-5533},
	shorttitle = {Why {Do} {New} {Technologies} {Complement} {Skills}?},
	url = {https://doi.org/10.1162/003355398555838},
	doi = {10.1162/003355398555838},
	number = {4},
	urldate = {2025-06-22},
	journal = {The Quarterly Journal of Economics},
	author = {Acemoglu, Daron},
	month = nov,
	year = {1998},
	pages = {1055--1089},
}

@article{romer_endogenous_1990,
	title = {Endogenous {Technological} {Change}},
	volume = {98},
	issn = {0022-3808},
	url = {https://www.jstor.org/stable/2937632},
	number = {5},
	urldate = {2025-06-23},
	journal = {Journal of Political Economy},
	author = {Romer, Paul M.},
	year = {1990},
	pages = {S71--S102},
}

@article{wang_scientific_2023,
	title = {Scientific discovery in the age of artificial intelligence},
	volume = {620},
	copyright = {2023 Springer Nature Limited},
	issn = {1476-4687},
	url = {https://www.nature.com/articles/s41586-023-06221-2},
	doi = {10.1038/s41586-023-06221-2},
	language = {en},
	number = {7972},
	urldate = {2025-06-23},
	journal = {Nature},
	author = {Wang, Hanchen and Fu, Tianfan and Du, Yuanqi and Gao, Wenhao and Huang, Kexin and Liu, Ziming and Chandak, Payal and Liu, Shengchao and Van Katwyk, Peter and Deac, Andreea and Anandkumar, Anima and Bergen, Karianne and Gomes, Carla P. and Ho, Shirley and Kohli, Pushmeet and Lasenby, Joan and Leskovec, Jure and Liu, Tie-Yan and Manrai, Arjun and Marks, Debora and Ramsundar, Bharath and Song, Le and Sun, Jimeng and Tang, Jian and Veličković, Petar and Welling, Max and Zhang, Linfeng and Coley, Connor W. and Bengio, Yoshua and Zitnik, Marinka},
	month = aug,
	year = {2023},
	pages = {47--60},
}

@misc{nordhaus_are_2015,
	address = {Rochester, NY},
	type = {{SSRN} {Scholarly} {Paper}},
	title = {Are {We} {Approaching} an {Economic} {Singularity}? {Information} {Technology} and the {Future} of {Economic} {Growth}},
	shorttitle = {Are {We} {Approaching} an {Economic} {Singularity}?},
	url = {https://papers.ssrn.com/abstract=2658259},
	doi = {10.2139/ssrn.2658259},
	language = {en},
	urldate = {2025-06-23},
	publisher = {Social Science Research Network},
	author = {Nordhaus, William D.},
	month = sep,
	year = {2015},
}

@article{weitzman_recombinant_1998,
	title = {Recombinant {Growth}},
	volume = {113},
	issn = {0033-5533},
	url = {https://dx.doi.org/10.1162/003355398555595},
	doi = {10.1162/003355398555595},
	language = {en},
	number = {2},
	urldate = {2025-06-23},
	journal = {The Quarterly Journal of Economics},
	author = {Weitzman, Martin L.},
	month = may,
	year = {1998},
	pages = {331--360},
}

@article{karabarbounis_global_2014,
	title = {The {Global} {Decline} of the {Labor} {Share}},
	volume = {129},
	issn = {0033-5533},
	url = {https://doi.org/10.1093/qje/qjt032},
	doi = {10.1093/qje/qjt032},
	number = {1},
	urldate = {2025-06-23},
	journal = {The Quarterly Journal of Economics},
	author = {Karabarbounis, Loukas and Neiman, Brent},
	month = feb,
	year = {2014},
	pages = {61--103},
}

@techreport{oecd_framework_2024,
	title = {Framework for {Anticipatory} {Governance} of {Emerging} {Technologies}},
	url = {https://www.oecd.org/en/publications/framework-for-anticipatory-governance-of-emerging-technologies_0248ead5-en.html},
	language = {en},
	number = {165},
	urldate = {2025-06-24},
	institution = {OECD},
	author = {{OECD}},
	month = apr,
	year = {2024},
}

@techreport{ilo_economics_2018,
	title = {The economics of artificial intelligence: {Implications} for the future of work},
	shorttitle = {The economics of artificial intelligence},
	url = {https://www.ilo.org/publications/economics-artificial-intelligence-implications-future-work},
	language = {en},
	urldate = {2025-06-24},
	institution = {ILO},
	author = {{ILO}},
	month = oct,
	year = {2018},
}

@incollection{schwab_fourth_2024,
	title = {The {Fourth} {Industrial} {Revolution}: what it means, how to respond},
	isbn = {978-1-80220-881-8},
	shorttitle = {The {Fourth} {Industrial} {Revolution}},
	url = {https://www.elgaronline.com/edcollchap/book/9781802208818/book-part-9781802208818-8.xml},
	language = {eng},
	urldate = {2025-06-24},
	booktitle = {Handbook of {Research} on {Strategic} {Leadership} in the {Fourth} {Industrial} {Revolution}},
	publisher = {Edward Elgar Publishing},
	author = {Schwab, Klaus},
	month = jul,
	year = {2024},
	note = {Section: Handbook of Research on Strategic Leadership in the Fourth Industrial Revolution},
	pages = {29--34},
}

@techreport{cazzaniga_gen-ai_2024,
	type = {Working paper},
	title = {Gen-{AI}: {Artificial} {Intelligence} and the {Future} of {Work}},
	shorttitle = {Gen-{AI}},
	url = {https://EconPapers.repec.org/RePEc:imf:imfsdn:2024/001},
	number = {2024/001},
	urldate = {2025-06-24},
	institution = {International Monetary Fund},
	author = {Cazzaniga, Mauro and Jaumotte, Florence and Li, Longji and Melina, Giovanni and Panton, Augustus and Pizzinelli, Carlo and Rockall, Emma and Tavares, Marina},
	month = jan,
	year = {2024},
}

@article{autor_fall_2020,
	title = {The {Fall} of the {Labor} {Share} and the {Rise} of {Superstar} {Firms}},
	volume = {135},
	issn = {0033-5533},
	url = {https://doi.org/10.1093/qje/qjaa004},
	doi = {10.1093/qje/qjaa004},
	number = {2},
	urldate = {2025-06-24},
	journal = {The Quarterly Journal of Economics},
	author = {Autor, David and Dorn, David and Katz, Lawrence F and Patterson, Christina and Van Reenen, John},
	month = may,
	year = {2020},
	pages = {645--709},
}

@article{solow_contribution_1956,
	title = {A {Contribution} to the {Theory} of {Economic} {Growth}},
	volume = {70},
	issn = {0033-5533},
	url = {https://www.jstor.org/stable/1884513},
	doi = {10.2307/1884513},
	number = {1},
	urldate = {2025-06-24},
	journal = {The Quarterly Journal of Economics},
	author = {Solow, Robert M.},
	year = {1956},
	note = {Publisher: Oxford University Press},
	pages = {65--94},
}

@book{acemoglu_introduction_2008,
	address = {New Jersey},
	title = {Introduction to {Modern} {Economic} {Growth}},
	isbn = {978-1-4008-3577-5},
	language = {Anglais},
	publisher = {Princeton University Press},
	author = {Acemoglu, Daron},
	year = {2008},
}

@article{aghion_model_1992,
	title = {A {Model} of {Growth} {Through} {Creative} {Destruction}},
	volume = {60},
	issn = {0012-9682},
	url = {https://www.jstor.org/stable/2951599},
	doi = {10.2307/2951599},
	number = {2},
	urldate = {2025-06-25},
	journal = {Econometrica},
	author = {Aghion, Philippe and Howitt, Peter},
	year = {1992},
	pages = {323--351},
}

@book{ljungqvist_recursive_2018,
	address = {Cambridge, Massachusetts},
	title = {Recursive {Macroeconomic} {Theory}, fourth edition},
	isbn = {978-0-262-03866-9},
	language = {Anglais},
	publisher = {The MIT Press},
	author = {Ljungqvist, Lars and Sargent, Thomas J.},
	year = {2018},
}

@article{adjemian_dynare_2021,
	title = {Dynare: {Reference} {Manual} {Version} 4},
	shorttitle = {Dynare},
	url = {https://ideas.repec.org//p/cpm/dynare/001.html},
	language = {en},
	urldate = {2025-06-25},
	journal = {Dynare Working Papers},
	author = {Adjemian, Stéphane and Bastani, Houtan and Juillard, Michel and Karamé, Fréderic and Maih, Junior and Mihoubi, Ferhat and Mutschler, Willi and Perendia, George and Pfeifer, Johannes and Ratto, Marco and Villemot, Sébastien},
	month = mar,
	year = {2021},
}

@article{rumelhart_learning_1986,
	title = {Learning representations by back-propagating errors},
	volume = {323},
	copyright = {1986 Springer Nature Limited},
	issn = {1476-4687},
	url = {https://www.nature.com/articles/323533a0},
	doi = {10.1038/323533a0},
	language = {en},
	number = {6088},
	urldate = {2025-07-01},
	journal = {Nature},
	author = {Rumelhart, David E. and Hinton, Geoffrey E. and Williams, Ronald J.},
	month = oct,
	year = {1986},
	pages = {533--536},
}

@book{james_introduction_2013,
	address = {Cham, Switzerland},
	title = {An {Introduction} to {Statistical} {Learning}},
	isbn = {978-3-031-39189-7},
	shorttitle = {An {Introduction} to {Statistical} {Learning}},
	language = {Anglais},
	publisher = {Springer International Publishing AG},
	author = {James, Gareth and Witten, Daniela and Hastie, Trevor and Tibshirani, Robert and Taylor, Jonathan},
	year = {2013},
}

@book{bishop_pattern_2006,
	address = {New York},
	title = {Pattern {Recognition} and {Machine} {Learning}},
	isbn = {978-0-387-31073-2},
	language = {Anglais},
	publisher = {Springer-Verlag New York Inc.},
	author = {Bishop, Christopher M.},
	year = {2006},
}

@book{todaro_economic_2020,
	address = {Hoboken},
	title = {Economic {Development}},
	isbn = {978-1-292-29115-4},
	language = {English},
	publisher = {Pearson},
	author = {Todaro, Michael and Smith, Stephen},
	year = {2020},
}

@article{rodrik_premature_2016,
	title = {Premature deindustrialization},
	volume = {21},
	issn = {1573-7020},
	url = {https://doi.org/10.1007/s10887-015-9122-3},
	doi = {10.1007/s10887-015-9122-3},
	language = {en},
	number = {1},
	urldate = {2025-07-05},
	journal = {Journal of Economic Growth},
	author = {Rodrik, Dani},
	month = mar,
	year = {2016},
	pages = {1--33},
}

@incollection{herrendorf_chapter_2014,
	series = {Handbook of {Economic} {Growth}},
	title = {Chapter 6 - {Growth} and {Structural} {Transformation}},
	volume = {2},
	url = {https://www.sciencedirect.com/science/article/pii/B9780444535405000069},
	urldate = {2025-07-05},
	booktitle = {Handbook of {Economic} {Growth}},
	publisher = {Elsevier},
	author = {Herrendorf, Berthold and Rogerson, Richard and Valentinyi, Akos},
	editor = {Aghion, Philippe and Durlauf, Steven N.},
	month = jan,
	year = {2014},
	pages = {855--941},
}

@article{lewis_economic_1954,
	title = {Economic {Development} with {Unlimited} {Supplies} of {Labour}},
	volume = {22},
	issn = {1467-9957},
	url = {https://onlinelibrary.wiley.com/doi/abs/10.1111/j.1467-9957.1954.tb00021.x},
	doi = {10.1111/j.1467-9957.1954.tb00021.x},
	language = {en},
	number = {2},
	urldate = {2025-07-05},
	journal = {The Manchester School},
	author = {Lewis, W. Arthur},
	year = {1954},
	pages = {139--191},
}

@article{hidalgo_product_2007,
	title = {The {Product} {Space} {Conditions} the {Development} of {Nations}},
	volume = {317},
	url = {https://www.science.org/doi/10.1126/science.1144581},
	doi = {10.1126/science.1144581},
	number = {5837},
	urldate = {2025-07-05},
	journal = {Science},
	author = {Hidalgo, C. A. and Klinger, B. and Barabási, A.-L. and Hausmann, R.},
	month = jul,
	year = {2007},
	pages = {482--487},
}

@book{rostow_stages_1991,
	address = {Cambridge},
	edition = {3rd},
	title = {The {Stages} of {Economic} {Growth}: {A} {Non}-{Communist} {Manifesto}},
	isbn = {978-0-521-40070-1},
	shorttitle = {The {Stages} of {Economic} {Growth}},
	url = {https://www.cambridge.org/core/books/stages-of-economic-growth/9CB46055035A1915509CE15A57848A07},
	urldate = {2025-07-05},
	publisher = {Cambridge University Press},
	author = {Rostow, W. W.},
	year = {1991},
}

@incollection{harrod_essay_1972,
	address = {London},
	title = {An {Essay} in {Dynamic} {Theory}},
	isbn = {978-1-349-01494-1},
	url = {https://doi.org/10.1007/978-1-349-01494-1_13},
	language = {en},
	urldate = {2025-07-05},
	booktitle = {Economic {Essays}},
	publisher = {Palgrave Macmillan UK},
	author = {Harrod, Roy},
	editor = {Harrod, Roy},
	year = {1972},
	doi = {10.1007/978-1-349-01494-1_13},
	pages = {254--277},
}

@article{domar_capital_1946,
	title = {Capital {Expansion}, {Rate} of {Growth}, and {Employment}},
	volume = {14},
	issn = {0012-9682},
	url = {https://www.jstor.org/stable/1905364},
	doi = {10.2307/1905364},
	number = {2},
	urldate = {2025-07-05},
	journal = {Econometrica},
	author = {Domar, Evsey D.},
	year = {1946},
	pages = {137--147},
}

@article{farrell_weaponized_2019,
	title = {Weaponized {Interdependence}: {How} {Global} {Economic} {Networks} {Shape} {State} {Coercion}},
	volume = {44},
	issn = {0162-2889},
	shorttitle = {Weaponized {Interdependence}},
	url = {https://doi.org/10.1162/isec_a_00351},
	doi = {10.1162/isec_a_00351},
	number = {1},
	urldate = {2025-07-05},
	journal = {International Security},
	author = {Farrell, Henry and Newman, Abraham L.},
	month = jul,
	year = {2019},
	pages = {42--79},
}

@misc{brynjolfsson_artificial_2017,
	type = {Working {Paper}},
	series = {Working {Paper} {Series}},
	title = {Artificial {Intelligence} and the {Modern} {Productivity} {Paradox}: {A} {Clash} of {Expectations} and {Statistics}},
	shorttitle = {Artificial {Intelligence} and the {Modern} {Productivity} {Paradox}},
	url = {https://www.nber.org/papers/w24001},
	doi = {10.3386/w24001},
	urldate = {2025-07-05},
	publisher = {National Bureau of Economic Research},
	author = {Brynjolfsson, Erik and Rock, Daniel and Syverson, Chad},
	month = nov,
	year = {2017},
	doi = {10.3386/w24001},
}

@techreport{imf_world_2025,
	title = {World {Economic} {Outlook}, {April} 2025: {A} {Critical} {Juncture} amid {Policy} {Shifts}},
	shorttitle = {World {Economic} {Outlook}, {April} 2025},
	url = {https://www.imf.org/en/Publications/WEO/Issues/2025/04/22/world-economic-outlook-april-2025},
	language = {ENG},
	urldate = {2025-07-06},
	institution = {IMF},
	author = {{IMF}},
	month = apr,
	year = {2025},
}

@article{griliches_hybrid_1957,
	title = {Hybrid {Corn}: {An} {Exploration} in the {Economics} of {Technological} {Change} - {The} {Econometric} {Society}},
	volume = {25},
	shorttitle = {Hybrid {Corn}},
	url = {http://www.econometricsociety.org/publications/econometrica/1957/10/01/hybrid-corn-exploration-economics-technological-change},
	language = {en},
	number = {4},
	urldate = {2025-07-09},
	journal = {Econometric Society},
	author = {Griliches, Zvi},
	year = {1957},
	pages = {501--522},
}

@article{nelson_technical_1993,
	title = {Technical {Innovation} and {National} {Systems}},
	url = {https://cir.nii.ac.jp/crid/1360581630976408320},
	doi = {10.1093/oso/9780195076165.003.0001},
	urldate = {2025-07-09},
	journal = {National Innovation Systems},
	author = {Nelson, Richard R. and Nathan, Rosenberg},
	month = jun,
	year = {1993},
	note = {Publisher: Oxford University PressNew York, NY},
	pages = {3--22},
}

@article{gomes_eagle_2012,
	title = {The {EAGLE}. {A} model for policy analysis of macroeconomic interdependence in the euro area},
	volume = {29},
	issn = {0264-9993},
	url = {https://www.sciencedirect.com/science/article/pii/S0264999312000958},
	doi = {10.1016/j.econmod.2012.04.002},
	number = {5},
	urldate = {2025-07-09},
	journal = {Economic Modelling},
	author = {Gomes, S. and Jacquinot, P. and Pisani, M.},
	month = sep,
	year = {2012},
	pages = {1686--1714},
}

@techreport{anderson_getting_2013,
	title = {Getting to {Know} {GIMF}: {The} {Simulation} {Properties} of the {Global} {Integrated} {Monetary} and {Fiscal} {Model}},
	shorttitle = {Getting to {Know} {GIMF}},
	language = {ENG},
	urldate = {2025-07-09},
	institution = {IMF},
	author = {Anderson, Derek and Hunt, Benjamin L. and Kortelainen, Mika and Kumhof, Michael and Laxton, Douglas and Muir, Dirk V. and Mursula, Susanna and Snudden, Stephen},
	year = {2013},
}

@techreport{oecd_blueprint_2023-1,
	title = {A blueprint for building national compute capacity for artificial intelligence},
	url = {https://www.oecd.org/en/publications/a-blueprint-for-building-national-compute-capacity-for-artificial-intelligence_876367e3-en.html},
	number = {350},
	urldate = {2025-07-11},
	institution = {OECD},
	author = {{OECD}},
	month = feb,
	year = {2023},
}

@misc{tritsaris_interdisciplinary_2019,
	title = {Interdisciplinary collaboration in research networks: {Empirical} analysis of energy-related research in {Greece}},
	shorttitle = {Interdisciplinary collaboration in research networks},
	url = {http://arxiv.org/abs/1805.04882},
	doi = {10.48550/arXiv.1805.04882},
	urldate = {2025-07-12},
	publisher = {arXiv},
	author = {Tritsaris, Georgios A. and Siddiqi, Afreen},
	month = mar,
	year = {2019},
	note = {arXiv:1805.04882 [cs]},
}

@article{barro_convergence_1991,
	title = {Convergence {Across} {States} and {Regions}},
	volume = {1991},
	issn = {0007-2303},
	url = {https://www.jstor.org/stable/2534639},
	doi = {10.2307/2534639},
	number = {1},
	urldate = {2025-07-12},
	journal = {Brookings Papers on Economic Activity},
	author = {Barro, Robert J. and Sala-I-Martin, Xavier and Blanchard, Olivier Jean and Hall, Robert E.},
	year = {1991},
	pages = {107--182},
}

@incollection{arrow_economic_1972,
	address = {London},
	title = {Economic {Welfare} and the {Allocation} of {Resources} for {Invention}},
	isbn = {978-1-349-15486-9},
	url = {https://doi.org/10.1007/978-1-349-15486-9_13},
	language = {en},
	urldate = {2025-07-12},
	booktitle = {Readings in {Industrial} {Economics}: {Volume} {Two}: {Private} {Enterprise} and {State} {Intervention}},
	publisher = {Macmillan Education UK},
	author = {Arrow, K. J.},
	editor = {Rowley, Charles K.},
	year = {1972},
	pages = {219--236},
}

@article{stiglitz_markets_1989,
	title = {Markets, {Market} {Failures}, and {Development}},
	volume = {79},
	issn = {0002-8282},
	url = {https://www.jstor.org/stable/1827756},
	number = {2},
	urldate = {2025-07-12},
	journal = {The American Economic Review},
	author = {Stiglitz, Joseph E.},
	year = {1989},
	pages = {197--203},
}

@article{comin_exploration_2010,
	title = {An {Exploration} of {Technology} {Diffusion}},
	volume = {100},
	issn = {0002-8282},
	url = {https://www.aeaweb.org/articles?id=10.1257/aer.100.5.2031},
	doi = {10.1257/aer.100.5.2031},
	language = {en},
	number = {5},
	urldate = {2025-07-12},
	journal = {American Economic Review},
	author = {Comin, Diego and Hobijn, Bart},
	month = dec,
	year = {2010},
	pages = {2031--2059},
}

@book{banerjee_poor_2011,
	title = {Poor economics: {A} radical rethinking of the way to fight global poverty},
	shorttitle = {Poor economics},
	url = {https://scholar.google.com/scholar?cluster=6324759133315472718&hl=en&oi=scholarr},
	urldate = {2025-07-12},
	publisher = {Public Affairs},
	author = {Banerjee, Abhijit V. and Duflo, Esther},
	year = {2011},
}

@book{kuznets_national_1941,
	title = {National {Income} and {Its} {Composition}, 1919-1938, {Volume} {I}},
	url = {https://www.nber.org/books-and-chapters/national-income-and-its-composition-1919-1938-volume-i},
	urldate = {2025-07-13},
	publisher = {NBER},
	author = {Kuznets, Simon and Epstein, Lillian and Jenks, Elizabeth},
	year = {1941},
}
\bibliographystyle{apalike}
}

\appendix
\chapter{Numerical Results}
Additional numerical results for Chapter~\ref{ch:modeling} are provided below:

\begin{figure}[H]
    \centering
    \includegraphics[width=1\linewidth]{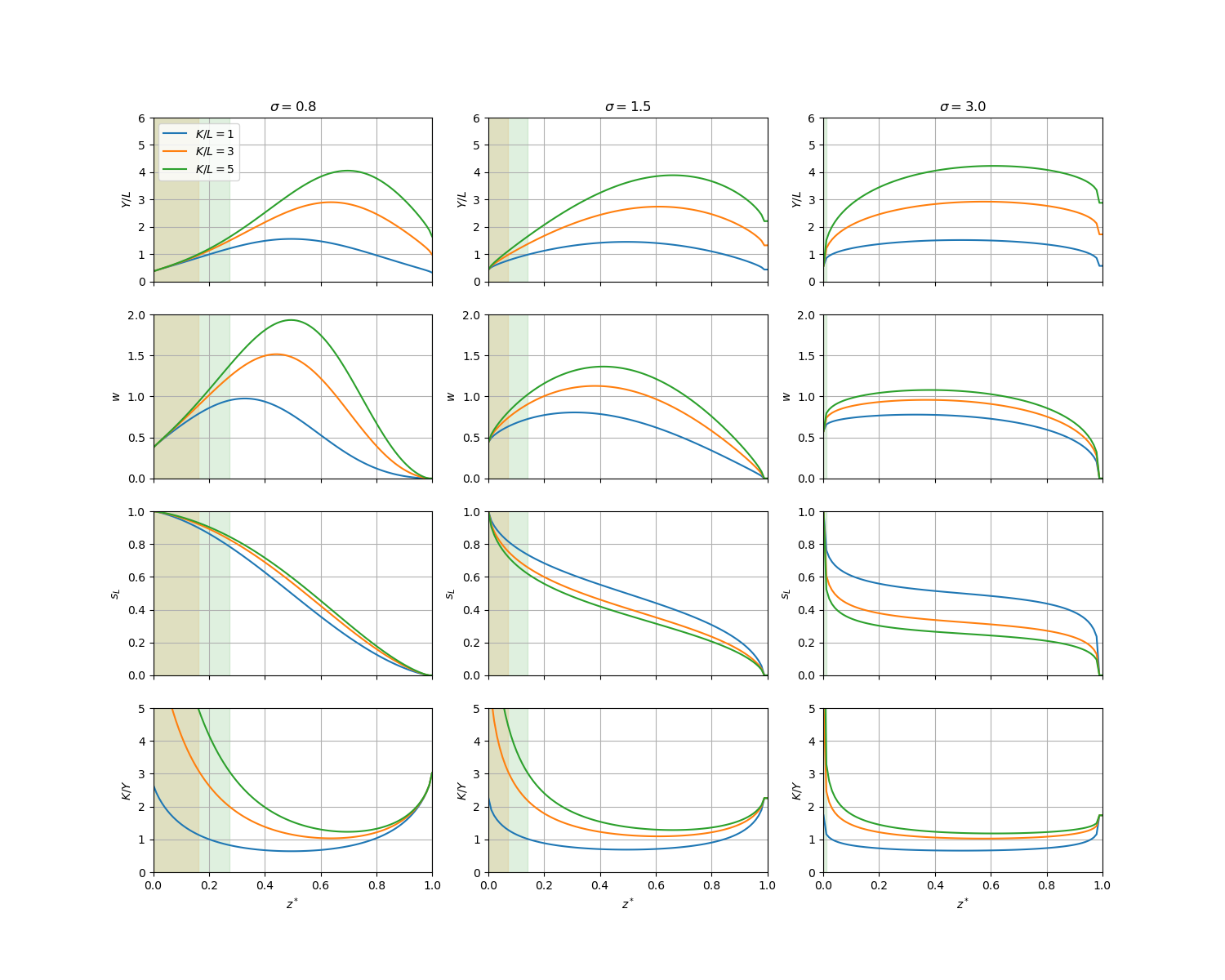}
    \caption{Numerical calculation of key variables of the baseline model.}
    \label{fig:1statics_2}
\end{figure}

\begin{figure}
    \centering
    \includegraphics[width=1\linewidth]{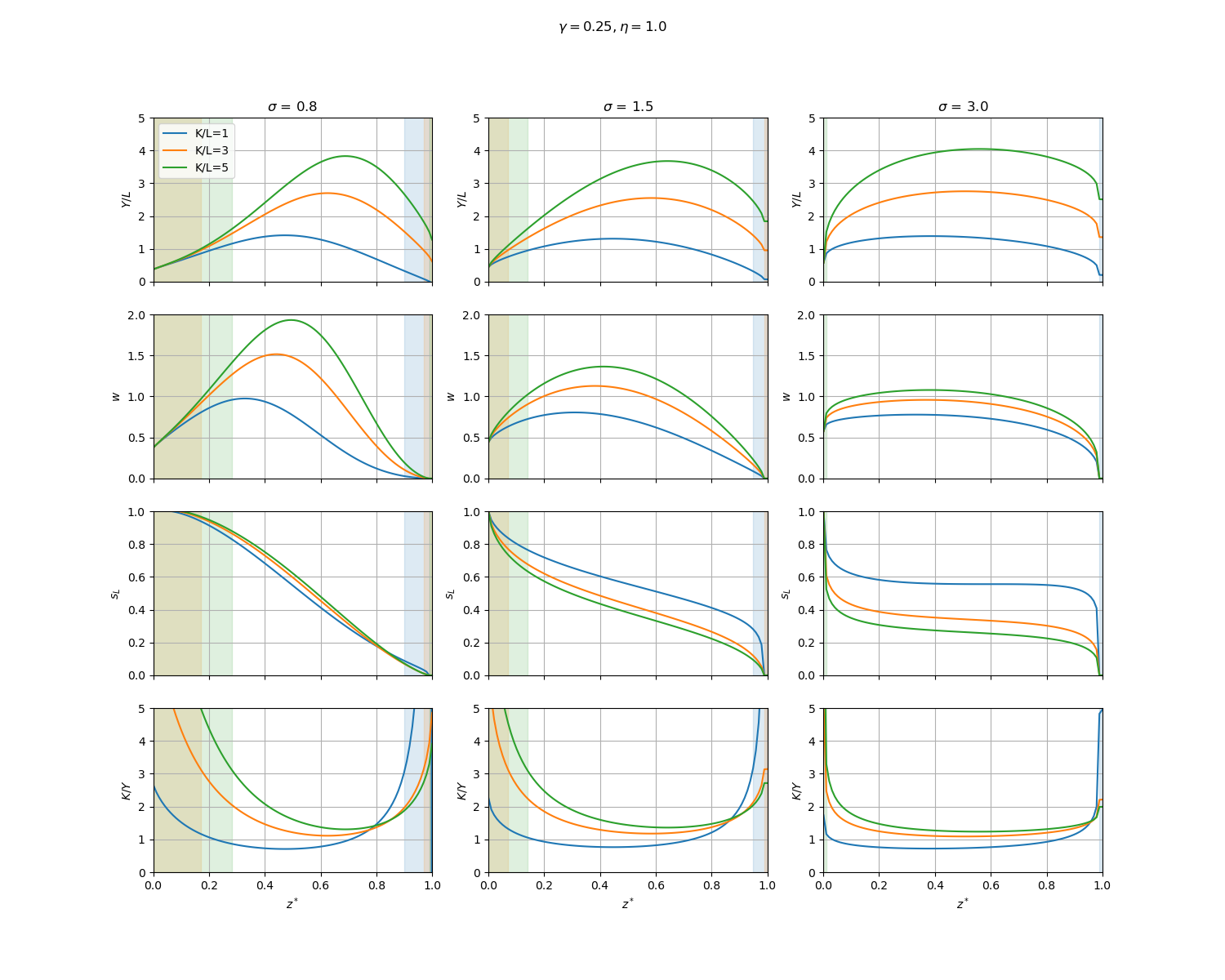}
    \caption{Numerical calculation of key variables of the model with frictions.}
    \label{fig:2statics_2}
\end{figure}

\begin{figure}
    \centering
    \includegraphics[width=1\linewidth]{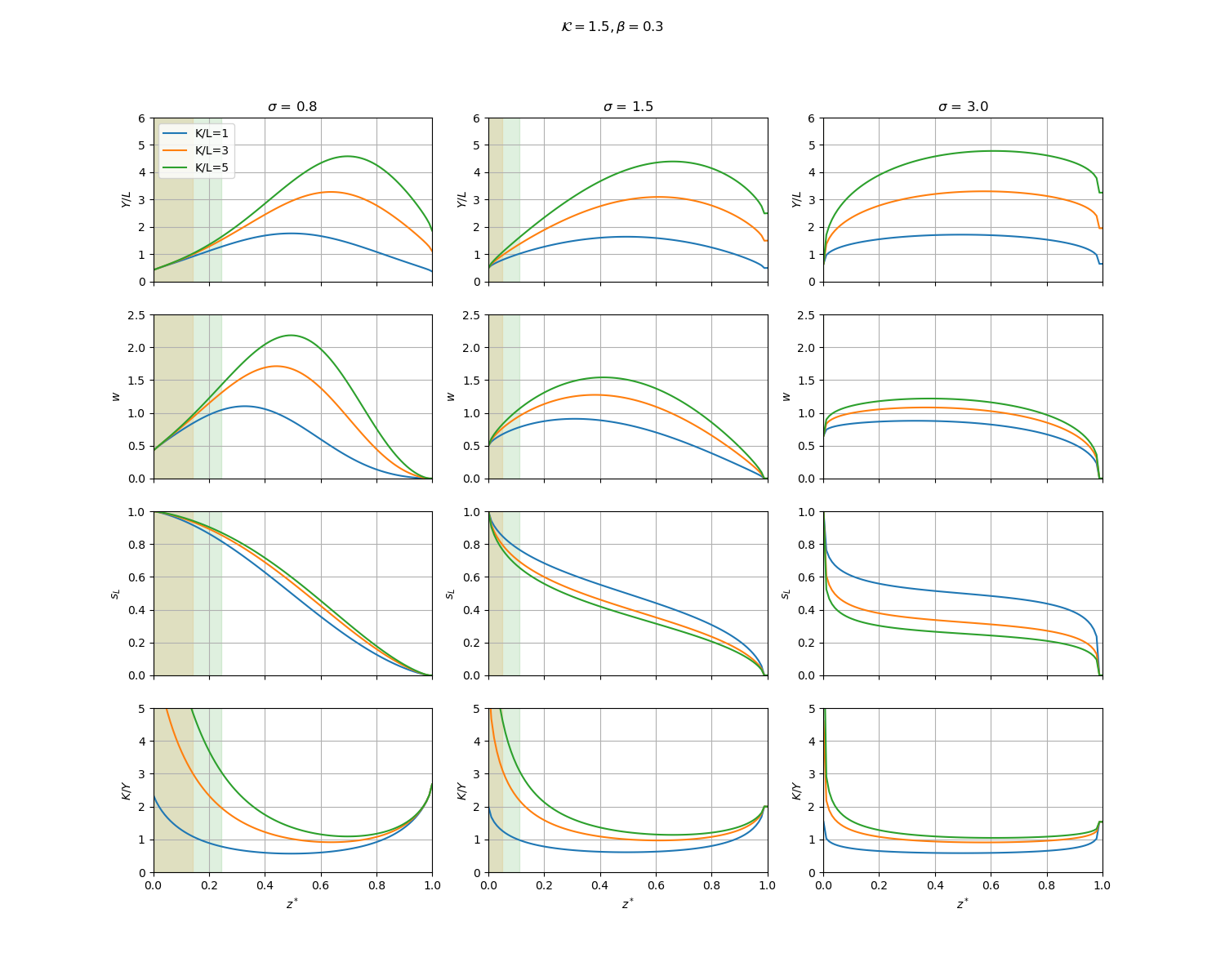}
    \caption{Numerical calculation of key variables of the model with knowledge accumulation.}
    \label{fig:3statics_2}
\end{figure}

\begin{figure}
    \centering
    \includegraphics[width=1\linewidth]{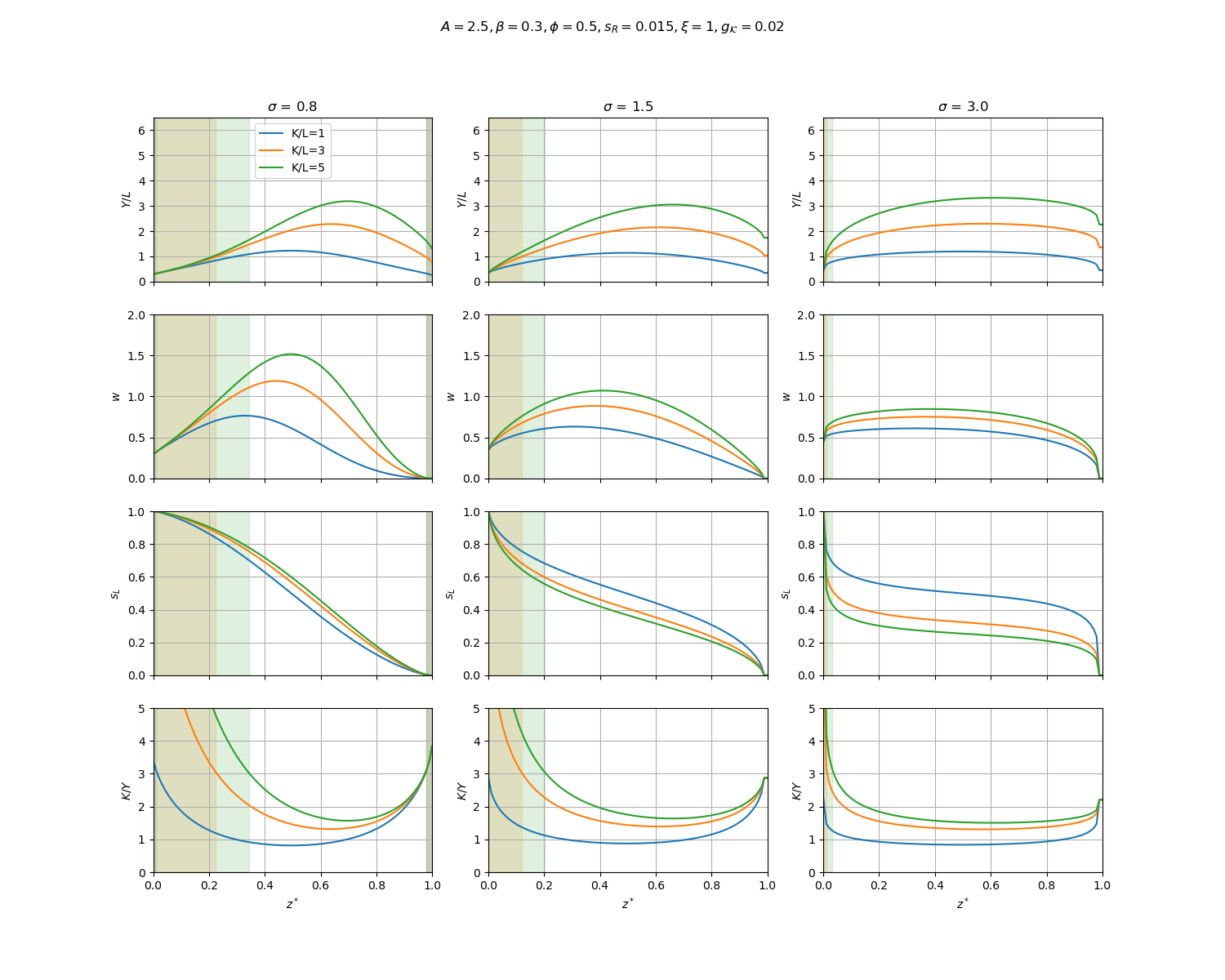}
    \caption{Numerical calculation of key variables of the model with GPT.}
    \label{fig:4statics_2}
\end{figure}

\begin{figure}
    \centering
    \includegraphics[width=1\linewidth]{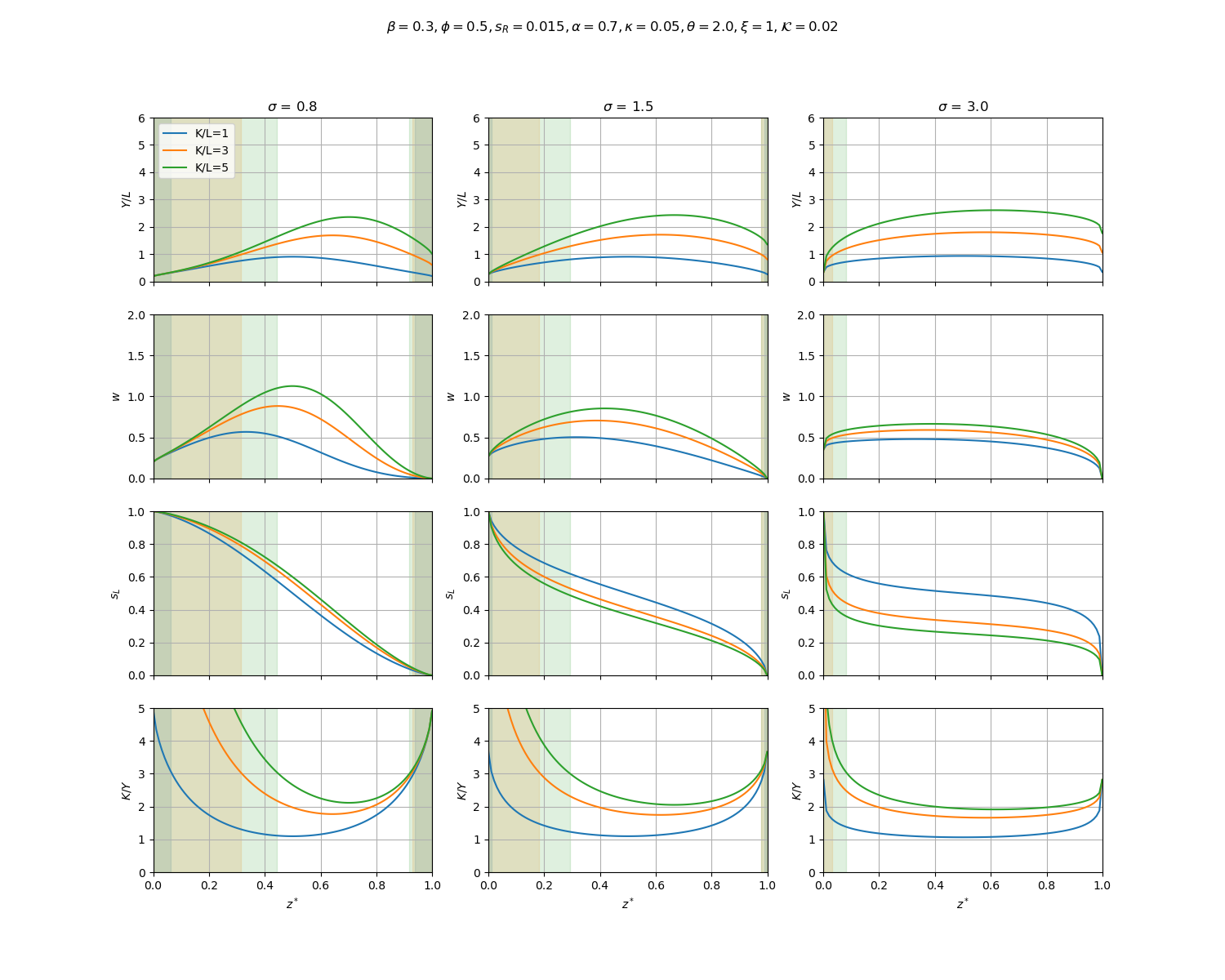}
    \caption{Numerical calculation of key variables of the model with GPT and knowledge accumulation costs.}
    \label{fig:5statics_2}
\end{figure}

\begin{figure}
    \centering
    \includegraphics[width=1\linewidth]{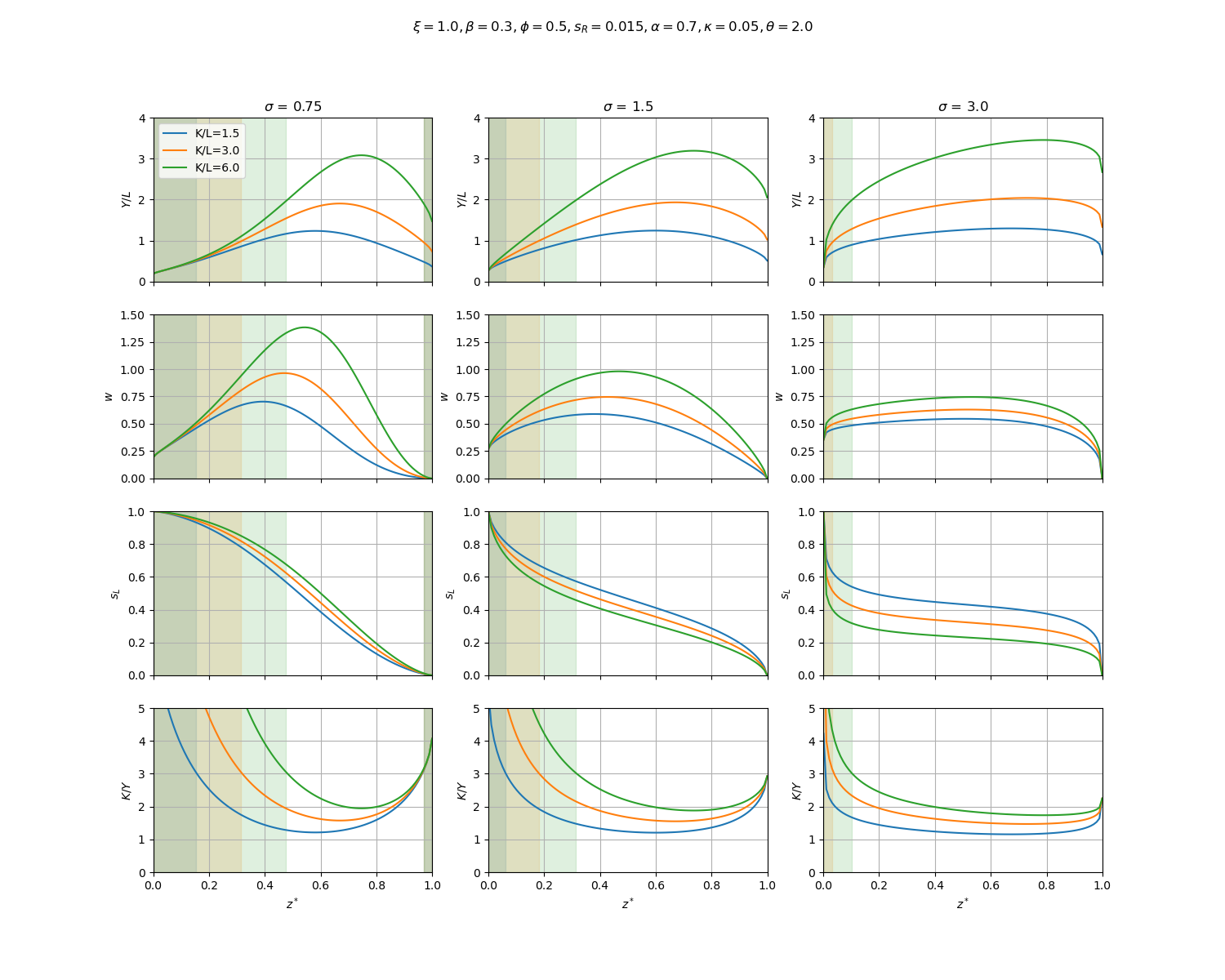}
    \caption{Numerical calculation of key variables of the model with GPT, knowledge accumulation costs, and adaptive knowledge generation.}
    \label{fig:6statics_2}
\end{figure}

\end{document}